\documentclass[aps,pre,twocolumn,floatfix,superscriptaddress]{revtex4-1}

\usepackage{amsmath}  
\usepackage{amsfonts} 
\usepackage{graphicx} 
\usepackage{gensymb}
\usepackage{physics}
\usepackage{xcolor}
\usepackage{soul}

\newcommand{\pres}{{(\text{pre})}}
\newcommand{\ext}{{(\text{ext})}}

\newcommand{\bph}{\boldsymbol{\hat \phi} }
\newcommand{\epsilonb}{\boldsymbol{\epsilon}}

\usepackage{tensor} 

\newcommand{\NN}{\text{NN}}
\newcommand{\NNN}{\text{NNN}}
\newcommand{\VV}{\mathcal{V}}

\newcommand{\ZZ}{\mathbb{Z}}
\newcommand{\BB}{\mathcal{B}}
\newcommand{\CC}{\mathcal{C}}
\newcommand{\GG}{\mathcal{G}}
\newcommand{\regA}{\mathcal{A}}
 
\newcommand{\sigmab}{\boldsymbol{\sigma}}
\newcommand{\taub}{\boldsymbol{\tau}}
\newcommand{\Pib}{\boldsymbol{\Pi}}

\newcommand{\scr}{r}
\newcommand{\scrb}{{\vb r}} 
\newcommand{\scrbh}{\vb{\hat r}}

\DeclareMathOperator{\Grad}{Grad}
\DeclareMathOperator{\Div}{Div} 

\DeclareMathOperator{\gradr}{grad}
\DeclareMathOperator{\divr}{div} 
\DeclareMathOperator{\curlr}{curl} 
\DeclareMathOperator{\Curl}{Curl}

\newcommand{\pressure}{\raisebox{-0.25\height}{\includegraphics{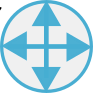}}}
\newcommand{\torque}{\raisebox{-0.25\height}{\includegraphics{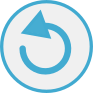}}}
\newcommand{\strone}{\raisebox{-0.25\height}{\includegraphics{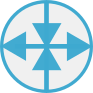}}}
\newcommand{\strtwo}{\raisebox{-0.25\height}{\includegraphics{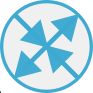}}}

\newcommand{\dilation}{\raisebox{-0.2\height}{\includegraphics{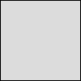}}}
\newcommand{\rotation}{\raisebox{-0.2\height}{\includegraphics{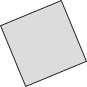}}}
\newcommand{\uone}{\raisebox{-0.15\height}{\includegraphics{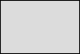}}}
\newcommand{\utwo}{\raisebox{-0.2\height}{\includegraphics{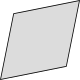}}}

\begin{document}

\title{
Topological defects in solids with odd elasticity
}

\author{Lara Braverman}
\thanks{These authors contributed equally to this work.}
\affiliation{James Franck Institute, The University of Chicago, Chicago, Illinois 60637, USA}
\affiliation{Department of Physics, The University of Chicago, Chicago, Illinois 60637, USA}

\author{Colin Scheibner}
\thanks{These authors contributed equally to this work.}
\affiliation{James Franck Institute, The University of Chicago, Chicago, Illinois 60637, USA}
\affiliation{Department of Physics, The University of Chicago, Chicago, Illinois 60637, USA}

\author{Bryan VanSaders}
\thanks{These authors contributed equally to this work.}
\affiliation{James Franck Institute, The University of Chicago, Chicago, Illinois 60637, USA}

\author{Vincenzo Vitelli}
\email{vitelli@uchicago.edu}
\affiliation{James Franck Institute, The University of Chicago, Chicago, Illinois 60637, USA}
\affiliation{Department of Physics, The University of Chicago, Chicago, Illinois 60637, USA}
\affiliation{Kadanoff Center for Theoretical Physics, The University of Chicago, Chicago, Illinois 60637, USA}

\begin{abstract}

Crystallography typically studies
collections of point particles whose interaction forces are the gradient of a potential. 
Lifting this assumption generically gives rise in the continuum limit to a form of elasticity with additional moduli known as odd elasticity.
We show that such odd elastic moduli modify the 
strain induced by topological defects and their interactions, even reversing the stability of, otherwise, bound dislocation pairs.
Beyond continuum theory, isolated dislocations can self propel via microscopic work cycles active at their cores that compete with conventional Peach-Koehler forces caused, for example, by an ambient torque density.  
We perform molecular dynamics simulations isolating active plastic processes and discuss their experimental relevance to solids composed of spinning particles, vortex-like objects, and robotic metamaterials.
\end{abstract}

\maketitle

\begin{figure}[t!]
    \centering
    \includegraphics[width=\columnwidth]{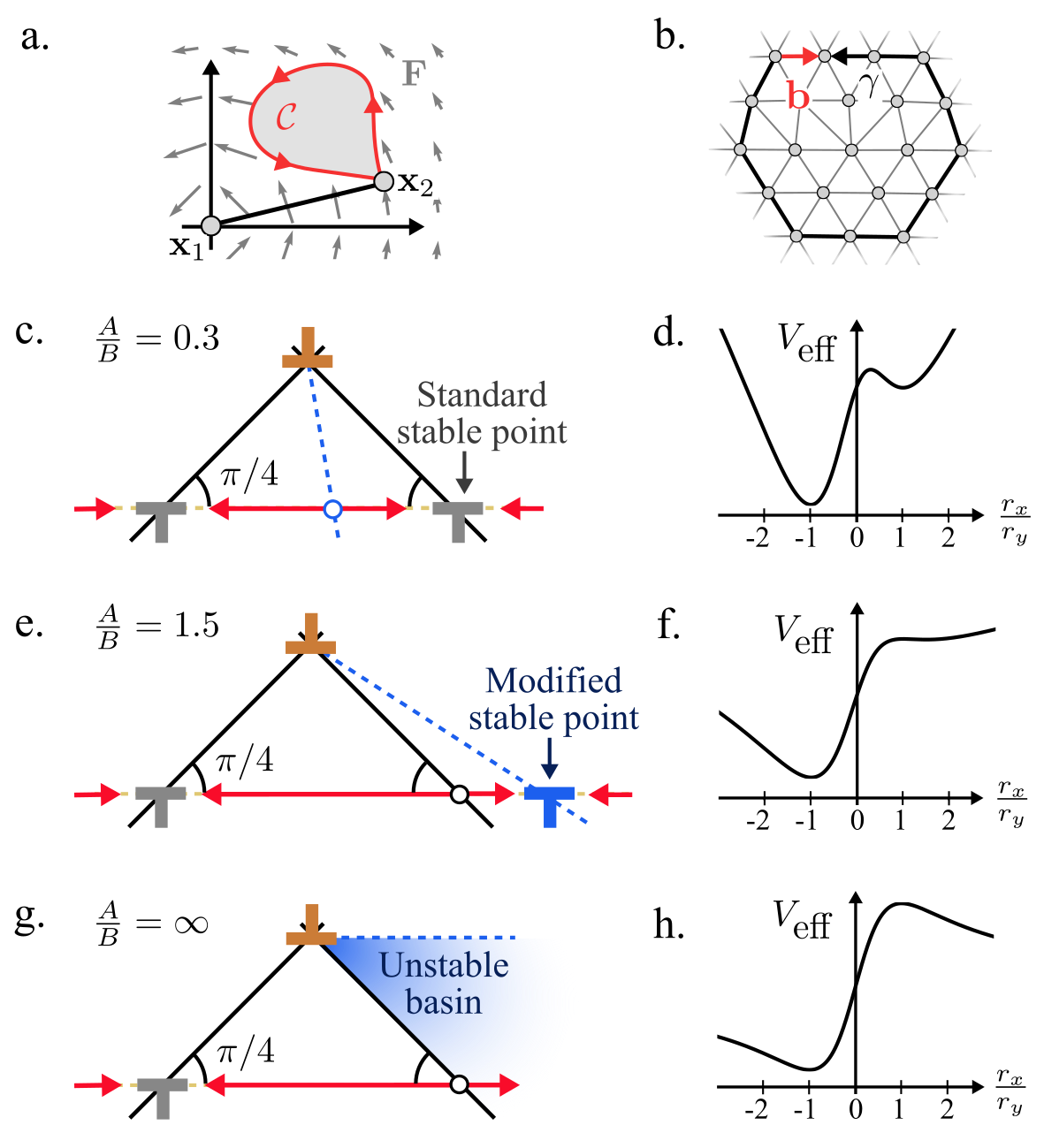}
    \caption{{\bf Odd elasticity modifies dislocation interactions and their stability.}~{\bf a.}~A particle at point $\vb x_1$ exerts a force $\vb F$ on a particle at point $\vb x_2$. This force is nonconservative, so nonzero work is done along the closed cycle $\mathcal{C}$. 
    {\bf b.}~A dislocation is defined by a Burgers vector $\vb b$ that represents the offset from what would otherwise be a closed loop $\gamma$. 
    {\bf c-d.}~ An orange dislocation is held stationary while a second anti-aligned dislocation is free to move along its glide plane subject to the Peach-Koehler force (red arrows). When $\abs{A/B}<1$, the free dislocation has two stable points located along rays forming an angle $\pi/4$ with the glide plane. 
    The effective potential $V_\text{eff}$ as a function of the horizontal ($r_x$) and vertical ($r_y$) distance between the dislocations.  
    {\bf e-f.}~When $A/B>1$, the rightmost stable equilibrium moves outward beyond $\pi/4$. {\bf g-h}.~When $A/B = \infty$, only one stable equilibrium position exists and the shaded region is an unstable basin.
    }
    \label{fig:edge}
\end{figure}

Topological defects are local singularities in an order parameter that have global consequences at large scales~\cite{Nelson2002, Weertman1964,Chaikin1995,Paulose2015,mietke2020anyonic, Pretko2018fracton,Bowick2009,Nelson1979,Wachtel2016Electrodynamic}. 
In active systems, topological defects exhibit distinctive properties such as self-propulsion or non-reciprocal interactions~\cite{shankar2020topological,Giomi2013,Duclos2020topological,shankar2020topological,Colen2021machine, vafa2020multidefect, pearce2020programming,Thijssen2020binding,Rouzaire2021Defect,Chardac2021Topology,Fruchart2021phase,Whitfield2017,Kole2021Layered, Maitra2020,gupta2020active, kumar2014flocking, VanSaders2019,VanSaders2021Sculpting,digregorio2021unified,Poncet2021When,Yan2015,Bililign2021Motile,Poncet2021When,vanZuiden2016}. 
In the study of crystalline defects, it is often assumed that a potential energy governs the interactions between the constituent particles. This assumption, however, need not hold in driven and active solids. For example, Fig.~\ref{fig:edge}a shows a nonconservative interaction force\textemdash one in which the work done between any two configurations depends on the path taken.
Such microscopic forces generically give rise in the continuum limit to odd elasticity: additional moduli that break the major symmetry of the elastic modulus tensor \cite{scheibner2020,scheibner2020non}. 
Experimental signatures of odd elasticity have been reported  in solids made of spinning embryos~\cite{tan2021development} and colloids~\cite{Bililign2021Motile} with hydrodynamic interactions, and robotic metamaterials~\cite{Chen2021Realization,brandenbourger2021active}. 
Likewise, 
gyroscopic matter~\cite{Wang2015,zhao2020gyroscopic,Carta2014dispersion,Hassanpour2014dynamics,Carta2017deflecting,Nash2015,Mitchell2018,Mitchell2018Realization,Mitchell2018Nature,Brun2012vortex} and vortex-like objects~\cite{Sonin1987vortex, Gifford2008dislocation,Nguyen2020Fracton,Moroz2018effective, Fetter2009rotating,Blatter1994vortices,Tkachenko1968elasticity,tkachenko1966vortex,tkachenko1966stability}, e.g. skyrmions~\cite{Benzoni2021Rayleigh,Huang2020melting,Ochoa2017Gyrotropic,Muhlbauer2009skyrmion,Yu2010Real,Brearton2021Magnetic}, can 
exhibit
a special case of odd elastic dynamics (see S.I.).

\emph{Crystallography without potentials\textemdash} 
A typical starting place in crystallography is  
a collection of point particles at positions $\underline{\vb x} = (\vb x^1, \vb x^2, \dots, \vb x^N) $ interacting via forces that are the gradient of a potential: 
\begin{align} 
\vb F^\alpha (\underline{\vb x}) = - \pdv{V (\underline{\vb x})}{\vb x^\alpha } \label{eq:potential}
\end{align}
However, in general, Eq.~(\ref{eq:potential}) need not be be valid. Experimentally relevant~\cite{vanZuiden2016,Bililign2021Motile,Yan2015,Grzybowski2000Dynamic,tan2021development,Han2021Fluctuating,Yeo2015Collective,Scholz2018Rotating,Goldman1967Brenner,Aragones2016} counterexamples include pairwise forces of the form $\vb F^\alpha (\underline{\vb x}) = \sum_{\beta \neq \alpha} \vb F (\vb x^\alpha - \vb x^\beta) $ where $\vb F (\vb r)$ depends only on particle separation: 
\begin{align}
\vb F (\vb r) = F^\parallel(r) \, \hat {\vb r} - F^\perp (r) \, \bph  \label{eq:pair}
\end{align}
Here, $\vb r$ is the relative coordinate between two interacting particles, $\bph =- \boldsymbol{\epsilon} \cdot \hat{\bf r}$, and $\boldsymbol{\epsilon}$ is the antisymmetric tensor. We will henceforth focus on first order dynamics $\dot {\vb x}^\alpha = \vb F^\alpha$, which arise in overdamped media, as well as in
vortex~\cite{Sonin1987vortex,  Gifford2008dislocation,Nguyen2020Fracton,Moroz2018effective, Fetter2009rotating,Blatter1994vortices,Tkachenko1968elasticity,tkachenko1966vortex,tkachenko1966stability} and gyroscopic~\cite{Wang2015,zhao2020gyroscopic,Carta2014dispersion,Hassanpour2014dynamics,Carta2017deflecting,Nash2015,Mitchell2018,Mitchell2018Realization,Mitchell2018Nature,Brun2012vortex} crystals in which the forces  $\vb F^\alpha = \epsilonb \cdot \pdv{V}{\vb x^\alpha}$ are transverse to potential gradients, see S.I.\S{S1A}. 
Subject to first order dynamics, the quantity $ P \equiv \dot{\vb x}^\alpha \cdot \vb F^\alpha (\underline{\vb x})$ is greater than zero for trajectories satisfying the equations of motion~\footnote{While the quantity $P$ corresponds to the power exerted by the interaction forces in purely mechanical systems, this physical interpretation does not carry over in gyroscopic or vortex crystals~\cite{scheibner2020non}}.
Of particular interest here are interparticle forces such that $W_\mathcal{C} = \oint_\mathcal{C} P \, \dd t \neq 0$ for closed contours $\mathcal{C}$ (see Fig.~\ref{fig:edge}a). Notice that $\nabla \times \vb F =0$ is equivalent to requiring that $W_\mathcal{C} =0$ for all contractible loops $\mathcal{C}$. In Newtonian mechanics, 
$W_\mathcal{C}$ has the interpretation of energy, and $\nabla \times \vb F \neq 0$ is equivalent to violating Maxwell-Betti reciprocity (MBR)~\cite{Truesdell1963Meaning}, which means that the linear response matrix between force and displacement is no longer symmetric, see S.I.\S{S1A}.

\begin{figure}[t!]
    \centering
    \includegraphics[width=0.48\textwidth]{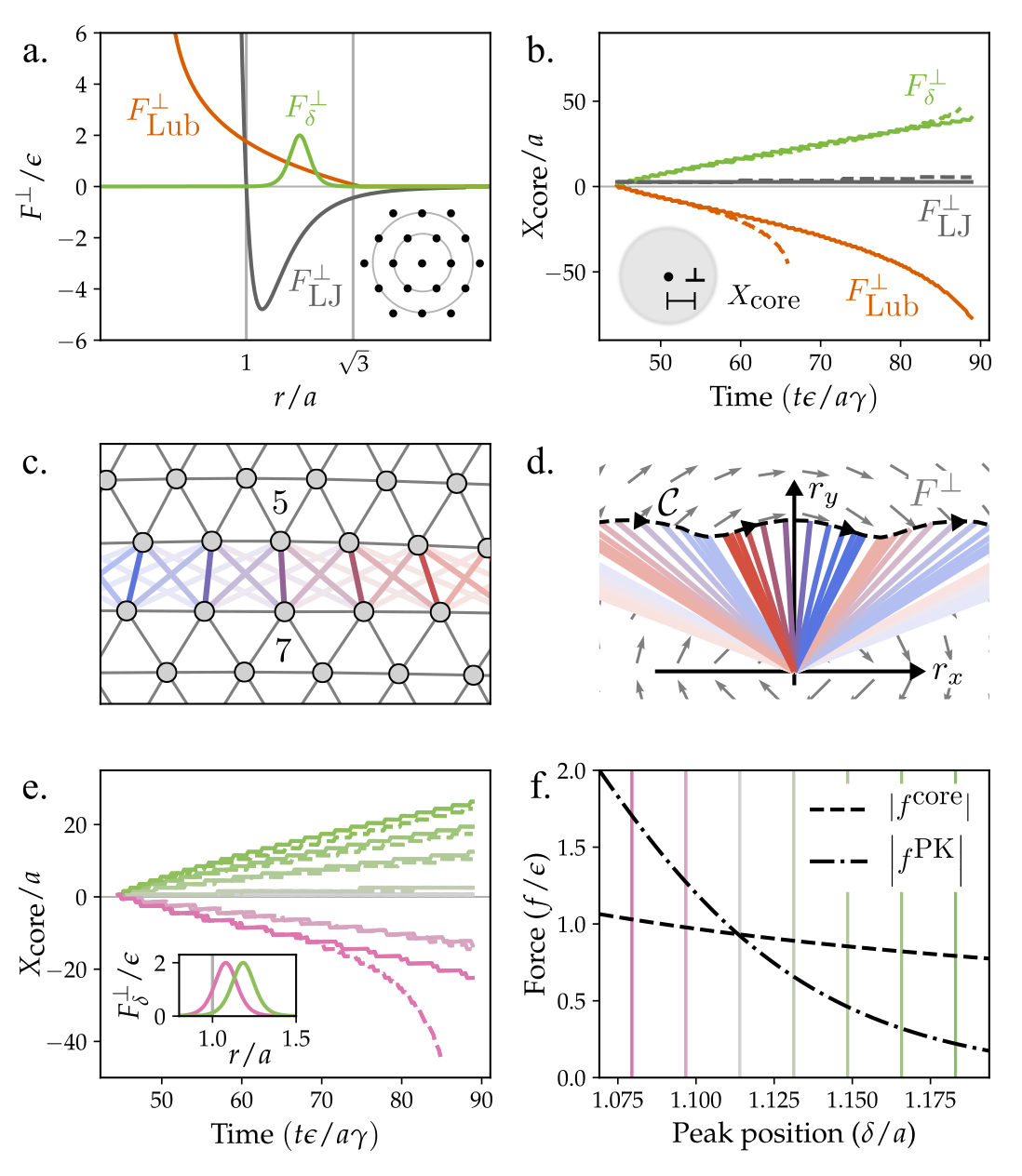}
    \caption{  {\bf Dislocations self propel via active work cycles at their cores.} ~{\bf a.}~Three transverse interactions $F^\perp_\text{LJ}$ (purple), $F^\perp_\text{Lub}$ (teal), and $F^\perp_\delta$ (orange), with the neighbor shells highlighted by grey lines. Inset: A hexagonal lattice with first and second neighbor shells highlighted. 
    {\bf b.}~Particles are arranged in a free floating circular cluster with a single dislocation located at the center, and the dislocation position is tracked as a function of time.  Simulations are performed with clusters of radius $R=50$ (dashed) and $R=100$ (solid). See Supplemental Movie S1.
    {\bf c}.~Bonds crossing the glide plane of a dislocation are highlighted. Hue indicates the bond's position in real space (blue: left, red: right). Opacity indicates the length of the bond (nearest neighbors darkest). 
    {\bf d.}~The highlighted bonds are plotted with their bases aligned. As the dislocation moves one unit cell to the right,  
    the tops of the bonds traces out a contour $\mathcal{C}$ (black dashed). 
    The gray arrows depict the interaction force field. See Supplemental Movie~S2.
    {\bf e.}~The interaction $F^\perp_\delta$ is varied by changing the location $\delta$ of its peak (pink: smaller $\delta$, green: larger $\delta$). For each value of $\delta$, the dislocation's position is tracked as a function of time. 
    {\bf f.}~ The magnitude of the Peach-Koehler force $f^\text{PK}$ and the active core force $f^\text{core}$ as a function of the peak position $\delta$. The vertical lines represent the values of $\delta$ used in the simulation. The direction change of the dislocation motion coincides with the crossover between $f^\text{core}$ and $f^\text{PK}$. 
    }
    \label{fig:motion}
\end{figure}

\emph{Continuum theory\textemdash}
In the continuum, we describe the state of the solid via a continuous displacement field $\vb u (\vb r)$ and we assume that the coarse-grained forces may be expressed as $f_j = \partial_i \sigma_{ij}$, where $\sigma_{ij}$ is the Cauchy stress tensor (see Ref.~\cite{Poncet2021When} for a treatment of dislocations that lifts this assumption). 
We expand the Cauchy stress tensor $\sigma_{ij}$ in terms of the displacement gradient $u_{mn}$ 
to obtain
$\sigma_{ij} =\sigma^0_{ij} +C_{ijmn} u_{mn}$. 
Here, $C_{ijmn}$ denotes the elastic modulus tensor and $\sigma^0_{ij}$ is the stress prior to deformation. In 2D isotropic crystals, the linearized stress-strain relationship is summarized by the pictorial equation~\cite{scheibner2020}:
\begin{align}
    \raisebox{-0.5\height}{\includegraphics{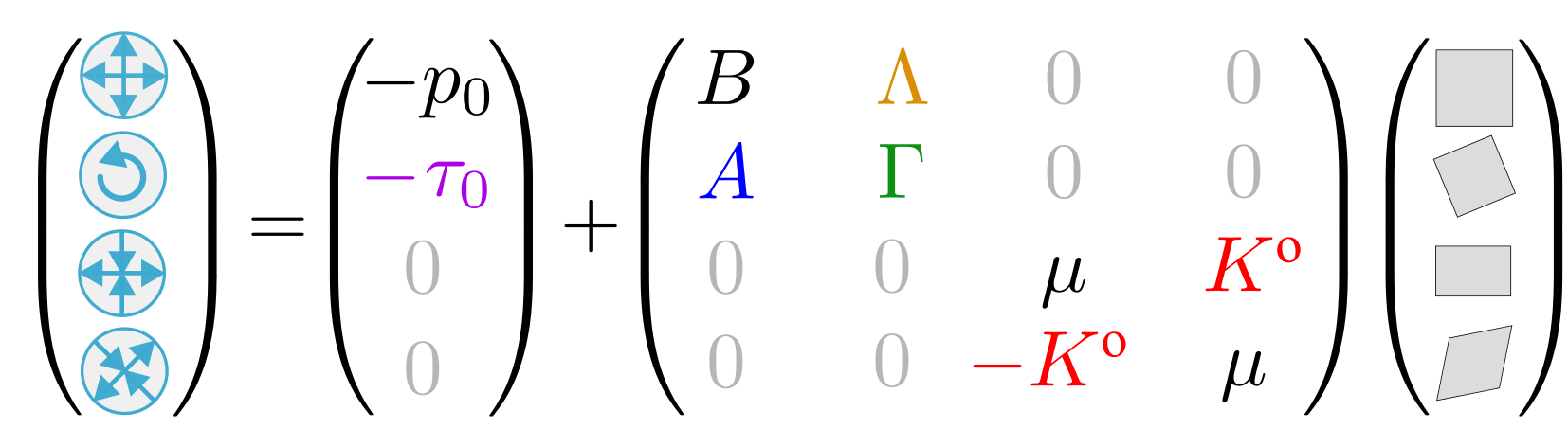}} \label{eq:el}
\end{align}
See S.I.\S{S}1C for tensor notation. 
In Eq.~(\ref{eq:el}), $p_0$ and $\tau_0$ are mechanically interpreted in terms of the pressure and torque density associated with $\sigma^0_{ij}$. 
The matrix in Eq.~(\ref{eq:el}) corresponds to $C_{ijmn}$ and has three diagonal components, the bulk $B$, shear $\mu$, and rotational $\Gamma$ moduli, and three off-diagonal moduli $\Lambda$, $A$, and $K^o$. 
The antisymmetric contributions to the matrix in Eq.~(\ref{eq:el}) are what we refer to as \emph{odd elastic moduli}~\cite{scheibner2020}.  
The counterpart of $P = \dot{\vb x}^\alpha \cdot \vb F^\alpha $ in the continuum is $P=  \int S_{ij} \dot u_{ij} \dd^2 r$, where $S_{ij}$ is the first Piola-Kirchhoff tensor~\cite{zubov2008nonlinear,Storm2005Nonlinear,Lubensky2002Symmetries,antman2013non,Ogden2001nonlinear}. 
Writing $S_{ij} = S^0_{ij} + \tilde C_{ijmn} u_{mn}$, 
we have
\begin{equation}
\tilde C_{ijmn} =C_{ijmn} + \sigma^0_{ij} \delta_{mn}- \sigma^0_{mj} \delta_{in} 
\end{equation}
In the continuum, the Maxwell-Betti reciprocity theorem states the internal forces must be non-conservative if 
$\tilde C_{ijmn} \neq \tilde C_{mnij}$~\cite{Truesdell1963Meaning}. 
Odd elasticity ($C_{ijkl}\neq C_{klij}$) coincides with broken MBR ($\tilde C_{ijkl} \neq \tilde C_{klij}$) when no ambient stress is present ($\sigma^0_{ij} =0$). In terms of the moduli in Eq.~(\ref{eq:el}) the condition for MBR, $\tilde C_{ijmn} = \tilde C_{mnij}$, reads:
\begin{equation}
2K^o = A - \Lambda = 2 \tau_0 \label{eq:cond}
\end{equation}
Notice from Eq.~(\ref{eq:cond}) that odd elasticity can arise even when MBR holds (i.e. the microscopic forces are conservative) provided that $\tau_0$ is nonvanishing. 
For instance, the transverse microscopic force $F^\perp (r) \propto \frac1r$ is curl free, i.e. conservative and, nonetheless, gives rise to $A$ and $K^o$ (S.I.\S{S}1E).
In this case $A$ and $K^o$ can be detected from static stress-strain measurements, but the work they generate during strain cycles must be cancelled by $\tau_0$. 
The distinction between $C_{ijkl}$ and $\tilde C_{ijkl}$ vanishes when no ambient stress $\sigma_{ij}^0$ is present. 
See~\S{S1E} for how crystals with purely transverse interactions, such as lattices of vortex-like objects~\cite{Sonin1987vortex,  Gifford2008dislocation,Nguyen2020Fracton,Moroz2018effective, Fetter2009rotating,Blatter1994vortices,Tkachenko1968elasticity,tkachenko1966vortex,tkachenko1966stability,Benzoni2021Rayleigh,Huang2020melting,Ochoa2017Gyrotropic,Muhlbauer2009skyrmion,Yu2010Real,Brearton2021Magnetic} or gyroscopes~\cite{Wang2015,zhao2020gyroscopic,Carta2014dispersion,Hassanpour2014dynamics,Carta2017deflecting,Nash2015,Mitchell2018,Mitchell2018Realization,Mitchell2018Nature,Brun2012vortex}, can be mathematically cast as a special case of odd elasticity with $B=\mu=0$.

\emph{Microscopics\textemdash}To relate the moduli to the microscopic transverse forces, we linearize Eq.~(\ref{eq:pair}) about the lattice spacing $a$:
$F^\perp (r) = F^\perp_0 - k^a (r-a)$.
The resulting odd elastic moduli for a hexagonal lattice read 
\begin{align} 
A \approx \frac{\sqrt{3}}{2} \qty(k^a + \frac{F^\perp_0}a ) \quad 
K^o \approx \frac{\sqrt{3}}4  \qty(k^a - \frac{3 F^\perp_0}a  )
\end{align}
along with an ambient torque density $\tau_0 = \sqrt{3} F_0^\perp/a$, see S.I.\S{S}1D and Refs.~\cite{Poncet2021When, scheibner2020}. Additionally, 
the modulus $A$ arises whenever the full torque density $\tau =\epsilon_{ij} \sigma_{ij}/2$ couples to local dilation $\partial_i u_{i} =- \delta \rho /\rho_0$, namely $A = \dv{\tau}{\rho} \rho_0$.
The forces in Eq.~(\ref{eq:pair}) depend only on $r$ and, therefore, cannot contribute to $\Gamma$ and $\Lambda$ which couple to solid body rotations. However, $\Gamma$ and $\Lambda$ can arise in response to external fields or interactions with a substrate~\cite{Nelson1979}, see S.I.\S{S}1E for examples.  
We henceforth set $\Lambda=\Gamma=0$, see S.I.\S{S}2C for a general treatment~\cite{Note0}.

\footnotetext[0]{See the Supplemental Material for additional discussion, including Refs.~\cite{Born1954dynamical,Fruchart2020Symmetries,Nelson1987,Anderson2008,Glaser2015,Jones1924,Kim2013microhydrodynamics,Yukawa1935,yang2012generalized}}

\emph{Continuum solutions\textemdash} 
Topological defects are singularities where $u_i (\vb r)$ becomes multi-valued, e.g. the dislocation in Fig.~\ref{fig:edge}b is defined by the Burgers vector $b_i$ 
\begin{align}
   b_j= \oint_\gamma \partial_i u_j \dd r_i   \label{eq:disl}
\end{align}
where $\gamma$ is a counterclockwise contour around the dislocation. We solve $\partial_i \sigma_{ij}(\vb r)=0 $  together with Eq.~(\ref{eq:disl}) to obtain static solutions of the dislocation displacement field (S.I.\S{S}2C):
\begin{align}
     \vb u_{\text{disl}} = \frac1{2\pi} \bigg\{ & \phi \, \vb b +  \frac{1-\nu}2  \log(r) \, \boldsymbol{\epsilon} \cdot \vb b  -\frac{1+\nu}{2}  (\vb{\hat r} \vdot  \vb b) \,  \boldsymbol{\hat \phi}     \nonumber \\
     & -\nu^o \qty[  \log (r) \, \vb b +  (\boldsymbol{\hat  \phi } \vdot \vb b)  \,  \boldsymbol{ \hat \phi}   ]
     \bigg \}  \label{eq:solb}
\end{align}
where $r$ and $\phi$ are polar coordinates about the defect. 
The elastic properties enter only through (i) a modified Poisson's ratio, $\nu$, (S.I.\S{S}2C) and (ii) a purely non-reciprocal \emph{odd ratio}~\cite{scheibner2020}
\begin{align}    
    \nu^o =& \frac{ B K^o - A \mu }{\mu(B+ \mu) + K^o (A+ K^o) } \label{eq:nuo}
\end{align}
The effect of $\nu^o$ in Eq.~(\ref{eq:solb}) is to globally rotate the local shear axis by an angle $\delta \alpha$
\begin{align}
    \delta \alpha = - \frac12 \arctan(\frac{2 \nu^o}{1+\nu} ) \label{eq:shearrot}
\end{align}
See Fig.~S5-6 for an illustration and numerical validation. 
S.I.\S{S}2B provides similar results for point defects and isolated disclinations, which have recently been observed in experiments~\cite{tan2021development} of  spinning embryos interacting via transverse forces, cf.~Eq.~(\ref{eq:pair}).

\begin{figure*}[t!]
    \centering
    \includegraphics[width=\textwidth]{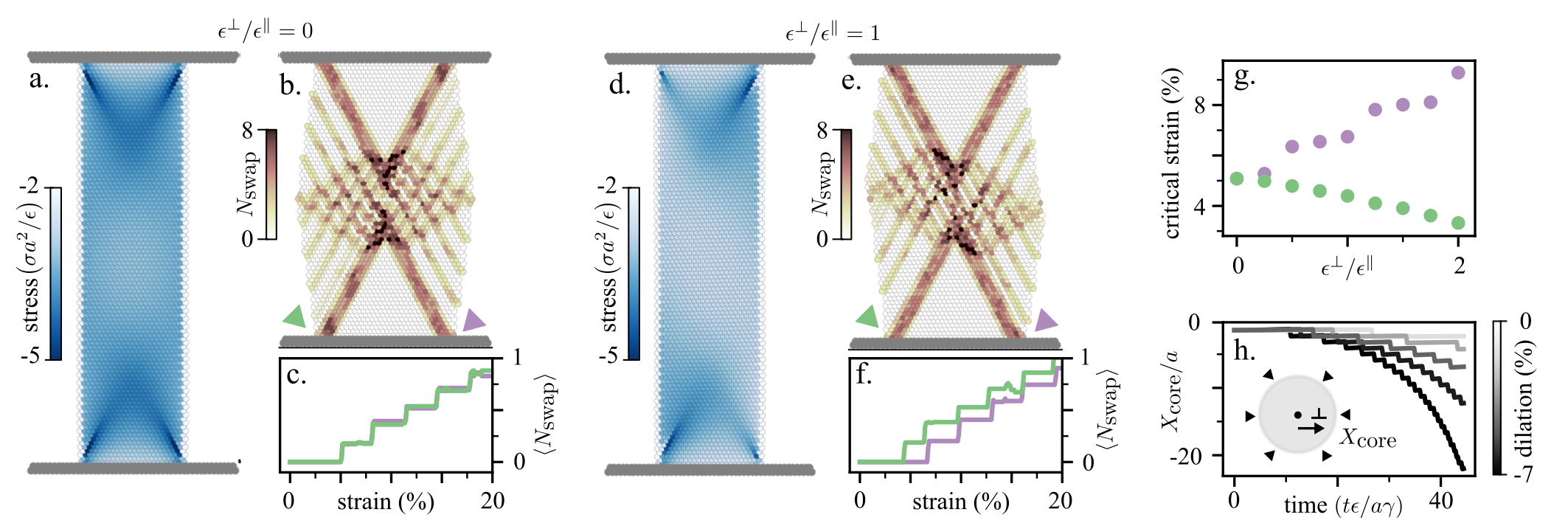}
    \caption{{\bf Active plasticity in odd elastic media.}
    \textbf{a-c.}~The compression of a solid beam with a standard LJ interaction of strength $\epsilon^\parallel$. 
    {\bf d-f.}~The same numerical experiment with the addition of transverse lubrication forces of magnitude $\epsilon^\perp$. See Supplemental Movies S3-4.
    (a., d.)~show the
    per-particle stress, resolved on the $[11]$ and $[\bar{1} 1]$ glide planes and summed, immediately prior to the first dislocation nucleation. 
    (b., e.)~After significant plastic deformation, we color particles by the cumulative number of neighbor changes in their first coordination shell ($N_\text{swap}$).
    (c., f.)~ $ N_\text{swap} $ averaged in the vicinity of the lower-left- and lower-right-hand corners as a function of strain. The strain is $\Delta h/h$ where $h$ is the height of the beam.  
    {\bf g}. The critical strain at first nucleation for the bottom left-hand (green) and right-hand (pink) corners. 
    \textbf{h.} 
    A disk of particles interacting via radial and transverse LJ forces is subject to compression (negative dilation). 
    At higher compression, a torque density is induced throughout the cluster which drives dislocation motion via the PK force. See the S.I.\S{S}5 for additional simulation details. 
    }
    \label{fig:plastic}
\end{figure*}

\emph{Dislocation interactions\textemdash}
The modified stress field alters dislocation interactions. Consider first the work done in quasistatically moving a test dislocation by $\delta X_i$ through a pre-existing stress field $\sigma_{ij}^\pres$ obeying $\partial_i \sigma^\pres_{ij} =0$.  Regardless of the material's constitutive properties, the work done by $\sigma^\pres_{ij}$ 
is given by the Peach-Koehler (PK) formula $\delta W = f_m^\text{PK} \delta X_m$,  where $f_m^\text{PK} = \epsilon_{mi} \sigma^\pres_{ij} b_j$ (S.I.\S{S}3A).  The continuum interaction between two dislocations is the PK force experienced by the test dislocation as a result of the stress field generated by the other.
In Fig.~\ref{fig:edge}c-h, we examine the interaction between two antiparallel dislocations in the presence of odd elastic moduli. 
The modulus $A$ provides a nontrivial modification:
\begin{align}
    f^\text{PK} (r_x) = \frac{(1-\nu) b_1 b_2 }{\pi r^4 } ( r_x^2-r_y^2 ) ( Br_x +Ar_y )  \label{eq:fpk}
\end{align}
where $f^\text{PK}$ is the PK force projected onto the glide plane of the dislocation. 
When $A=0$, the dislocations obey the classic result: their separation vector forms an angle $\pi/4$ with respect to their glide planes~\cite{Weertman1964,Landau7}. 
When $0 < A/B < 1$, the mechanically stable positions remain the same while their basins of attraction change. When $A/B >1$, the right equilibrium point moves out beyond the $\pi/4$ angle. When $A/B \to \infty$, the rightmost basin becomes unstable. When the Burgers vectors are not parallel, the dislocation interactions are non-reciprocal in the sense of being non-mutual: their forces are not equal and opposite (S.I.\S{S}3B). 

\emph{Dislocation motion from PK forces\textemdash}~While the continuum theory provides insights at long length scales, whether and in what directions dislocations actually move depends on microscopic details. To illustrate this, we perform overdamped molecular dynamics simulations of particles interacting with a radial force $F^\parallel (r)$ given by  a Lennard-Jones (LJ) force and  three different realizations of the transverse force $F^\perp(r)$ in Eq.~(\ref{eq:pair}), see Fig.~\ref{fig:motion}a-b and Movie~S1.

Consider first a transverse interaction $F^\perp_\text{LJ} (r)$, which like $F^\parallel$, is an LJ force: a dislocation introduced to the center
remains stationary as in a passive crystal. The reason is that the total force on any particle is simply a rotation of the force due to $F^\parallel$. Hence the static configuration guaranteed by energy minimization for $F^\parallel$ remains static when $F^\perp_\text{LJ}$ is introduced. Next we introduce $F^\perp_\text{Lub}$, which is a monotonically decreasing function of a single sign shown in Fig.~\ref{fig:motion}a, generically representative of hydrodynamic, lubrication and frictional forces between self-spinning particles~\cite{Aragones2016,Yan2015,tan2021development,Grzybowski2000Dynamic,Goldman1967Brenner}.  
For $F^\perp_\text{Lub}$, the dislocation travels at a near constant speed to the {left}. 
Since $F^\perp_\text{Lub}$ is nonzero at the first neighbor shell, it produces an ambient torque density $\tau_0$. 
Upon setting 
$\sigma^\pres_{ij}= \tau_0 \epsilon_{ij}$, the direction of dislocation motion follows the standard PK force expression $f^\text{PK}_i = \epsilon_{ij} \sigma_{jk}^\pres b_k = - \tau_0 b_i $, in agreement with the experiments and analysis of spinning-colloids crystals in Refs.~\cite{Bililign2021Motile,Poncet2021When}.

{\it Dislocation self-propulsion from microscopic work cycles at their cores\textemdash}~
We now show that not all mechanisms of dislocation propulsion can be captured by continuum considerations. 
The continuum PK force is derived from a coarsegrained approximation to the work done during dislocation motion.
However, when a potential is not well defined, contributions to $\delta W$ from short length scales need not average out during dislocation motion. In some cases, they can even overcome continuum predictions.
To illustrate this, we use as a probe the force $F^\perp_\delta(r)$ narrowly peaked at a tunable interparticle distance $r=\delta$ and 
with the same sign as $F^\perp_\text{Lub}$. 
However, when the peak $\delta$ lies half way between the first and second neighbor shells, the dislocation now travels to the right, the opposite direction of $F^\perp_\text{Lub}$ (see Fig.~\ref{fig:motion}ab). 
Since the force $F^\perp_\delta$ is negligible at the first and second neighbor shells, the odd moduli $A$ and $K^o$ as well as the ambient torque density
$\tau_0$ are vanishingly small. 
This suggests that the underlying mechanism of dislocation motility evades the standard continuum explanation in terms of PK forces provided in the previous paragraph. 

As we now show, this dislocation motility is a form of self-propulsion associated with microscopic work cycles acting at dislocation cores. 
We first highlight all the bonds that straddle the glide plane  (Fig.~\ref{fig:motion}c) and align their bases so that they are viewed in the space of their relative coordinates $(r_x,r_y)$ (Fig.~\ref{fig:motion}d). Crucially, as
the dislocation moves by a single unit cell, each highlighted bond assumes the position of its neighbor to the right. Next, we concatenate all the individual bond trajectories into a single contour $\mathcal{C}$ (dashed line) that begins at $r_x = -\infty$ and ends at $r_x =\infty$.  
The total work done when each of the bonds moves a short distance is then equivalent to that of a single bond traveling the entire contour, c.f. Fig.~\ref{fig:edge}a.
Notice that if the force falls off faster than $1/r$, then the contour may be closed in the upper half plane. 
Similar to the single-bond cycle shown in Fig. \ref{fig:edge}a, the corresponding work $W_\text{glide}$ reads
\begin{align}
    W_\text{glide} \approx \oint_\mathcal{C} \vb F \cdot \dd \vb r  = \int_\mathcal{A} \nabla \times \vb F \; \dd^2 r  \label{eq:curl}   
\end{align}
where $\mathcal{A}$ is the upper half plane enclosed by $\mathcal{C}$ (S.I.\S{S}4 and Movie~S2).

Since $W_\text{glide}$ is associated with motion through one lattice spacing, the corresponding force on the dislocation reads $ f^\text{core} = \frac1a W_\text{glide} $ and it is directed along the glide plane. 
In principle, the detailed shape of $\mathcal{C}$ depends on protocol and dynamics. However, a useful first approximation is to take $\mathcal{C}$ to be the line at $r_y = \frac{\sqrt{3}}{2}a$, giving:
\begin{align}
    f^\text{core} = \frac1a W_\textrm{glide} \approx \eval{\frac1a \int_{-\infty}^\infty F^\perp_x \dd r_x}_{r_y = \frac{ \sqrt{3}}2 a} \label{eq:coreapprox}
\end{align}
In Fig.~\ref{fig:motion}e we perform simulations with $F^\perp_\delta$ but we vary the parameter $\delta$, which sets the location of the central peak. 
At small $\delta$, there is significant overlap between $F^\perp_\delta$ and the first neighbor shell, giving rise to a large ambient torque density $\tau_0 \approx  \sqrt 3 F^\perp(a)/a$ and the corresponding PK force $f^\text{PK}$. 
Fig.~\ref{fig:motion}f shows the relative magnitudes of $f^\text{core}$ and $f^\text{PK}$ as a function of $\delta$. 
The  crossover in dominant force coincides with the sign reversal in dislocation speed corroborating our theoretical derivation of $f^\text{core}$. While the sign reversal is a dramatic effect that occurs under specific conditions, $f^\text{core}$ is generically present for all non-conservative microscopic interaction forces.
Solids whose microscopic interactions violate Newton's third law can also display spontaneous dislocation motion \cite{Poncet2021When}.

\emph{Active plasticity\textemdash}~Finally, we examine the effects of odd elasticity on plastic deformation. 
In Fig.~\ref{fig:plastic}a, we perform a simple uniaxial compression of a solid interacting via a transverse lubrication force (see also Supplemental Movies~S3-4). Before the first dislocation nucleates, odd elasticity biases the stress distribution (Fig.~\ref{fig:plastic}a,d). At the end of the compression, the permanently deformed shape of the beam breaks all mirror symmetries (Fig.~\ref{fig:plastic}b,e). 
The change in final shape arises because the biased stress distribution favors dislocation nucleation from the upper-right- and lower-left-hand corners (Fig.~\ref{fig:plastic}c,f). Empirically, we find that introducing transverse forces generally lowers the plastic yield strain at which the first dislocation nucleates~(Fig.~\ref{fig:plastic}g).
In Fig.~\ref{fig:plastic}h, we consider a single dislocation in the center of a disk. 
In a passive medium, a uniform compression induces an isotropic stress $-B (\delta \rho/ \rho)  \delta_{ij}$ with an associated $f^\text{PK}_i =  -B (\delta \rho /\rho) \epsilon_{ij} b_j$ in the climb direction. This typically results in no motion or defect splitting.
The odd elastic solid in Fig.~\ref{fig:plastic}h features a $F^\perp_\text{LJ}$ which induces no dislocation motion in the absence of external stresses (recall Fig.~\ref{fig:motion}b). 
However, due to the odd modulus $A$, an area change gives rise to a torque density $\tau = A (\delta \rho /\rho)$, which in turn promotes motion along the glide plane via $f^\text{PK}_i = - \tau b_i$. 
 
To summarize, we studied how defect strains, interactions and motility are modified in systems for which the interactions are more general than standard pairwise, potential forces. 

\begin{acknowledgments}
V.~V.~acknowledges support from the Simons Foundation
and the University of Chicago Materials Research Science
and Engineering Center, which is funded by the National
Science Foundation under Grant No.~DMR-2011854.
B.~V.~S.~acknowledges support from the Kadanoff-Rice
Postdoctoral Fellowship. V.~V.~and B.~V.~S.~acknowledge
support by the Complex Dynamics and Systems Program
of the Army Research Office under Grant No.~W911NF19-1-0268. 
C.~S.~acknowledges support from the
Bloomenthal Fellowship and the National Science
Foundation Graduate Research Fellowship under Grant
No.~1746045. L.~B.~acknowledges support from the
Heising-Simons Scholarship and the James Franck
Institute Undergraduate Summer Research Award. Some
of us benefited from participation in the KITP program on
Symmetry, Thermodynamics, and Topology in Active
Matter supported by Grant No.~NSF PHY-1748958. This
work was completed in part with resources provided
by the University of Chicago Research Computing
Center. The authors would like to thank M.~Han, M.~Fruchart, A.~Poncet, D.~Bartolo, and W.~Irvine for helpful
conversations.
\end{acknowledgments}

\clearpage

\onecolumngrid 

\renewcommand{\theequation}{S\arabic{equation}}
\setcounter{equation}{0}
\setcounter{figure}{0}
\renewcommand{\thetable}{S\arabic{table}}  
\renewcommand{\thefigure}{S\arabic{figure}}
\renewcommand\figurename{Fig.}
\renewcommand{\thesection}{S\arabic{section}}

\begin{center}
 \Large
{\bf Supplementary Information }
\normalsize
\end{center}

\medskip

In \S\ref{sec:el}, we detail the relationship between odd elasticity, nonconservative microscopic forces, and broken Maxwell-Betti reciprocity (MBR). While broken MBR naturally gives rise to odd elasticity, nonlinear distinctions between the Cauchy and Piola-Kirchhoff stress tensors reveal that odd elasticity can arise in solids obeying MBR in the presence of an ambient stress. We provide examples with simple, solvable models as well as explicit coarsegrainging procedures. In \S\ref{sec:solutions}, we derive the stress and strain fields surrounding dislocations, disclinations, and point defects in solids with odd elasticity. This section identifies experimentally measurable signatures of odd elasticity that arise in the shear strain surrounding topological defects. 
In \S\ref{sec:dislint}, we study the elastic interaction between dislocations. We derive the modified interaction force between dislocations with parallel glide planes. Moreover, we show that when the glide planes are not parallel, the forces between dislocations need not be equal and opposite. In \S\ref{sec:core}, we provide a calculation scheme for the active core force, which is not captured by standard continuum considerations. In \S\ref{sec:num} we detail our numerical methodology and in \S\ref{sec:coarse} we provide background on the coarsegraining procedures that connect microscopic variables to continuum fields.

\section{Odd elasticity and broken Maxwell-Betti reciprocity} \label{sec:el}

\subsection{Microscopic interactions and Maxwell-Betti reciprocity}
\label{sec:potmicro}
Here we introduce the notion of nonconservative forces in the context of microscopic interactions. We model a crystal as a collection of point particles with coordinates $\vb x^\alpha(t)$, $\alpha = 1,2, \dots, N$. 
For each particle, there exists a function $\vb F^\alpha ( \underline{\vb x} )$ which we refer to as the force. 
For a given trajectory $\CC$,  we define $W_\CC \equiv \int_\CC P \, \dd t $ where $P = \dot{\vb x}^\alpha \cdot \vb F^\alpha$. 
In the spirit of Hodge decomposition, one can classify the interparticle forces based on the value of $W_\CC$ for closed contours $\CC$.  
If $W_\CC =0$ for all closed contours $\CC$, we say that  $\vb F^{\alpha}$ is \emph{exact}. In this case, there exists a globally single-valued potential $V(\underline{\vb x})$ such that $\vb F^{\alpha} =- \pdv{V}{\vb x^\alpha}$. Secondly, if $W_\CC =0$ for all contractible loops $\CC$, we say that $\vb F^{\alpha}$ is \emph{closed}. In this case, $V(\underline{\vb x})$ exists locally, but may be globally multivalued. An example of $\vb F^\alpha$ that is closed but not exact is 
\begin{align}
\vb F^\alpha = \sum_{\beta \neq \alpha } \vb F (\vb x^\alpha - \vb x^\beta) \label{eq:pairwise} 
\end{align} 
with $\vb F (\vb r) = \frac1r \bph$. In this case, one can introduce the multivalued potential 
\begin{align} 
V(\underline{\vb x}) =  \sum_{\alpha < \beta} \arctan( \frac{x^\alpha -x^\beta}{ y^\alpha -y^\beta} ) 
\end{align} 
Notice that $W_\CC =  w(\CC)$, the sum of the number of times any pairs of particles winds about each other (see also \S\ref{sec:microexample2}). Recall that the exterior derivative of any closed form must vanish. For pairwise interactions in Eq.~(\ref{eq:pairwise}), the exterior derivative vanishing is equivalent to $\nabla \times \vb F =0$. 
While we will typically concentrate on pairwise forces for simplicity, the distinction between exact, closed, and not closed forces can be made for any force that is a function of particle positions. 
It is useful to consider the linearization $\vb F^\alpha  = \vb F^0_\alpha + \vb M_{\alpha \beta} \cdot \delta \vb x^\beta $ about some configuration $\underline{\vb x}^0$. 
If the linear response matrix $\vb M_{\alpha \beta}$ is symmetric (i.e. $\vb M_{\alpha \beta} = \vb M_{\beta \alpha}^T$), then one says that the force obeys Maxwell-Betti reciprocity (MBR) microscopically.   
If the force obeys MBR for all linearization points $\underline{\vb x}^0$, then $\vb F^\alpha (\underline{\vb x})$ must be closed, i.e. a locally well-defined potential must exist.  
In \S\ref{sec:cauchypiola} we detail the continuum formulation of the MBR. 

We will concentrate on first order dynamics of the form
\begin{align}
    \dot{\vb x}^\alpha = \vb F^{\alpha} (\underline{\vb x}) \label{eq:dyn1}
\end{align}
Given the first order dynamics, $W_\CC > 0$ for all trajectories satisfying the equations of motion. 
In many settings, such as particles interacting with electromagnetic forces in an overdamped environment, $\vb F^\alpha$ corresponds to a Newtonian force while the quantity $P$ corresponds to the physical power exerted by the interparticle forces. 
However, so long as $\vb F^\alpha$ prescribes the velocity $\dot{\vb x}^\alpha$, our analysis will still apply.  
For example, systems with purely transverse interactions, such as simple models of gyroscopic lattices~\cite{Wang2015,zhao2020gyroscopic,Carta2014dispersion,Hassanpour2014dynamics,Carta2017deflecting,Nash2015,Mitchell2018,Mitchell2018Realization,Mitchell2018Nature,Brun2012vortex}, vortices in ideal fluids~\cite{Sonin1987vortex,  Gifford2008dislocation,Nguyen2020Fracton,Moroz2018effective, Fetter2009rotating,Blatter1994vortices,Tkachenko1968elasticity,tkachenko1966vortex,tkachenko1966stability}, and
skyrmions~\cite{Ochoa2017Gyrotropic,Benzoni2021Rayleigh,Huang2020melting,Muhlbauer2009skyrmion,Yu2010Real,Brearton2021Magnetic}, can be modeled by a Lagrangian of the form 
\begin{align} 
L =  \frac{ \dot{\vb x}^\alpha \cdot \epsilonb \cdot \vb x^\alpha}{2} -V(\underline{\vb x}) \label{eq:perplag}
\end{align} 
For example, $V(\underline{\vb x}) =  \sum_{\alpha < \beta} \log |\vb x^\alpha - \vb x^\beta |$ for interacting vortices and  $V(\underline{\vb x}) = \frac12 \sum_{\alpha < \beta } |\vb x^\alpha - \vb x^\beta |^2$ for gyroscopes connected by springs.
The equations of motion corresponding to the variation of the action $S = \int L \dd t$ are typically written as $\epsilonb \cdot \dot{\vb x}^\alpha = - \pdv{V}{\vb x^\alpha}$. Moving the $\epsilonb$ tensor to the right-hand side yields Eq.~(\ref{eq:dyn1}) with $\vb F^\alpha =  \epsilonb \cdot \pdv{V}{\vb x^\alpha} $. In the case of vortices or gyroscopes $\vb F^\alpha$ does not correspond to a Newtonian force, $P$ does not correspond to physical energy per unit time, and $\vb F^\alpha$ is the level curve, not the gradient, of the potential energy. 
In \S\ref{sec:gryoel}, we discuss the continuum formulation of such media. 
Forces of the form $\vb F^\alpha = \epsilonb \cdot \pdv{V}{\vb x^\alpha} $ or $\vb F^\alpha = -\pdv{V}{\vb x^\alpha }$ are only special cases: more general forces generically occur in systems with robotic or hydrodynamic interactions. We note that the conditions for static equilibrium ($\vb F^{\alpha} =0$) is independent of the form of the dynamical equations of motion.

Finally, we comment that there is a second common usage of the word \emph{reciprocity} that is conceptually distinct from MBR. 
Forces decomposed as $\vb F^\alpha = \sum_{\beta} \vb F^{\alpha \beta} $ are sometimes called \emph{reciprocal} if they respect ``Newton's third law": $\vb F^{\alpha \beta} = - \vb F^{\beta \alpha }$.
We will refer to this ``Newton's third law" property as \emph{mutuality}. It is readily seen that nonmutual forces necessarily violate MBR, but not all forces that violate MBR are nonmutual. 
The continuum theory below applies to microscopic interactions that are mutual, enabling the definition of a stress tensor.

\subsection{Elastic moduli and nonlinear considerations: Cauchy versus Piola-Kirchhoff stress tensors } \label{sec:cauchypiola}

\begin{figure}[t!]
    \centering
    \includegraphics[width=0.8 \textwidth]{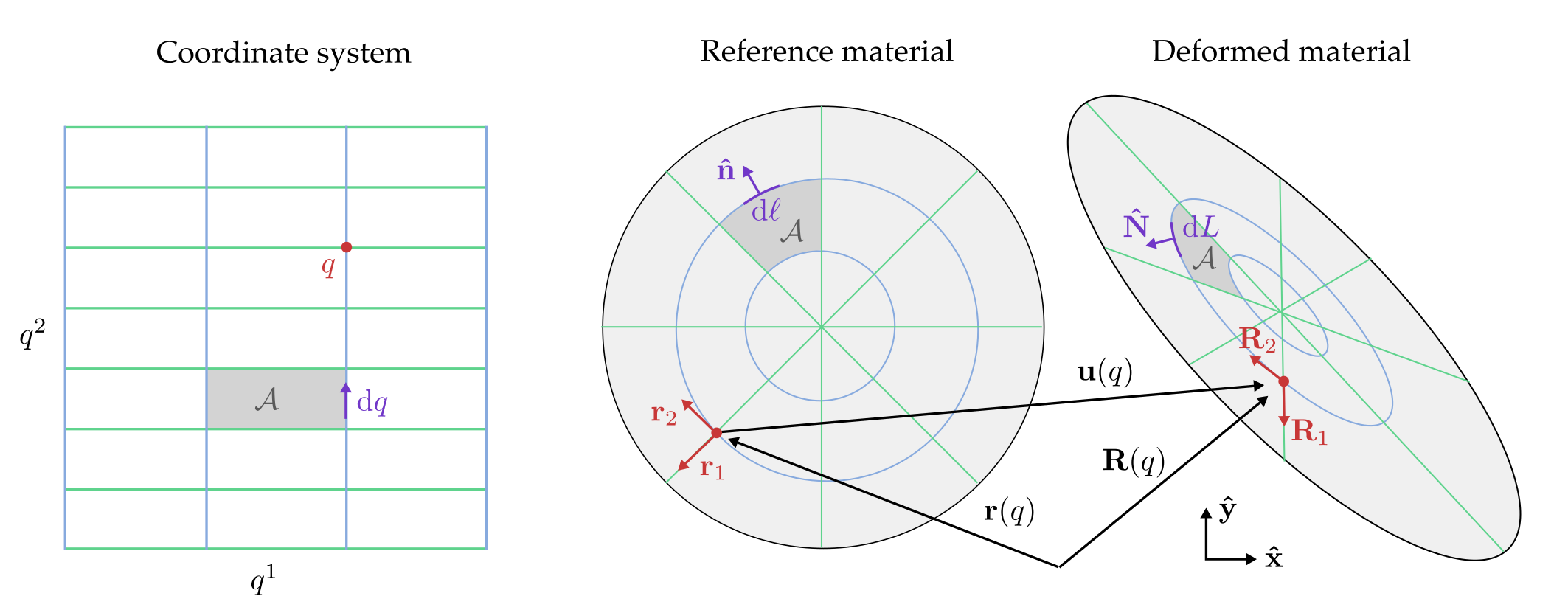}
    \caption{ {\bf Kinematics of nonlinear elasticity.} We illustrate the geometric quantities  for nonlinear elasticity introduced in \S\ref{sec:el}. }
    \label{fig:nonlinear}
\end{figure}

In this section, we detail the relationship in the continuum between broken Maxwell-Betti reciprocity and the elastic modulus tensor. 
We begin by introducing the geometric notation necessary for nonlinear elasticity~\cite{zubov2008nonlinear,Storm2005Nonlinear,Lubensky2002Symmetries,antman2013non,Ogden2001nonlinear}. As shown in Fig.~\ref{fig:nonlinear}, we first introduce a set of coordinates $q =(q^1,q^2)$ such that each point $q$ labels a unique element of material. The deformation is defined by introducing two maps: $\vb r(q) $ represents the position of the element $q$ prior to deformation (i.e. in the \emph{reference material}), and $\vb R (q,t)$ represents the position of the element after deformation (i.e. in the \emph{deformed material}). Here, $\vb r(q) = r_x (q) \vb{\hat x} + r_y(q) \hat{\vb y}$ and $\vb R(q,t) = R_x(q,t) \vb{\hat x} + R_y (q,t) \vb{\hat y} $ are simply vectors in the Cartesian $x$-$y$ plane. 
The velocity field is given by $\vb v (q, t) = \dot {\vb R} (q,t)$ and the displacement field is given by $\vb u(q,t) = \vb R(q,t) - \vb r(q)$. For each $q$, it is convenient to introduce two sets of basis vectors $\vb r_i (q) = \pdv{\vb r}{q^i}$ and $\vb R_i(q) = \pdv{\vb R}{q^i}$ as well as their duals $\vb R^i$ and $\vb r^i$, defined by $\vb R^i \cdot \vb R_j = \vb r^i \cdot \vb r_j = \delta\indices{^i_j}$. The dual vectors allow us to define derivatives with respect to the reference and deformed material. For an arbitrary tensor $\vb \Phi (q)$, we define the gradient and divergence in the deformed material as $\Grad \vb \Phi = \vb R^i \otimes \pdv{\vb \Phi}{q^i}$ and $\Div{\vb \Phi} = \vb R^i \cdot \pdv{\vb \Phi}{q^i} $, respectively. The gradient and divergence in the reference material are defined as 
$\gradr{\vb \Phi} = \vb r^i \otimes \pdv{\vb \Phi}{q^i}$ and $\divr{\vb \Phi} = \vb r^i \cdot \pdv{\vb \Phi}{q^i} $, respectively. A sum over repeated indices is implied. 
The local strain and rotation is encoded in the deformation gradient tensor $\vb J = \gradr \vb R= \vb r^i \otimes \vb R_i = \vb 1 + \gradr \vb u$, where $\vb 1$ is the identity tensor. 

The goal of elasticity theory is to relate the deformation $\vb J$ to the flow of momentum within the material. To describe the flow of momentum, one can take an infintesimal line element of material, described by $\dd q^i$, and ask: how much momentum flows through the line element in a given amount of time? By asking this question for all possible line elements, one can reconstruct the current of momentum, known as stress. To encode the stress as a tensor, one must embed the line element $\dd q^i$ in either the reference or deformed material. 
Given an infinitesimal line element $\dd q^i$, 
its length in the deformed material is $ \dd L =\sqrt{ G_{ik} \dd q^i \dd q^k}$ and its length in the refernce material is $\dd \ell = \sqrt{ g_{ik} \dd q^i \dd q^k }$, where $G_{ik} = \vb R_i \cdot \vb R_k$ and $g_{ik} = \vb r_i \cdot \vb r_k$. Likewise, the normal vectors in the deformed and reference material are given by $\vb{\hat N } = \boldsymbol{\epsilon} \cdot \frac{ \vb r_i  \dd q^i }{\dd L}$ and $\vb{\hat n} = \boldsymbol{\epsilon} \cdot \frac {\vb R_i \dd q^i}{\dd \ell}$, respectively. Here, $\epsilonb=\vb{\hat  x} \otimes \vb { \hat  y}  - \vb{\hat y} \otimes \vb { \hat x}$ is the antisymmetric tensor, which implements a rotation clockwise by $\pi/2$. 
Given the total momentum $\vb P(\dd q^i)$ flowing across an arbitrary line element $\dd q^i$ in a time $\delta t$, one can unambiguously define two distinct tensors, $\sigmab$ and $\vb S$, by requiring $-\vb P = \delta t \; \vb{\hat N} \cdot \sigmab \; \dd L   = \delta t \; \vb{\hat n} \cdot \vb S \; \dd \ell $ for all $\dd q^i$. Here $\sigmab $ is the Cauchy stress tensor and describes the flow of momentum in the deformed material. By contrast $\vb S $ is the first Piola-Kirchhoff stress, which represents the corresponding flow through the reference material. In general the two tensors differ from each other because the line element $\dd q^i$ undergoes stretching and rotation when the material is deformed. 

From the definitions of $\sigmab$ and $\vb S$, the total force on a patch of material $\regA$ is given by
\begin{align}
    \mathbb{F} (\regA) =& \oint_{\partial \regA }  \vb{\hat N}  \cdot \sigmab \; \dd L = \oint_{\partial \regA }  \vb{\hat n} \cdot \vb S \; \dd \ell  \\
    =& \int_\regA \Div \sigmab \; \dd^2 R = \int_\regA \divr \vb S \; \dd^2 r \label{eq:FD}
\end{align}
where in the last line we have applied Stokes' theorem and  $\dd^2 R = \sqrt{\det G_{ij} \vb R^i  \otimes \vb R^j } \dd^2 q $ and $\dd^2 r = \sqrt {\det  g_{ij} \vb r^i \otimes \vb r^j } \dd^2 q$. From Eq.~(\ref{eq:FD}), we identify the force per unit area in physical and material space, $\vb f$ and $\tilde{\vb f}$ respectively, as 
\begin{align} 
\vb f= \Div  \sigmab \quad \text{and} \quad 
\tilde{\vb f} = \divr \vb S
\end{align} 
Finally, the power exerted by this patch of material $\mathcal{A}$ is given by: 
\begin{align}
    \dv{W (\mathcal{A}) }{t} =& \int_\mathcal{A} \vb f \cdot \vb v \; \dd^2 R  =- \int_\mathcal{A}  \sigmab :  \Grad \vb v  \; \dd^2 R + \oint_{\partial \mathcal{A} } \vb{ \hat  N} \cdot \sigmab \cdot \vb v  \; \dd L \label{eq:cauchy} \\
    =&\int_\mathcal{A} \tilde{\vb f} \cdot \vb v \; \dd^2 r = -\int_\mathcal{A}  \vb S : \gradr \vb v \; \dd^2 r + \oint_{\partial \mathcal{A}} \vb{\hat n} \cdot \vb S \cdot \vb v \; \dd \ell  \label{eq:piola} 
\end{align}
Equations~(\ref{eq:cauchy}-\ref{eq:piola}) are exact to all orders in deformation. We now introduce a constitutive relations in which the stress is written to linear order in the displacement gradient $\gradr \vb u$.   To do so, we introduce two distinct tensors, $\vb C$ and $\tilde{\vb C}$ defined such that
\begin{align} 
\sigmab =& \sigmab^0 + \vb C : \gradr \vb u + \order{\varepsilon^2} \\
\vb S =& \vb S^0 +\tilde {\vb C} : \gradr \vb u+\order{\varepsilon^2} . 
\end{align} 
where $\varepsilon$ represents the typical size of $\gradr \vb u$.
Here $\sigmab^0$ and $\vb S^0$ are the Cauchy and Piola-Kirchhoff stresses present before any deformation is applied. Notice that $\vb C$ and $\vb {\tilde C} $ are both linear response tensors; However, since $\gradr \vb v = \dot{ \gradr \vb u}$, Eq.~(\ref{eq:piola}) states that $\vb S$ and $\gradr \vb u$ are energy conjugates. Hence, the Maxwell-Betti reciprocity theorem states that the material is compatible with a potential energy if and only if the tensor $\tilde{\vb C}$ is symmetric about its major axis. Explicitly, we may write $\tilde{ \vb C} =  \tilde C_{ijkl} \vb r^i \otimes \vb r^j \otimes \vb r^k \otimes \vb r^l$, in which case the condition for Maxwell-Betti reciprocity is
\begin{align}
\tilde C_{ijkl} = \tilde C_{klij} \label{eq:symcond}
\end{align}
If $\tilde C_{ijkl} \neq \tilde C_{klij}$, then cycles of deformation can be performed which extract energy from the medium despite starting and ending in the same configuration~\cite{scheibner2020}. Typically in linear elasticity, the distinction between $\vb S$ and $\sigmab$ can be ignored due to the smallness of the strain. However, in the presence of an ambient stress (i.e. $\sigmab^0$ or $\vb S^0$), this is not the case. 
The Piola transformation states $J \vb J^{-T} \cdot \sigmab =  \vb S  $, where $J = \det \vb J$ and $\vb J^{-T} $ is the inverse transpose. Thus, to linear order in $\varepsilon $
\begin{align}
    \vb S = \vb \sigmab^0 + \sigmab^0  \divr \vb u - \gradr \vb u^T \cdot \sigmab^0   +  \vb C : \gradr \vb u   +\order{\varepsilon^2}
\end{align}
Hence, to leading order in $\varepsilon$, one finds $\vb S^0= \sigmab^0$ and 
\begin{align}
    \tilde C_{ijkl} = C_{ijkl} + \sigma^0_{ij} g_{kl} - \sigma^0_{kj} g_{ l i} \label{eq:tenrel}
\end{align}
where $\sigmab^0 = \sigma^0_{ij} \vb r^i \otimes \vb r^j$, and $\vb C = C_{ijkl} \vb r^i \otimes \vb r^j \otimes \vb r^k \otimes \vb r^l$. Notably, in the presence of an ambient stress $\sigmab^0$, the relationship $\tilde C_{ijkl} = \tilde C_{klij}$ does not imply $C_{ijkl} = C_{klij}$. Hence, in the presence of an ambient stress, the tensor $C_{ijkl} $ can be asymmetric when $\tilde C_{ijkl}$ is symmetric. 
Although $\tilde {\vb  C } $ is the tensor whose symmetry implies Maxwell-Betti reciprocity, $\vb C$ is often experimentally more relevant since the Cauchy stress $\sigmab$ corresponds to the current of momentum in the deformed material and does not rely on a reference state. 

In \S\ref{sec:tensor}, we will systematically enumerate the content of $C_{ijkl}$ and $\tilde C_{ijkl}$ for isotropic 2D materials. As we shall see in \S\ref{sec:micromac}, the inclusion of an ambient stress is important since generic microscopic interactions of interest have a nonvanishing torque density or pressure.  
In \S\ref{sec:examples}, we will present the paradigmatic case of particles interacting via transverse $\frac1r$ forces, in which $C_{ijkl}$ contains antisymmetric moduli, but Maxwell-Betti reciprocity is respected due to an ambient torque density. But first, to clarify the meaning of the above symbols, we provide a simple concrete example.

\medskip 

{\bf Example 1. } Here we provide a pedagogical example to illustrate the distinction between the Piola-Kirchhoff and the Cauchy stress. Let us use the coordinate system $\vb r(q) = q^x \vb{\hat x} + q^y \vb{\hat y}$, in which case $\vb r_x = \vb{\hat x} $ and $\vb r_y= \vb{\hat y} $. We will consider a deformation that consists of a pure dilation set by $\lambda$ and a counter-clockwise rotation through an angle $\theta$ :
\begin{align}
    \vb R( q) = (1+\lambda) (q^x \cos \theta + q^y \sin \theta ) \vb {\hat x} + (1 +\lambda) (q^y \cos \theta -q^x \sin \theta ) \vb {\hat y}. 
\end{align}
Writing $\vb J = J_{ij} \vb r^i \otimes \vb r^j$, the deformation gradient tensor is given by:
\begin{align}
    \mqty( J_{xx} & J_{xy} \\ J_{yx} & J_{yy} )= (1+\lambda )\mqty( \cos \theta & -\sin \theta \\ \sin \theta & \cos \theta ) 
\end{align}
The displacement gradient is given by $\gradr \vb u = \vb J -\vb 1$, or 
\begin{align}
    \mqty( (\gradr  u)_{xx} & (\gradr u)_{xy} \\ (\gradr u)_{yx} & (\gradr u)_{yy} )= (1+\lambda )\mqty( \cos \theta & -\sin \theta \\ \sin \theta & \cos \theta ) - \mqty (  1 & 0 \\ 0 & 1) = \mqty( \lambda & - \theta \\ \theta & \lambda ) +\order{\varepsilon^2}
\end{align}
where we have taken $\lambda$ and $\theta$ to be of order $\varepsilon$ and $\gradr \vb u = (\gradr u)_{ij} \vb r^i \otimes \vb r^j$.  
Suppose that the medium has an ambient Cauchy stress before deformation: $\sigmab = \sigmab^0 = - p_0 \vb 1 + \tau_0 \boldsymbol{\epsilon} $, or concretely:
\begin{align}
\mqty( \sigma^0_{xx} & \sigma^0_{xy} \\ \sigma^0_{yx} & \sigma^0_{yy} ) =  \mqty( - p_0 & \tau_0 \\ -\tau_0 & -p_0 )
\end{align}
where $\sigmab =\sigma_{ij} \vb r^i \otimes \vb r^j$. 
Then the Piola-Stress is given by $\vb S = J \vb J^{-T} \cdot \sigmab  $, or explicitly:
\begin{align} 
\mqty(S_{xx} & S_{xy} \\ S_{yx} & S_{yy} ) =& (1+\lambda ) \mqty( \cos \theta & -\sin \theta \\ \sin \theta & \cos \theta ) \mqty( - p_0 & \tau_0 \\ -\tau_0 & -p_0 ) \\
\approx & \mqty( - p_0 & \tau_0 \\ -\tau_0 & -p_0 ) + \tau_0 \mqty(  \theta  & \lambda \\ -\lambda & \theta  ) + p_0 \mqty( - \lambda  & - \theta  \\ \theta  & -\lambda  ) + \order{\varepsilon^2}
\end{align} 
where $\vb S = S_{ij} \vb r^i \otimes \vb r^j$.
Even if we take $C_{ijkl}=0$, the ambient stress nonetheless gives a nominal dependence of $\vb S$ on the $\gradr \vb u$. Indeed, one finds that:
\begin{align}
    \tilde C_{ijkl} = - \frac12 p_0 (\delta_{ij} \delta_{kl} + \epsilon_{ij} \epsilon_{kl} ) + \frac12  \tau_0 (  \delta_{ij} \epsilon_{kl} -\epsilon_{ij} \delta_{kl} ) \label{eq:specific} 
\end{align}
Notice that the term proportional to $\tau_0$ in Eq.~(\ref{eq:specific}) is antisymmetric under $ij \leftrightarrow kl$. Consequently, one can extract energy by a closed cycle of deformations. This is physically intuitive because the torque density (corresponding to the antisymmetric part of $\sigmab$) is constant no matter what the state of dilation. Hence, one can dilate the solid and effectively increase the lever arm by which it does work on its surroundings when it is rotated.  Then one can return the solid to its initial radius and rotate it back with the material absorbing less energy than it put out. 

\medskip

\subsection{Isotropic 2D elastic modulus tensor} \label{sec:tensor}
Here we present the general form of the elastic modulus tensor $\vb C$ and $\tilde {\vb C} $ in two-dimensional isotropic materials. 
As illustrated in Eq.~(\ref{eq:el}) of the main text, we will use the following basis for two-by-two matrices:
\begin{align}
\taub^0 =& \vb 1  &  \taub^1=& -\boldsymbol{\epsilon} &  \taub^2 =& \vb{ \hat x} \otimes \vb{\hat  x} - \vb{\hat  y} \otimes \vb{\hat  y} & \taub^3 =& \vb{\hat  x} \otimes \vb{\hat  y} + \vb{\hat  y} \otimes \vb{\hat  x}  
\end{align}
Using the $\taub^\alpha$, we define $u^\alpha = \taub^\alpha : \gradr \vb u$ and $\sigma^\alpha = \taub^\alpha : \sigmab$.  In this basis, the elastic modulus tensor may be expressed as a four-by-four matrix $C^{\alpha \beta} = \frac12 \tau^\alpha_{ij} C^{ijmn} \tau^\beta_{mn}$. Thus, the constitutive equations may be written as:
\begin{align}
    \mqty( \sigma^0 \\ \sigma^1 \\ \sigma^2 \\ \sigma^3) = \mqty( \sigma^0_0 \\ \sigma^1_0 \\ \sigma^2_0 \\ \sigma^3_0)+
    \mqty( C^{00} & C^{01} & C^{02} & C^{03} \\ 
    C^{10} & C^{11} & C^{12} & C^{13} \\ 
    C^{20} & C^{21} & C^{22} & C^{23} \\ 
    C^{30} & C^{31} & C^{32} & C^{33} \\ 
    )
    \mqty( u^0 \\ u^1 \\ u^2 \\ u^3). 
\end{align}
where $\sigma_0^\alpha$ corresponds to the ambient stress. 
The pictorial notation in the main text corresponds to the physical interpretation of each component: $u^0$ (\dilation) represents dilation, $u^1$ (\rotation) represents rotation,  $u^2$ (\uone) represents shear strain with an axis of elongation in the $x$-direction, and $u^3$ (\utwo) represents shear strain with an axis of elongation at a positive $45^\circ$ angle. Similarly, $\sigma^0$ (\pressure) represents the isotropic stress (negative of pressure), $\sigma^1$ (\torque) represents an internal torque density, and $\sigma^2$ (\strone) and $\sigma^3$ (\strtwo) represent shear stresses.

Notice that $C^{\alpha \beta }$ has 16 independent components when no physical restrictions are imposed.  
If the Cauchy stress is required to be symmetric, then $\sigma^1_0=C^{1 \alpha} =0$ for all $\alpha$. This is the case when there are no internal sources of angular momentum. If no stress is induced by solid-body rotations, then $C^{\alpha 1} =0$ for all $\alpha$.  
Next, we consider the role of isotropy, which states that the elastic modulus tensor is invariant under rotations. Upon a passive rotation of the coordinate system through angle $\theta$, the elastic modulus tensor transforms as 
\begin{align}
 C'_{ijmn} = \mathcal{R}_{ik} \mathcal{R}_{jl} \mathcal{R}_{mp} \mathcal{R}_{nq}C^{klpq} \quad  \text{ where } \quad 
\boldsymbol{\mathcal{R}} = \cos \theta \vb 1 +  \sin \theta \boldsymbol{\epsilon} . 
\end{align}
Hence $C^{\alpha \beta}$ transforms as $C'^{\alpha \beta} = R^{\alpha \gamma} R^{\beta \sigma} C^{\gamma \sigma}$, where 
\begin{align}
    R^{\alpha \beta} = \mqty( 1 & 0 & 0 & 0 \\ 0 & 1 & 0 & 0 \\
                        0 & 0 & \cos 2 \theta & \sin 2 \theta \\
                        0& 0 & -\sin 2 \theta & \cos 2 \theta ). 
\end{align} 
Requiring that $C'^{\alpha \beta} = C^{\alpha \beta}$ for all $\theta$ implies that $C^{\alpha \beta}$ must take the form:
\begin{align}
    C^{\alpha \beta} = 2\mqty( B & \Lambda & 0 & 0 \\ A & \Gamma & 0 & 0 \\
                                0 & 0 & \mu & K^o \\ 0 & 0 & - K^o & \mu).
\end{align} 
Finally, we note that $C_{ijmn} = \frac12 \tau^\alpha_{ij} \tau^\beta_{mn} C^{\alpha \beta}$. Thus, in standard Cartesian coordinates, the elastic modulus tensor reads:
\begin{align}
    C_{ijmn} =& B \delta_{ij} \delta_{mn} + \mu (\delta_{im} \delta_{jn} + \delta_{jm} \delta_{in} - \delta_{ij} \delta_{mn}) - A \epsilon_{ij} \delta_{mn} + K^o E_{ijmn} + \Gamma \epsilon_{ij} \epsilon_{mn} - \Lambda \delta_{ij} \epsilon_{mn} \label{eq:bigten}
\end{align}
where 
\begin{align}
    E_{ijmn} = \frac12 ( \epsilon_{im} \delta_{jn} + \epsilon_{in} \delta_{jm} + \epsilon_{jm} \delta_{in} +\epsilon_{jn}\delta_{im} ) .
\end{align}
Requiring that the ambient stress be isotropic imposes $\sigma^2_0 = \sigma^3_0 =0$. 
Finally, Maxwell-Betti reciprocity corresponds to the condition  $\tilde C^{\alpha \beta} = \tilde C^{\beta \alpha }$ where $\tilde C^{\alpha \beta} = \frac12 \tau^{\alpha}_{ij}  \tilde C^{ijmn} \tau^{\beta}_{mn}$ and is related to $C^{\alpha \beta}$ via
\begin{align}
    \tilde C^{\alpha \beta} = C^{\alpha \beta} + 
    \mqty( \sigma^0_0 & - \sigma^1_0 & - \sigma^2_0 & - \sigma^3_0 \\
    \sigma^1_0 & \sigma^0_0 & \sigma^3_0 & - \sigma^2_0 \\
    \sigma^2_0 & -\sigma^3_0 & -\sigma^0_0 & - \sigma^1_0 \\
    \sigma^3_0 & \sigma^2_0 & \sigma^1_0 & - \sigma^0_0 \\
        )
\end{align}
With our conventions, $\sigma_0^0 = -2 p_0$ and $\sigma^1_0 = - 2 \tau_0$.

\subsection{Coarsegraining lattices} \label{sec:micromac}
Here we provide a coarsegraining procedure for the lattices containing a single particle per unit cell, most notably the hexagonal lattice. We consider a homogeneous strain $\vb J = \vb 1 + \vb U$, where $\vb U$ is a constant tensor. Working in Cartesian coordinates, the Cauchy stress is given by 
\begin{align}
    \sigma_{ij} = \frac{-1}{2 J V_\text{cell}   } \sum_{\alpha } J_{ki} \scr_k^{\alpha} F_{j} \left[ \vb J^T \cdot \scrb^{\alpha} \right] \label{eq:cstress} 
\end{align}
where $\scrb^\alpha$ is the position of particle $\alpha$ in the undeformed lattice, $\vb F(\scrb)$ is the interparticle force law, $V_\text{cell}$ is the area of a unit cell.  (Equation~(\ref{eq:cstress}) can be obtained from Eq.~(\ref{eq:coarsestress}) upon using a smoothing function $g(\vb R) =1/V_\text{system}$ where $V_\text{system}$ is the area of the solid). 
The elastic modulus tensor is given by:
\begin{align}
    C_{ijmn} =& \eval{\pdv{\sigma_{ij}}{U_{mn}}}_{\vb U=0} \\
    =&\frac{-1}{2V_\text{cell}} \sum_\alpha \qty[  \eval{F_j }_{\scrb^\alpha} \qty(\delta_{in} \scr^\alpha_m - \scr_i^\alpha \delta_{mn} ) +  \eval{\partial_n F_j}_{\scrb^\alpha} r^\alpha_i \scr^\alpha_m ] \label{eq:ccoarse} 
\end{align}
where we have used the fact that $J \approx 1 + \Tr \vb U$. 
Next we work in the nearest neighbor approximation in which we only include the lattice sites $\scrb^\alpha$ nearest to the origin. For the hexagonal lattice, this includes the six points $\scrb^\alpha = a\qty(\pm 1, 0)^T, a\qty(\pm \frac12, \pm \frac{\sqrt{3}}2 )^T$ where $a$ is the lattice spacing.  We are often interested in forces of the form of Eq.~(\ref{eq:pair}), repeated here:
\begin{align}
\vb F (\scrb ) = F^\parallel(\scr) \scrbh  - F^\perp (\scr)\bph  
\end{align}
We can then define  the quantities $F^{\parallel(\perp)}_0 =  F^{\parallel(\perp)}(a) $ and $k = - \eval{\pdv{F^\parallel}{\scr}}_a$ and $k^a =- \eval{ \pdv{F^\perp}{\scr}}_a$. In this case, Eq.~(\ref{eq:ccoarse}) becomes:
\begin{align}
    C_{ijmn} = \frac{-1}{ \sqrt{3} a^5 } \sum_{\alpha } a^2( F_0^\parallel \delta_{jl} + F^\perp_0 \epsilon_{jl}  ) (\delta_{in} \scr_l^\alpha \scr_m^\alpha -  \delta_{mn} \scr_i^\alpha \scr_l^\alpha   ) 
    + ( F^\parallel_0 \epsilon_{jl} -F^\perp_0 \delta_{jl}  ) \epsilon_{nk} \scr^\alpha_k \scr^\alpha_l \scr^\alpha_i \scr^\alpha_m 
    - a (k \delta_{jl} +k^a \epsilon_{jl} )  \scr_n^\alpha \scr_l^\alpha \scr_i^\alpha \scr_m^\alpha \label{eq:express} 
\end{align} 
where we have used the fact that $V_\text{cell} = \frac{\sqrt 3}2 a^2$. Directly placing Eq.~(\ref{eq:express}) into the definition $C^{\alpha \beta} = \frac12 \tau^\alpha_{ij} C^{ijkl} \tau_{mn}^\beta$ yields
\begin{align}
    C^{\alpha \beta} = 2 \mqty( \frac{\sqrt3}2 \qty( k+\frac{F^\parallel_0}{a} )  & 0 & 0 & 0 \\
    \frac{\sqrt3}{2} \qty(k^a+ \frac{F^\perp_0}{a}) & 0 & 0 & 0 \\
    0 & 0 & \frac{\sqrt{3}}4  \qty( k- \frac{3 F^\parallel}a ) &  \frac{\sqrt{3}}4 \qty( k^a-\frac{3 F^\perp}a  ) \\
    0 & 0 &  -\frac{\sqrt{3}}4  \qty(k^a- \frac{3 F^\perp}a  ) &\frac{\sqrt{3}}4 \qty( k- \frac{3 F^\parallel}a )
    )
\end{align}
which implies that the moduli are:
\begin{align}
    B = \frac{\sqrt3}2 \qty(k+ \frac{F^\parallel_0}{a}  )  
    \quad A=\frac{\sqrt3}{2} \qty( k^a + \frac{F^\perp_0}{a} ) 
    \quad \mu = \frac{\sqrt{3}}4 \qty( k-\frac{3 F^\parallel}a  )
    \quad K^o =\frac{\sqrt{3}}4  \qty( k^a - \frac{ 3 F^\perp}a  ) \label{eq:modulipre}
\end{align}
Notice that when $F^\parallel =F^\perp =0$, an additional symmetries known as the Cauchy relations emerges: $C_{ijmn} = C_{mjin}$, see, e.g., Ref.~\cite[Sec~III]{Born1954dynamical}. Setting $C_{ijmn} - C_{mjin}=0$, we find that the Cauchy relations in isotropic media imply:
\begin{align}
0=&B+\Gamma- 2\mu    \\ 
0=&A-2 K^o - \Lambda 
\end{align}
Finally, note that 
\begin{align}
    \tilde C^{\alpha \beta} = 2 \mqty( \frac{\sqrt3}2 k   &  \frac{\sqrt 3}2 \frac{F^\perp}a & 0 & 0 \\
    \frac{\sqrt3}{2} k^a & -\frac{\sqrt 3}2 \frac{F^\parallel}a & 0 & 0 \\
    0 & 0 & \frac{\sqrt{3}}4  \qty( k- \frac{F^\parallel}a ) &  \frac{\sqrt{3}}4 \qty(k^a- \frac{ F^\perp}a  ) \\
    0 & 0 &  -\frac{\sqrt{3}}4  \qty(k^a- \frac{  F^\perp}a  ) &\frac{\sqrt{3}}4 \qty( k- \frac{F^\parallel}a )
    ) \label{eq:tildecmat}
\end{align}
Similarly, the ambient stress is given by $\eval{\sigma_{ij}}_{\vb U=0}$, which yields $p_0 = \sqrt{3} F^\parallel_0/a $ and $\tau_0 =  \sqrt{3} F^\perp_0/a$. For crystals with multiple atoms per unit cell, see~\cite{scheibner2020,Fruchart2020Symmetries}.

\subsection{Illustrations with exactly solvable models} \label{sec:examples}

\subsubsection{Microscopic models for $\Lambda$ and $\Gamma$}
In the main text, we focus on interactions of the form of Eq.~(\ref{eq:pair}), which give rise to the moduli $B$, $\mu$, $A$, and $K^o$. For completeness, here we discuss example models that give rise to the moduli $\Lambda$ and $\Gamma$.  
Both these moduli imply that stress is induced by rotations. Since rotations do not change interparitcle distances, interparticle forces that depend only on the interparticle distance $r$ are not enough to give rise to $\Gamma$ and $\Lambda$. To give rise to $\Lambda$ and $\Gamma$, the particles must be able to gauge their orientation with respect to their embedded space, which requires interaction with a substrate, external field, or surrounding medium.  

For example, the modulus $\Gamma$ can arise due to interactions with a patterned potential on a substrate, as considered in Ref.~\cite{Nelson1979}, where it is referred to as $\gamma_R$. When interactions with a substrate are present, conservation of linear momentum among the particles is often only approximate since the substrate can absorb momentum as it mediates interactions between particles. However, $\Gamma$ can also arise when linear momentum conservation between particles is exact. For example, consider a collection of magnetic dipoles $\vb d^\alpha$ situated on a triangular lattice. Each of the dipoles are connected by springs with a bending stiffness that depends on the angle between the dipole and the bond.  Explicitly, the system is defined by the following potential energy:
\begin{align}
    V (\underline{ \vb d} , \underline{ \vb x } ) =  \sum_{\alpha } \qty{   H   \vb{\hat  y} \cdot \vb { \hat d^\alpha}  + \sum_{\beta \in N(\alpha) }   \frac{\kappa }{2}    \qty[ \theta^{\alpha \beta} - \theta_0^{\alpha \beta}  ]^2  } 
\end{align}
where $N(\alpha)$ are the neighbors of $\alpha$,  $\theta^{\alpha \beta} = \arccos(\vb{ \hat d}^{\alpha } \cdot \scrbh^{\alpha \beta }  )$ is the angle between the dipole $\vb d^\alpha$ and the bond vector $\scrb^{\alpha \beta} = \vb x^\alpha - \vb x^\beta $,
$\theta_0^{\alpha \beta}$ is the angle formed between the bond $\scrb^{\alpha \beta}$ and $\vb{\hat  y}$ when the lattice is undeformed, and $H$ is the strength of the magnetic field in the $\vb{\hat y}$ direction. By construction, this system is conservative and only depends on relative positions $\scrb^{\alpha \beta}$, hence linear momentum is exactly conserved among the particles. In the limit that $H \gg \kappa $, the dipoles will be frozen in the $\hat y$ direction and we can ignore their orientational dynamics. In this case, the effective potential energy becomes 
\begin{align} 
V_\text{eff}(\underline{ \vb x} ) = \frac {\kappa } { 2} \sum_{\alpha} \sum_{ \beta \in N(\alpha) }  [ \theta^{\alpha \beta} - \theta_0^{\alpha \beta}]^2 
\end{align}
In the continuum limit,
one then finds $\Gamma \propto \kappa $. Likewise, to realize the coupling $\Lambda$, one could consider an engineered spring that senses the quantity $\theta^{\alpha \beta}$ using electronics and then induces linear tension along the spring. This would couple rotations to pressure in a way that exactly conserves linear momentum.

\medskip 

\subsubsection{Odd elastic moduli without broken Maxwell-Betti reciprocity} \label{sec:microexample2}
The force $\vb F(\scr )$ in Eq.~(\ref{eq:pair}) generically violates MBR when $F^\perp(\scr) \not \equiv 0$ since $\nabla \times \vb F = \frac1\scr \pdv{\scr} (\scr F^\perp) $. However, as noted in \S\ref{sec:potmicro}, an illustrative counterexample is $F^\perp (\scrb) = \frac 1 \scr$. In this case, $\vb F (\scr) $ is closed, i.e. it is locally the gradient of a potential $V (\scrb) = \arctan(\scr_y/\scr_x )$, so there is no infinitesimal deformation cycle that can extract work. Accordingly, the matrix $\tilde C^{\alpha \beta}$ in Eq.~(\ref{eq:tildecmat}) is manifestly symmetric since $k^a = F^\perp/a =  1/a^2 $, where $a$ is the lattice spacing.
Nonetheless, the transverse forces still give rise to odd elastic moduli since $A$ and $K^o$ in Eq.~(\ref{eq:modulipre}) are both nonzero. 
Although there is no infinitesimal cycle that gives extracts energy, there is a topologically nontrivial cycle which  extracts energy. This cycle consists of rotating a single bond (or the entire macroscopic solid) by $2 \pi $.  In the language of Hodge theory, the ability to extract work on topologically nontrivial cycles but not on contractible local cycles is a statement that $\vb F =\frac1\scr \bph $ is the harmonic function corresponding to a nontrivial homotopy classes of the punctured plane. 
Physically, the $1/\scr$ fall off is consistent with a constant torque exerted between particles: as the particles move further apart, the lever arm grows as $\scr$ so the force falls off as $1/\scr$.

\medskip 

{\bf Example 2.} It is instructive to see in a concrete setting how the work done by ambient stress cancels with the work done by the odd moduli. Consider again the parameterization of $\vb R(q,t)$ from Example 1, and consider a deformation protocol $\lambda = \cos t$ and $\theta = \sin t$, in which dilation and rotation are applied out of phase. Using the fact that $\Grad \vb v = \vb J^{-1} \cdot \dot{\gradr \vb u}$, one obtains:
\begin{align} 
\mqty( (\Grad \vb v)_\text{xx} &  (\Grad \vb v)_\text{xy} \\  (\Grad \vb v)_\text{yx} &  (\Grad \vb v)_\text{yy} ) = \frac{\dot \lambda}{1+ \lambda } + \dot \theta  \mqty(0 & -1 \\ 1 & 0 )
\end{align} 
The Cauchy stress is given by $\sigmab = \sigmab^0 + \vb C : \gradr \vb u+\order{\varepsilon^2} $,
\begin{align}
    \mqty(\sigma_{xx} & \sigma_{xy} \\ \sigma_{yx} & \sigma_{yy}) =  \mqty( 0 &  \tau_0 \\ -\tau_0 & 0 ) + A \mqty( 0 & -2\lambda \\ 2\lambda & 0  ) +\order{\varepsilon^2}
\end{align}
where $\tau^0 = A= \sqrt 3 /a^2  $. Consider a patch of material $\regA$ whose area in the reference material is $\abs{\regA} = \int_\regA \dd^2 r$. The elastic work done by $\regA$ over a single period is given by :
\begin{align} 
W(\regA) =& \int_0^{2\pi} \int_\regA  \sigmab : \Grad \vb v  \; \dd^2 R \; \dd t\\
=& \abs{\regA}  \int_0^{2 \pi } J \sigmab : \Grad \vb v  \; \dd t \\
=& \abs{\regA}  \int_0^{2 \pi } \qty[ - 2\tau_0 \dot \theta (1+ 2\lambda) + 4 A \lambda \dot \theta  ] \; \dd t \\
=& \abs{\regA} 4 \pi ( -  \tau_0  + A   )
\end{align} 
where in the final step we used the prescribed form of $\theta(t)$ and $\lambda(t)$. Notice that for $F^\perp(r) = 1/r$, one has $\tau_0 = A$. Thus, the contributions from $\tau_0$ and the elasticity $A$ cancel exactly.

\subsubsection{Continuum elasticity of media with transverse interactions}
\label{sec:gryoel}
Here we discuss how systems with transverse interactions (e.g. models of skyrmion and gyroscope lattices) can be cast mathematically as a special case of odd elasticity. 
The continuum analogue to Eq.~(\ref{eq:perplag}) is
\begin{align}
    L = \rho_0 \, \frac{\dot {\vb u} \cdot \epsilonb \cdot \vb u}2 - V(\vb J) 
\end{align}
A variation of the action $S = \int  \int  L \,  \dd^2 r \, \dd t $ yields the following equation of motion:
\begin{align}
     \rho_0 \, \epsilonb \cdot \dot{ \vb u} = \divr \bar {\vb S} \label{eq:eom1}
\end{align}
where $\bar {\vb S} = \pdv{V}{\vb J}$ is the derivative of the potential energy. Equation~(\ref{eq:eom1}) may also be written as
\begin{align}
     \rho \, \epsilonb \cdot \dot{ \vb u} = \Div \bar \sigmab
\end{align}
where $\rho = \frac{\rho_0}{J}$ and $\bar \sigmab = \frac1J \vb J^T \cdot \bar{\vb S}$. 
The stresses may be expanded as $\bar {\vb S} = \bar {\vb S}^0 +  \tilde{\bar {\vb C}} : \gradr \vb u$ and $\bar \sigmab = \bar \sigmab^0 + \bar {\vb C} : \gradr \vb u$.
Notice that $\tilde{\bar{\vb  C}}$ and $\bar {\vb S}^0$ cannot violate the condition for MBR in Eq.~(\ref{eq:symcond}) since $\bar{\vb S}$ is the derivative of a potential. However, we may write ${\vb S} = -\bar{\vb S} \cdot \epsilonb$ and $\sigmab =- \bar \sigmab \cdot \epsilonb $ and accordingly ${\vb S}^0 = -\bar{\vb S}^0 \cdot \epsilonb$, $\tilde C_{ijkl} = \epsilon_{jm} \tilde{ \bar C}_{imkl}$,
$\sigmab^0 = -\bar \sigmab^0 \cdot \epsilonb$, and $C_{ijkl} = \epsilon_{jm} \bar{C}_{imkl}$. With this notation, the equations of motion become 
\begin{align}
    \rho_0 \, \dot {\vb u} = \divr {\vb S}
\end{align}
or equivalently
\begin{align}
   \rho \,  \dot {\vb u} = \Div \sigmab. 
\end{align}
resembling standard first order dynamics. 
Notice that $\tilde{\vb C}$ generically violates MBR [Eq.~(\ref{eq:symcond})] and $\vb C$ generically features odd elastic moduli. We note, however, that the quantity $\sigmab : \Grad \vb v$ does not correspond to physical power (i.e. the rate of change of $V$) in this context. 

If the lattice is isotropic, then $\bar {\vb C}$ will consist of a bulk and shear modulus $\bar B$ and $\bar \mu$. By contrast, $\vb C$ contains the moduli $A= \bar B$ and $K^o = \bar \mu$. More generally, suppose $C_{ijkl} = R_{jm}(\theta) \tilde C_{imkl}$ with $\tilde C_{ijkl}$ containing only the moduli $\bar B$ and $ \bar \mu$. Then $C_{ijkl}$ consists of $\frac A B = \frac {K^o}{\mu} =\tan \theta $ and $\frac B {\bar B} = \frac \mu {\bar \mu} = \cos \theta$. For this set of moduli, $\nu^o =0$ [see Eq.~(\ref{eq:nuo}) in the main text], indicating that the effects of the odd elasticity vanish from the static strain distribution around dislocations, assuming appropriate boundary conditions are applied.

\section{Defects in isotropic 2D nonreciprocal crystals} \label{sec:solutions}

In this section, we explicitly solve for the continuum displacement, strain, and stress fields for disclinations, dislocation, and point defects in the presence of odd elastic moduli. In \S\ref{sec:mech}, we formulate precisely the general mathematical problem to be solved. 
In \S\ref{sec:disc}-\ref{sec:disl} we provide explicit solutions for dislocations and disclinations while setting $\Lambda = \Gamma=0$, as assumed in the main text. The effects of $A$ and $K^o$ are evident in the stress and strain fields. In \S\ref{sec:rot}, we then consider the modifications to the dislocation solution introduced by nonzero $\Lambda$ and $\Gamma$.  
In \S\ref{sec:int}, we provide solutions for point defects and local torques at the core of dislocations. Finally, in \S\ref{sec:airy}, we remark on the Airy stress function.

\subsection{Mechanical stability of topological defects} \label{sec:mech} 

\begin{figure}
    \centering
    \includegraphics[width=0.8\textwidth]{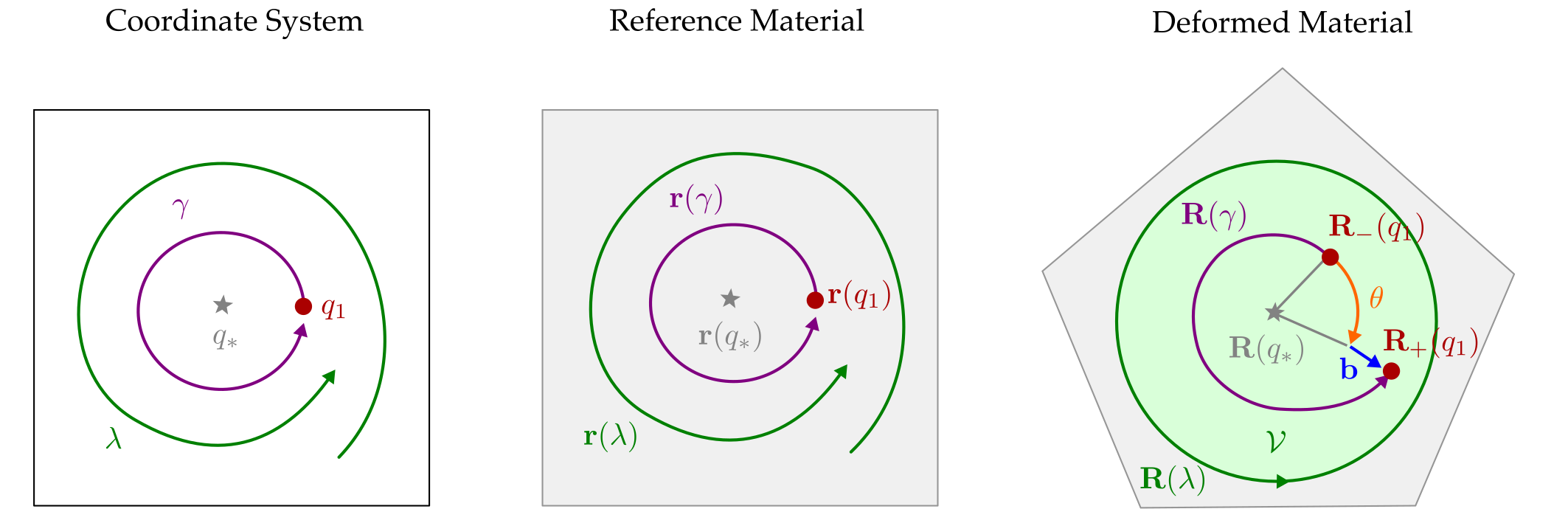}
    \caption{ {\bf Geometric properties of disclinations and dislocations.}~We illustrate the dislocation angle $\theta$ and Burgers vector $\vb b$, as well as the contours used in \S\ref{sec:mech}} 
    \label{fig:ddef}
\end{figure}

We now introduce the notion of a topological defect within nonlinear elasticity. As before, we will assume a continuous map $\vb r(q)$ from a coordinate system $q^i$ to the cartesian plane. We will now allow the map $\vb R(q)$ to be, in general, multivalued. We will require the metric induced by the deformation, $G_{ij}(q) = \vb R_i \cdot \vb R_j $, to be continuous and single valued everywhere except at a single point $q_*$ representing the location of the topological defect. Given the value of $\vb J $ at a point $q_1$, the value of $\vb J$ at point $q_2$ is given by:
\begin{align}
    \vb J (q_2) = \exp(\int_\gamma  \vb r_i \cdot \Pib (q) \dd q^i ) \cdot \vb J(q_1) ,
\end{align}
where $\gamma$ is the path through coordinate space connecting $q_1$ and $q_2$, the exponential is path ordered, and 
\begin{align}
    \Pib= \qty( \vb R^n \cdot \pdv{ \vb R_m}{q^k} - \vb r_m \cdot \pdv{\vb r^n}{q^k} ) \vb r^k \vb r^m \vb r_n 
\end{align}
see e.g. Ref.~\cite{zubov2008nonlinear} for a derivation. (One can show that $\Pib$ can be determined entirely in terms the metric $G_{ij}(q)$.) 
Since the deformed solid is embedded in the flat Cartesian plane,
we impose the compatibility constraint that $G_{ij}(q)$ have zero Riemann curvature.
Subject to this compatibility condition, one can show that~\cite{zubov2008nonlinear} 
\begin{align}
\exp(\int_\gamma  \vb r_i \cdot \Pib (q) \dd q^i ) = \vb 1
\end{align} 
whenever $\gamma$ is a closed loop that does not contain the defect $q_*$. If $\gamma$ encloses the defect, then the integral takes the form:
\begin{align}
\exp(\int_\gamma  \vb r_i \cdot \Pib (q) \dd q^i ) = e^{ -\theta \epsilonb } \label{eq:rot} 
\end{align} 
The right-hand side of Eq.~(\ref{eq:rot}) is a rotation matrix through angle $\theta$ representing the charge of a disinclination. The value of $\theta$ is independent of the precise path taken. 
Suppose $q_1 = q_2$ and let $\vb J_\pm$ represent $\vb J(q_1)$ as approached from the $q_1$ (${-}$) or $q_2$ (${+}$) side of $\gamma$. Then we have $\vb J_+ = e^{- \theta \epsilonb } \vb J_-$. Since $\vb J_\pm  = \gradr \vb R_\pm$, we have $\vb R_+ =  e^{-i\theta \epsilonb } \cdot \vb R_- + \vb b $, where $\vb b$ is a constant translation known as the Burgers vector. See Fig.~\ref{fig:ddef} for an illustration. 

To derive the displacement, strain, and stress field about a defect, one often minimizes an elastic energy subject the topological constraints imposed by the defect charges $\theta$ and $\vb b$. However, in absence of an elastic energy, we must begin directly from the definition of mechanical equilibrium. As described in \S\ref{sec:el}, we assume that linear momentum is conserved, and therefore the net force $\mathbb{F}$ on any patch of material is given by the flux of through its boundary. Let $\lambda$ denote a contour of the coordinate space such that $\vb R(\lambda)$ is a closed contour in the deformed material enclosing the physical material $\VV$. 
Then we require 
\begin{align}
    \mathbb F(\VV) = \oint_{\lambda } \vb{\hat   N} \cdot  \sigmab \;  \dd L =0 \label{eq:netForce}
\end{align}
By contrast, we will generally not assume that angular momentum is conserved, and therefore the net torque acting on the material enclosed by $\lambda$ is the sum of two contributions: the flux of angular momentum at its boundary and the sources integrated over its area:
\begin{align}
    \mathbb{T}(\VV) = -\oint_{\lambda } \vb { \hat N} \cdot  \sigmab \cdot \boldsymbol{ \epsilon} \cdot  \vb R \; \dd L + \int_{ \VV} \mathcal{T} \; \dd^2 R  \label{eq:netTorque}
\end{align}
where $\mathcal{T} (\vb R)$ is the torque density in the deformed solid.  Away from the core, $\mathcal{T} =-  \boldsymbol{ \epsilon}  : \sigmab \;  (\equiv -2\tau ) $. Hence, if $\VV$ does not contain the topological defect, then we may apply Stokes' theorem to Eq.~(\ref{eq:netForce}) and Eq.~(\ref{eq:netTorque}) to obtain:
\begin{align}
   \mathbb{F}(\VV ) = \int_\lambda \Div \sigmab \; \dd^2R  \quad \text{and} \quad 
     \mathbb{T}(\VV) =- \int_\lambda \Div \sigmab  \cdot \boldsymbol{ \epsilon} \cdot \vb R  \; \dd^2 R  \label{eq:stokeForce}
\end{align}
Since Eq.~(\ref{eq:stokeForce}) applies for all choices of $\VV$ that do not enclose the defect core, we find that a necessary condition for mechanical equilibrium is:
\begin{align}
\Div \sigmab =0  \quad \text{ away from the core} \label{eq:cond1} 
\end{align}
However, note that Eq.~(\ref{eq:cond1}) is necessary but not sufficient to ensure mechanical stability. When $\regA$ encloses the topological defect, we must directly apply Eq.~(\ref{eq:netForce}) and Eq.~(\ref{eq:netTorque}). In this case, Eq.~(\ref{eq:netTorque}) takes the form:
\begin{align}
\mathbb{T}(\VV) = -\oint_{\lambda } \vb {\hat N} \cdot \sigmab \cdot \boldsymbol{\epsilon} \cdot \vb R \; \dd L - \int_{\VV \setminus \vb R (q_*) }  \boldsymbol{\epsilon} : \sigmab \; \dd^2 R  +T^c  \label{eq:cond2}
\end{align}
where $\vb R(q_*)$ is the defect location and $T^c$ is the net torque associated with the core of the defect. For a well-posed problem, one can prescribe a stress $\sigmab$ on $\lambda$ that is compatible with $\mathbb{F}(\lambda)=0$. Then one must determine $\vb R(q)$ subject to Eq.~(\ref{eq:cond1}) and Eq.~(\ref{eq:cond2}). We note that the force balance conditions here are formulated in terms of the Cauchy stress $\sigmab$, i.e. the stress in the deformed material. To formulate the static requirements rigorously in the reference material, one must account for the fact that  $\vb r(\lambda)$ is in general not a closed contour in the reference material~\cite{zubov2008nonlinear}. Accordingly, the Piola-Kirchhoff stress $\vb S$ becomes multivalued, and the appropriate value of $\vb S$ is determined by continuity along $\vb r(\lambda)$.

\subsection{Disclinations} \label{sec:disc}

Here we provide an explicit solution for the displacement, strain and stress field surrounding a disclination in a medium with nonzero $B$, $\mu$, $A$, $K^o$. We will work in Cartesian coordinates $\vb r(q) = q^x \vb{\hat x} + q^y \vb{\hat y} $ so that the coordinate system $q^i$ and the components of $\vb r (q)$ are interchangeable and there is no distinction between raised and lowered indices. 
We will assume that $\gradr \vb u $ is of order $\varepsilon$ and we will work to linear order in $\varepsilon$. To leading order in $\varepsilon$, Eq.~(\ref{eq:rot}) for a crystal with $C_6$ symmetry may be written as
\begin{align}
    \frac{s \pi}{3} = \frac12 \oint_\gamma \epsilon_{ij} \partial_k \partial_i u_j \dd r_k \label{eq:top}
\end{align}
where $\gamma$ is an arbitrary counterclockwise path enclosing the disclination at the origin and $\theta = \pi s/3$ for an integer $s$. Equation~(\ref{eq:top}) implies that the disclination takes the form
\begin{align}
    u_i (\vb r) =-\frac{s}{6} \phi \epsilon_{ik} r_k + v_i (\vb r),
\end{align}
where $\phi$ is the polar angle and $v_i(\vb r)$ is a single-valued function. To linear order in $\varepsilon$, the force balance condition $\Div \sigmab=0$ simply takes the form $\partial_i \sigma_{ij} (\vb r)=0$.
Together with the constitutive relation $\sigma_{ij} (\vb r) = C_{ijkl} \partial_k u_l (\vb r)$, we obtain the following differential equation for $v_i (\vb r)$:
\begin{align}
    \Delta (\partial_i v_i) = \frac{s}{3} \qty(\frac{1-\nu}2) \delta (\vb r) \\
    \Delta (\partial_i \epsilon_{ij} v_j) = -\frac{s}{3} \nu^o \delta (\vb r) 
\end{align}
where $\nu$ and $\nu^o$ are, respectively, the Poisson's and odd ratios given by:
\begin{align}
\nu \equiv& \frac{\mu ( B- \mu) + K^o (A-K^o)}{ \mu (B+ \mu) +K^o (A+ K^o) }\\
\nu^o \equiv& \frac{ BK^o - A\mu }{ \mu (B+ \mu) +K^o (A+ K^o) }
\end{align}
For boundary conditions, we will require that $\hat r_i \sigma_{ij}=\hat r_i \sigma^0_{ij}$ along a circle of radius $R$. We obtain the solution
\begin{align}
    u_i (\vb r) = \frac{s}{6} \left\{ - \phi  \epsilon_{ij } x_j  + \frac{(1-\nu)}2 r_i \log(r/R_1)  +\nu^o \epsilon_{ij} r_j \log(r/R_2)    \right\}
\end{align}
Here, $R_1 = R_2= R/\sqrt{e}$ are constants introduced to ensure the stress boundary conditions are satisfied. 
Using $\sigma_{ij} = C_{ijmn} \partial_m u_n$, we obtain the stress:
\begin{align}
    \sigma_{ij} (\vb r) 
    =& (1-\nu) \frac{s}{12} \bigg\{ B \qty[  2 \log (r/R_1) \delta_{ij} -  \qty( \frac{2r_i r_j}{r^2} -\delta_{ij} ) ]-
    A \qty[ 2\log (r/R_2) \epsilon_{ij} + \frac{ r_i \epsilon_{jl}r_l + r_j \epsilon_{il }r_l } {r^2}  ]
        \bigg\}  +\sigma^0_{ij}
    \label{eq:discstress}
\end{align}
In enforcing the stability conditions, we have made a common simplification to replace Eq.~(\ref{eq:cond2}) with a simpler condition: $0=-\oint_{\partial \regA} \hat n_i \sigma_{ij} \epsilon_{jk} r_k \; \dd \ell - \int_\regA \epsilonb : \sigmab \; \dd^2 r$. This amounts to replacing $\vb R$ with $\vb r$ and ignoring the non-closed nature of the contour $\vb  r(\lambda)$. 
While not strictly consistent to linear order in $\varepsilon$,
this simplification is often made even in the solution of passive disclinations~\cite{Nelson1979} because it renders the problem tractable. 
Taking $\regA$ to be a circle of radius $R$, it is readily verified that Eq.~(\ref{eq:discstress}) satisfies the simplified boundary conditions.

We now interpret the effects of $K^o$ and $A$ on the stress and strain. First, we note that $K^o$ and $A$ only modify the pressure $ p (\vb r) = -\sigma_{ii} (\vb r)/2 = -(1-\nu) sB \log(r)/12$ through the coefficient $\nu$ and leave the functional form of $p (\vb r)$ unchanged. Furthermore, the ratio of the isotropic stress to the anitsymmetric stress $\tau (\vb r) = \epsilon_{ij} \sigma_{ij}(\vb r)/2$ is constant in space and is given by the ratio:
\begin{align}
    \frac{\tau (\vb r) }{p(\vb r)} = \frac{A}{B}
\end{align}
To characterize the shear stress, we introduce a local basis for traceless symmetric tensors: $S^{r}_{ij} = \frac{2 r_i r_j }{r^2} - \delta_{ij} $ and $S^\phi_{ij} = ( \epsilon_{ik} r_k r_j + \epsilon_{jk} r_k r_i)/r^2$. 
For $S^r_{ij}$, the axis of extension points along $\hat r_i$ and for  $S^\phi_{ij}$, the axis of extension points along $(\hat r_i-\hat \phi_i)/\sqrt{2}$. When $A=0$, Eq.~(\ref{eq:discstress}) shows that the shear stress is entirely proportional to $S^r_{ij}$. However, when $A$ is nonzero, the axis of shear is everywhere rotated through an angle $\delta \chi$ given by: 
\begin{align}
    \delta \chi (\vb r) = -\frac12 \arctan( \frac{S^\phi_{ij} (\vb r) \sigma_{ij} (\vb r) }{ S^r_{ij} (\vb r) \sigma_{ij} (\vb r) } ) = -\frac12 \arctan (\frac{A}{B})
\end{align}
We can apply a similar analysis to the displacement gradient $\partial_i u_j (\vb r)$. Like the pressure, $K^o$ and $A$ only modify the dilation $\partial_i u_i (\vb r) = (1-\nu ) s \log(r)/3$ through the constant $\nu$. Furthermore, $A$ and $K^o$ induce a local rotation of the axis of shear strain given by:
\begin{align}
    \delta \alpha (\vb r) = -\frac12 \arctan( \frac{S^\phi_{ij} (\vb r) \partial_i u_j (\vb r) }{ S^r_{ij} (\vb r) \partial_i u_j (\vb r) }) = -\frac12 \arctan({\frac{2 \nu^o}{1+\nu}} )
\end{align}
as appears in Eq.~(\ref{eq:shearrot}) of the main text. 

\subsection{Dislocations} \label{sec:disl}

Here we provide an explicit solution for the displacement and strain field surrounding a dislocation in a medium with nonzero $B$, $\mu$, $A$, $K^o$. As in \S\ref{sec:disc}, we will work in Cartesian coordinates $\vb r(q) = q^x \vb{\hat x} + q^y \vb{\hat y} $ and work to 
linear order in displacement gradient. 
The Burgers vector $b_i$ for a pure dislocation is given by:
\begin{align}
    \oint_\gamma \partial_i u_j \dd r_i = b_j \label{eq:burg}
\end{align}
Condition Eq.~(\ref{eq:burg}) implies that the solution for $u_i$ is of the form:
\begin{align}
 u_i (\vb r) = \frac{b_i}{2 \pi} \phi + v_i (\vb r)
\end{align}
where once again $v_i (\vb r) $ is a single-valued function. The requirement that $\partial_i \sigma_{ij} (\vb r) =0$ for $\vb r \neq 0$ yields the following differential equation for $v_i (\vb r)$:
\begin{align}
\Delta (\partial_i v_i) =& \nu b_i \epsilon_{ij}  \partial_j \delta (\vb r) \label{eq:divb} \\
\Delta (\partial_i \epsilon_{ij} v_j) =& (- b_i  +2 \nu^o \epsilon_{ij} b_j ) \partial_i \delta (\vb r) \label{eq:curlb}
\end{align}
As a boundary condition, we require $\sigma_{ij} \to \sigma_{ij}^0$ as $r \to \infty$. We obtain the resulting solution for the displacement field:
\begin{align}
u_i (\vb r) 
    =& \frac1{2\pi} \bigg \{ b_i \phi + \epsilon_{ik} b_k  \frac{(1-\nu)}{2} \log (r) + \frac{ (1+\nu)}{2} \frac{    \epsilon_{im} r_m b_n r_n }{r^2}  - \nu^o  \left [b_i \log(r) - \frac{ \epsilon_{im} r_m r_n \epsilon_{nk} b_k  }{r^2} \right ]\bigg\} \label{eq:udisl}
\end{align}
and the stress is given by:
\begin{align}
    \sigma_{ij} (\vb r)
    =& \frac{(1-\nu)}{2 \pi r^2} \bigg\{ B \left[  r_m \epsilon_{mn} b_n \delta_{ij} - b_k r_k \left( \frac{r_i \epsilon_{jm} r_m + r_j \epsilon_{im} r_m }{r^2} \right)  \right] - A  \left[  r_m \epsilon_{mn} b_n \epsilon_{ij} - b_k r_k \left (  \frac{2r_i r_j }{ r^2}  - \delta_{ij} \right )  \right]
    \bigg\} +\sigma_{ij}^0
    \label{eq:res}
\end{align}
When $A =0$, the shear stress is entirely proportional to $S^\phi_{ij} (\vb r)$. As with the disclination, when $A$ is nonzero we have $\tau (\vb r)/p(\vb r)=A/B$ and the shear stress is rotated by an angle $\delta \chi (\vb r) =- \arctan(A/B)/2$. Moreover, the local dilation is unmodified by $A$ and $K^o$ (except through the value of $\nu$) and the shear strain is locally rotated by $\delta \alpha (\vb r) =-\arctan[2\nu^o/(1+\nu)]/2$.

\subsection{Coupling to rotations} \label{sec:rot} 
In this section, we consider the role of the moduli $\Lambda$ and $\Gamma$. These moduli imply that solid-body rotations induce pressure and torque, respectively. 
Notice that disclinations explicitly require a large rotation in the bond angle field $\epsilon_{ij} \partial_i u_j$, and hence a linear coupling to rotations is not appropriate. However, 
dislocations can still be self-consistently treated using the linear coupling, since the bond-angle field $\epsilon_{ij} \partial_i u_j$ remains small for dislocations.

Using the full elastic modulus tensor in Eq.~(\ref{eq:bigten}), the generalization of Eqs.~(\ref{eq:divb}-\ref{eq:curlb}) is 
\begin{align}
\Delta (\partial_i v_i) =&  b_i (\nu \epsilon_{ij} -2 \gamma_1 \delta_{ij} ) \partial_j \delta (\vb r) \label{eq:gendivb}\\
\Delta (\partial_i \epsilon_{ij} v_j) =& b_i ( 2  \nu^o \epsilon_{ij} - \gamma_2 \delta_{ij} ) \partial_j \delta (\vb r) \label{eq:gencurlb}
\end{align}
where we now have the quantities:
\begin{align}
    \nu =& \frac{(B-\mu)(\mu+\Gamma)+(A-K^o)(K^o-\Lambda)}{(B+\mu)(\mu+\Gamma)+(A+K^o)(K^o-\Lambda)} \\
    \nu^o =& \frac{B K^o - A \mu }{(B+\mu)(\mu+\Gamma)+(A+K^o)(K^o-\Lambda)}\\
    \gamma_1 =& \frac{ K^o \Gamma + \Lambda \mu}{(B+\mu)(\mu+\Gamma)+(A+K^o)(K^o-\Lambda)} \\
    \gamma_2 =& \frac{(B + \mu) (\mu - \Gamma)+(A+K^o)(K^o+\Lambda)}{(B+\mu)(\mu+\Gamma)+(A+K^o)(K^o-\Lambda)}
\end{align}
Notice that $\gamma_1 \to 0$ and $\gamma_2 \to 1$ when $\Lambda, \Gamma \to 0$. Solving Eqs.~(\ref{eq:gendivb}-\ref{eq:gencurlb}) with $\Lambda, \Gamma \neq 0$ then yields the displacement field:
\begin{align}
    u_i (\vb r) = \frac1{2\pi} \left[ b_i \phi + \frac{(\gamma_2-\nu)}2 \epsilon_{ik} b_k \log(r) - (\nu^o + \gamma_1) b_i \log (r) + \frac{(\gamma_2 + \nu)}{2}  \frac{  r_k b_k \epsilon_{il} r_l}{r^2} +(\nu^o - \gamma_1) \frac{b_k r_k r_i}{r^2} \right] 
\end{align}
The strain field is characterized by the following three quantities. The dilation is given by:
\begin{align}
    \partial_m u_m = \frac{1}{2\pi r^2} \left[ (1-\nu) r_m \epsilon_{mk} b_k - 2 \gamma_1 b_m r_m \right]
\end{align}
The rotation is given by:
\begin{align}
    \epsilon_{mn} \partial_m u_n = \frac{-1}{2\pi r^2} \left[   (1+ \gamma_2) r_m b_m +2 \nu^o r_m \epsilon_{mn} b_n \right ] 
\end{align}
The shear strain is given by:
\begin{align}
\frac{\partial_m u_n + \partial_n u_m - \delta_{mn} \partial_i u_i}{2}= -\frac{1}{4 \pi r^2} \bigg \{ & \left[ 2 \nu^o b_k r_k + (\gamma_2 -1) b_j  \epsilon_{jk} r_k \right] \left( \frac{2 r_m r_n }{r^2} - \delta_{mn} \right) + \nonumber  \\
     & \left[ (1+\nu) b_k r_k + 2 \gamma_1 b_j \epsilon_{jk} r_k \right] \qty( \frac{r_m \epsilon_{nk}r_k + r_n \epsilon_{mk} r_k}{r^2} ) \bigg\}
\end{align}
The stress is characterized by the following three quantities. The isotropic stress: 
\begin{align}
    \sigma_{ii} = \frac{1}{\pi r^2} \left\{ [ B(1-\nu) + 2 \Lambda \nu^o ] r_m \epsilon_{mn} b_n - [2 B \gamma_1 - \Lambda (1+ \gamma_2)] r_m b_m \right\}
\end{align}
The torque density:
\begin{align}
    \epsilon_{ij} \sigma_{ij} = \frac{1}{\pi r^2}\left\{  -[ A (1- \nu) +2 \Gamma \nu^o ] r_m \epsilon_{mn} b_n + [2 A \gamma_1 - \Gamma (1+ \gamma_2) ]r_m b_n   \right\}  
\end{align}
And the shear stress:
\begin{align}
    \frac{\sigma_{ij} + \sigma_{ji} - \delta_{ij} \sigma_{mm}}{2} = -\frac1{2\pi r^2} \bigg\{  &\left[ (2\Gamma\nu^o - A(1-\nu)) b_k r_k  + (\Gamma(1+\gamma_2) + 2 A \gamma_1) b_j \epsilon_{jk} r_k \right] \qty( \frac{2 r_i r_j}{r^2} -\delta_{ij} ) + \nonumber  \\
    &  \left[ (B(1-\nu) -2 \Lambda \nu^o ) b_k r_k - (2 B\gamma_1+\Lambda (1+ \gamma_2)) b_j \epsilon_{jk} r_k \right ] \qty( \frac{r_i \epsilon_{jn} r_n + r_j \epsilon_{in} r_n }{r^2}) \bigg\}
\end{align}
Throughout, the qualitatively new terms introduced by $\Lambda$ and $\Gamma $ are the terms proportional to $b_m r_m$ in the trace and antisymmetric parts of the stress and strain, as well as the terms proportional to $b_j \epsilon_{jk} r_k$ in the symmetric traceless symmetric part of the stress and strain. 

\subsection{Local dilations and point torques} \label{sec:int}
\begin{figure}[t!]
    \centering
    \includegraphics[width=0.8\textwidth]{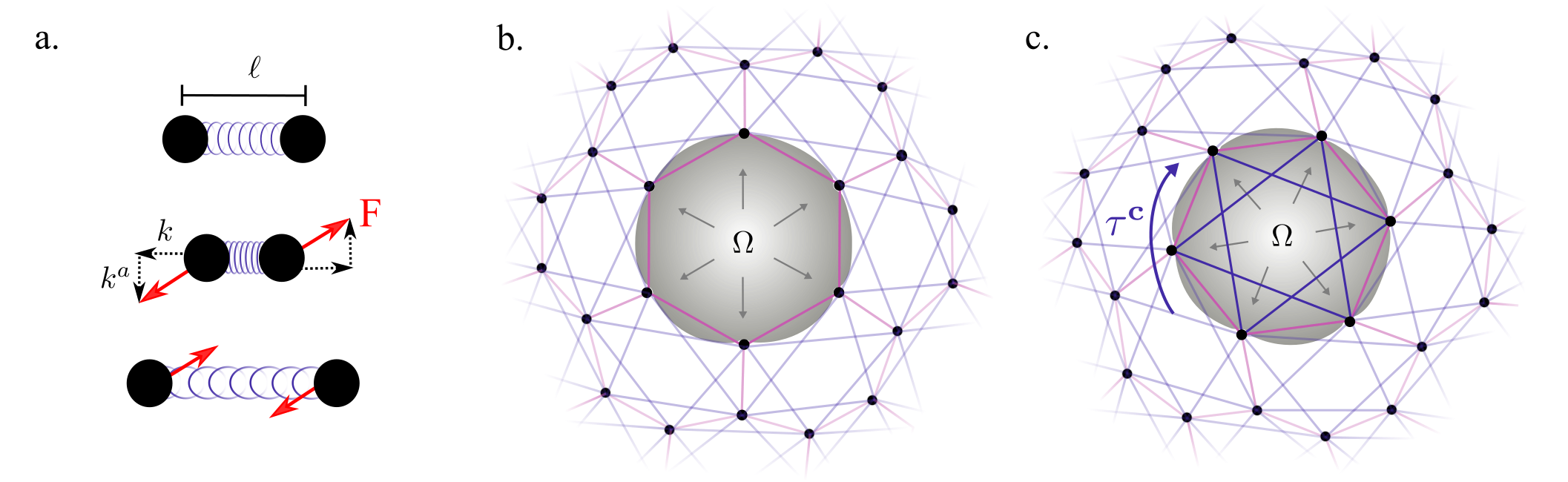}
    \caption{{\bf Generalized Hookean springs and point torques.}~{\bf a.} The generalized Hookean spring described in \S\ref{sec:FTS}. When compressed or elongated, the spring exhibits a pair of equal and opposite forces, whose magnitude $F$ is proportional to the spring's change in length. The radial component of the force is proportional to the spring constant $k$ and the transverse component is proportional to $k^a$. The transverse component results in a nonzero net torque. 
    {\bf b-c.} A microscopic point dilation $\Omega$ is induced by an expanding diaphragm (grey) in a lattice of active springs. In {\bf b}, the next-nearest-neighbor springs crossing the diaphragm are removed, and in {\bf c}~the next-nearest-neighbor springs are kept in place.  As a result, the torque $T^c$ at the center of the point defect is greater in {\bf c} than in {\bf b}.  }
    \label{fig:coretorque}
\end{figure}

We now address defects with local dilations and torques at their core. Unlike disclinations and dislocations, which require that the displacement field to be multivalued, local dilations and torques describe properties of the core itself that are felt at large distances. 
An interstitial and a local rotation at the core are defined by the differential equations
\begin{align}
    \partial_i u_i (\vb r) = \Omega \delta (\vb r) \label{eq:divint} \\
    \partial_i \epsilon_{ij} u_j (\vb r) = \varphi \delta (\vb r) \label{eq:curlint}
\end{align}
The corresponding solution is
\begin{align}
    u_i (\vb r) = \frac{\Omega}{2\pi} \frac{r_i}{r^2} + \frac{\varphi}{2\pi} \epsilon_{ij} \frac{r_j}{r^2} \label{eq:dispCirc}
\end{align}
The gradient of Eq.~(\ref{eq:dispCirc}) is given by
\begin{align}
    \partial_i u_j (\vb r) = \frac{\Omega}{2\pi r^2}  \qty( \frac{2 r_i r_j}{r^2} -\delta_{ij}) +  \frac{\varphi}{2\pi r^2} \qty( \frac{r_i \epsilon_{jk} r_k + r_j \epsilon_{ik}r_k}{r^2}) \label{eq:strainint}
\end{align}
Due to Eq.~(\ref{eq:divint}) and Eq.~(\ref{eq:curlint}), $\partial_i u_j$ in Eq.~(\ref{eq:strainint}) consists entirely of shear for $\vb r \neq 0$. Consequently, the stress corresponding to Eq.~(\ref{eq:strainint}) also consists entirely of shear:
\begin{align}
    \sigma_{ij} (\vb r) = \qty(\frac{\mu \Omega - K^o \varphi }{2\pi r^2})  \qty( \frac{2 r_i r_j}{r^2} -\delta_{ij}) +  \qty(\frac{\varphi \mu + K^o \Omega}{2\pi r^2} ) \qty( \frac{r_i \epsilon_{jk} r_k + r_j \epsilon_{ik}r_k}{r^2}) \label{eq:stressint}
\end{align}
Notice that $\partial_i \sigma_{ij} (\vb r) =0$ and $\epsilon_{ij} \sigma_{ij} (\vb r) =0 $ for all $\vb r \neq 0$. Using Eq.~(\ref{eq:netTorque}), we find that Eq.~(\ref{eq:stressint}) captures the response to a point torque at $\vb r=0$ whose magnitude is given by 
\begin{align}
    T^c = \varphi \mu + K^o \Omega \label{eq:tauc}
\end{align}
Hence, we find that the strain field due to a defect depends both on the volume change $\Omega$ of the core and the torque $T^c$ provided by the core.  In standard isotropic elasticity,  Eq.~(\ref{eq:tauc}) implies that $\varphi=0$,  and hence Eq.~(\ref{eq:dispCirc}) yields the familiar strain distribution about an interstitial. 
When $\mu=0$ and $K^o \neq 0$, Eq.~(\ref{eq:tauc}) implies that the core of a defect must experience a volume change when $T^c$ is nonzero in order to be mechanically stable. 

In \S\ref{sec:disc} and \ref{sec:disl}, we obtained the solutions for the disclination and dislocation for $\Omega=0$ and $T^c=0$. 
To incorporate nonzero $T^c$ and $\Omega$, one needs to add Eq.~(\ref{eq:dispCirc}) and Eq.~(\ref{eq:stressint}) to the displacement and stress, respectively, of the dislocation or disclination. For the disclination, in order to maintain stress free boundary conditions one need adjust 
\begin{align}
R_1=& R \exp{\frac{ 3\qty[(\mu^2 +(K^o)^2 )\Omega - K^o T^c ] }{\mu B(1-\nu) s R^2} - \frac12 } \\
R_2=& R \exp{\frac{3 T^c }{B(1-\nu) s R^2} -\frac12}
\end{align}

Finally, we note that $T^c$ and $\Omega$ are contingent on the microscopic details of the topological defect. As an illustration, Fig.~(\ref{fig:coretorque}) shows an inclusion in a lattice composed of bonds that exert torques when stretched or elongated, as described in \S\ref{sec:FTS}. In panel~b, the next-nearest-neighbor springs are removed, while in panel~c the next nearest neighbor springs are retained. Since each spring exerts a torque when elongated, the torque at the core will be greater in panel~c than in panel~b. This difference depends on the details of the microscopic construction and therefore must be provided phenomenologically.

\subsection{Airy stress function} \label{sec:airy} 
Isotropic 2D elasticity is often simplified by the introduction of an Airy stress function~\cite{Landau7,Nelson1987}. When $A = \Gamma =0$, the stress tensor $\sigma_{ij}$ is symmetric. Therefore, when the system is force balanced (i.e. $\partial_i \sigma_{ij} =0$), we may write the stress as $\sigma_{ij} = \epsilon_{jk} \epsilon_{il} \partial_k \partial_l \chi $, where $\chi$ is the Airy stress function. Moreover, when $\Lambda=0$, an invertible relationship exists between the symmetric stress $\sigma_{ij}$ and the symmetric strain $u^s_{ij} = (\partial_i u_j + \partial_j u_i)/2 $ given by:
\begin{align}
    u^s_{ij} = \frac{1}{E} \bigg\{ (1-\nu) \delta_{ij} \delta_{mn} + (1+\nu) ( \delta_{im} \delta_{jn} + \delta_{in} \delta_{jm} - \delta_{ij} \delta_{mn} ) -  2 \nu^o E_{ijmn}  \bigg\} \sigma_{mn}
\end{align}
where $E$ is the Young's modulus generalized to include $K^o$:
\begin{align}
    E = \frac{4 B  [\mu^2  + (K^o)^2 ]  }{ \mu (B+ \mu) +(K^o)^2 }
\end{align}
Next, we utilize the differential versions of Eq.~(\ref{eq:top}) and Eq.~(\ref{eq:burg})
\begin{align}
    \epsilon_{ij} \partial_{i} \partial_j u_k =&  \sum_\alpha b_k^{\alpha } \delta(\vb r - \vb r^\alpha) \\
    \epsilon_{ij} \partial_{i} \partial_j \epsilon_{lk} \partial_l u_k =& \frac{2 \pi}{3}  \sum_\alpha  s^\alpha \delta(\vb r - \vb r^\alpha)
\end{align}
where $b^\alpha_i$ and $s^\alpha$ are the charges associated with a defect at point $\vb r^\alpha $. By evaluating $\partial_k \epsilon_{ki} \partial_l \epsilon_{lk} u^s_{ij}$, we obtain an expression identical in form to that of standard isotropic 2D elasticity:
\begin{align}
    \Delta^2 \chi = E s(\vb r) \label{eq:airsi} 
\end{align}
where
\begin{align}
    s(\vb r) = \sum_\alpha \qty[ \frac{\pi}{3} s^\alpha + b^\alpha_i \epsilon_{ij} \partial_j ] \delta(\vb r - \vb r^\alpha )  
\end{align}
is the defect density. For a single disclination at the origin, we obtain $\Delta^2 \chi = E \pi s\delta (\vb r)/3 $, which yields a stress:
\begin{align}
    \sigma_{ij} = \frac{sE}{12} \qty[ \frac{\epsilon_{il} \epsilon_{jk} r_l r_k}{r^2} + \log(r) ] 
\end{align}
which agrees with Eq.~(\ref{eq:discstress}) upon setting $A=0$ and using $E = 2B(1-\nu)$. Similarly, for a single dislocation, we solve $\Delta^2 \chi = E b_i \partial_i \delta(\vb x) $ to obtain 
\begin{align}
    \sigma_{ij} = \frac{E}{4\pi r^2} \qty[ r_m \epsilon_{mn} b_n \delta_{ij} - b_k r_k \qty( \frac{r_i \epsilon_{jm} r_m + r_j \epsilon_{im} r_m }{r^2} ) ],
\end{align}
in agreement with Eq.~(\ref{eq:res}) with $A \to 0$.

\section{Dislocation interactions} \label{sec:dislint}

\subsection{Peach-Koehler force} \label{sec:PKforce}

\begin{figure}
    \centering
    \includegraphics{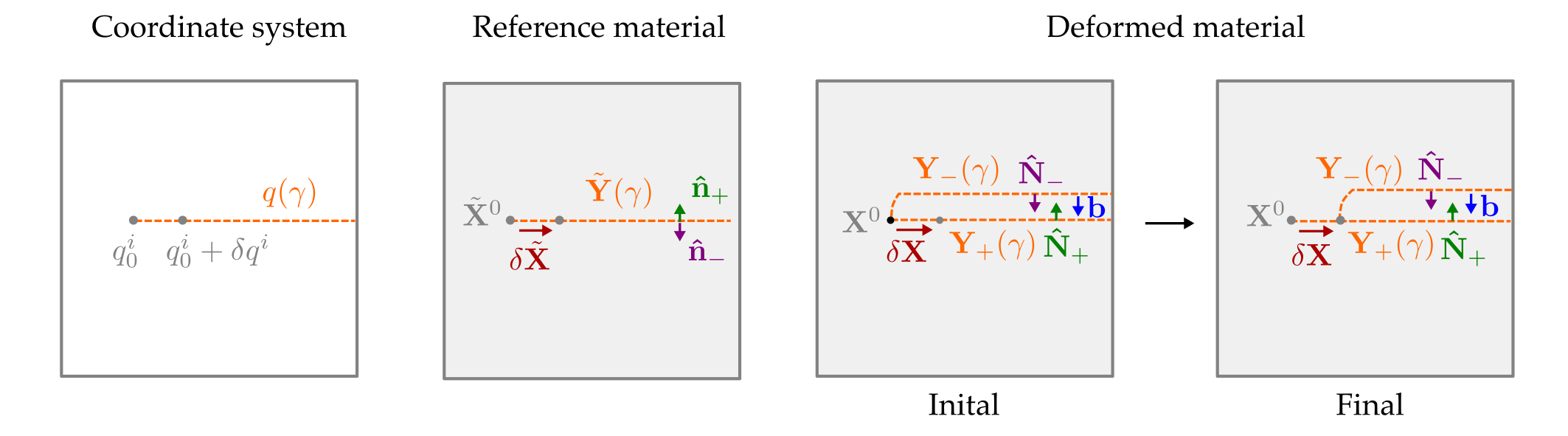}
    \caption{ {\bf Branch cuts for dislocation motion.}~Illustration of the branch cut construction used in \S\ref{sec:PKforce}-\ref{sec:mass}.} 
    \label{fig:branch}
\end{figure}

Here we provide a derivation of the Peach-Koehler formula in two dimensions.
Suppose that an isolated dislocation with Burgers vector $\vb b$ is subjected to a smooth external Piola-Kirchhoff stress $\vb S^\pres$ satisfying the equations for static equilibrium $\divr \vb  S^\pres = 0$.
We seek to compute the work $\delta W$ associated with moving the dislocation   infinitesimally from the point $q_0^i$ to the point $q^i_0+\delta q^i$ in coordinate space, see Fig.~\ref{fig:branch}. In the reference material, the dislocation starts at position $\tilde{\vb X^0 } = \vb r(q_0)$ and moves by an amount $\delta \tilde{\vb X} = \vb r_i \delta q^i$. We define a contour in coordinate space $q^i(\gamma)$ such that $\vb{\tilde Y}(\gamma) = \vb r[q(\gamma)] = \vb {\tilde X}^0 + \gamma \delta \vb{\tilde X}$.
Before the dislocation moves, the displacement field $\vb u$ has a discontinuity $\vb b$ across the entire branch cut  $\gamma \in [0,\infty)$. After the dislocation moves, the displacement field is continuous across the contour for $\gamma \in[0,1)$, but retains the discontinuity for $\gamma \in [1, \infty)$. Let $\delta \vb u$ be the change in the displacement field. The work done by the external stress $\vb S^\pres$ is 
\begin{align}
    \delta W =& -\int_\regA \vb  S^\pres : \gradr \delta  \vb u \; \dd^2 r \label{step1} \\
    =& -  \int_\regA  \divr (\vb S^\pres \cdot \delta \vb u)  \dd^2r  \label{step2} \\
    =& - \int_{\partial A}  \vb{\hat n} \cdot \vb S \cdot \delta \vb u \;  \dd \ell \label{step3} 
\end{align}
Between Eq.~(\ref{step1}) and Eq.~(\ref{step2}), we used the fact that $\divr \vb S^\pres=0$ and then we applied Stokes' theorem. Here $\regA$ is taken to be all of space, and $\partial \regA$ crucially consists of two oppositely oriented surfaces lying along the branch cut $\vb Y$. As shown in Fig.~\ref{fig:branch}, we will denote the normal vector and displacement field on either side of the branch cut by $\vb{\hat n}_\pm$ and $\vb u_\pm$, respectively.  
Thus we have:
\begin{align}
    \delta W =& \int_{\gamma > 0} [  \vb{\hat n}_+ \cdot \vb S^\pres  \cdot \delta \vb u_+ +\vb{\hat n}_- \cdot \vb S^\pres  \cdot \delta \vb u_- ]   \; \dd \ell  \\ 
    =&\int_{\gamma > 0} \vb{\hat n}_+ \cdot \vb S^\pres  \cdot  [ \vb b \theta(\gamma-1) -\vb b ]   \; \dd \ell \\
    =& -\int_{ \gamma \in (0,1]} \vb{ \hat n}_+ \cdot \vb S^\pres \cdot \vb b \; \dd \ell  
\end{align}
Where $\theta(x)$ is the heavyside function passing from $0$ to 1 at $x=0$. We have used the facts that $\hat{\vb n}_+= - \hat{\vb n}_-$ and that $\vb u_+ - \vb u_- = \vb b$ when the displacement field is discontinuous. 
Finally assuming that $\vb S^\pres$ is continuous near the dislocation, we may approximate the infinitesimal integral as $\int_{\gamma \in (0,1]} \dd \ell \approx \delta {\tilde X}$. Moreover, by construction $\vb{\hat  n} = \boldsymbol{\epsilon} \cdot \delta \tilde{\vb X}/\delta \tilde X $. Therefore, 
\begin{align}
    \delta W \approx \delta\tilde{  \vb X} \cdot \boldsymbol{ \epsilon} \cdot  \vb S^\pres \cdot \vb b
\end{align}
Hence, we may express the work as $\delta W = \delta \tilde{\vb  X} \cdot \tilde{\vb f}^\text{PK}$, where $\tilde{\vb f}^\text{PK}$ is the Peach-Koehler force on the dislocation with respect to the undeformed material. It is given by:  
\begin{align}
   \tilde{\vb f}^\text{PK} =   \boldsymbol{\epsilon} \cdot \vb S^{\pres} \cdot  \vb b \label{eq:peach}
\end{align}
We emphasize that $\delta \tilde{\vb X}$ is the motion of the dislocation in the reference material. The motion of the dislocation in the deformed material is given by $\delta \vb X = \vb R_i \delta q^i=\vb J^T \cdot \delta \tilde{\vb X}$. Therefore, one has:
\begin{align}
    \delta W = \delta \vb X \cdot \vb J^{-1} \cdot  \boldsymbol{\epsilon} \cdot \vb S^\pres \cdot \vb b =\delta \vb X \cdot \boldsymbol{\epsilon} \cdot \sigmab^\pres \cdot \vb b
\end{align}
where we have used the identities $-\boldsymbol{\epsilon} \cdot \vb J^{-1} \cdot \boldsymbol{\epsilon} = \frac1J \vb J^{T} $ and $\sigmab = \frac1J \vb J^T \cdot \vb S $. Thus we identify 
\begin{align}
   \vb f^\text{PK} =   \boldsymbol{\epsilon} \cdot \sigmab^{\pres} \cdot  \vb b \label{eq:peachdeformed}
\end{align}
as the Peach-Koehler force with respect to the motion $\delta \vb X$ in the deformed material. 
Notice first that the Peach-Koehler force is agnostic to the elasticity of the solid and relies only on the expression for the virtual work and the definition of a dislocation.
Secondly, the Peach-Koehler force is conservative in the sense that its curl vanishes:
\begin{align}
\curlr \tilde{\vb f}^\text{PK} = - \divr \vb S^\pres \cdot \vb b =0 \\
\Curl \vb f^\text{PK} = - \Div \sigmab^\pres \cdot \vb b =0
\end{align}
where we have used the assumption that $\divr \vb S^\pres=0$, the fact that $\divr \vb S = 0$ implies $\Div \sigmab =0$, and the definition $\Curl \boldsymbol{\Phi} = \Div \boldsymbol{\epsilon} \cdot \boldsymbol{\Phi}$ and $\curlr \boldsymbol{\Phi} = \divr \boldsymbol{\epsilon} \cdot \boldsymbol{\Phi}$ .  

Finally, we note that Eq.~(\ref{eq:peach}) does not directly predict the motion of the dislocation. The quantity $\vb f^\text{PK} \cdot \delta \vb X$ yields the work done when $\delta r$ is small but still larger than the microscopic lattice spacing. Forces associated with local bond rearrangements that are contingent on the microscopic structure of the dislocation can play a significant role in the dislocation motion. In \S\ref{sec:mass} we review the consequences of mass conservation that generally constrain the dislocation to move along its glide plane. In \S\ref{sec:core} we show that microscopic bond rearrangements can enable dislocations in active media to self propel.

\subsection{Nonmutual dislocation interactions}
We now compute the elastic interaction between two dislocations. Suppose a dislocation with Burgers vector $b_i$ is located at the origin and a dislocation with Burgers vector $d_i$ is located at point $r_i$. The force on $d_i$ due to $b_i$ is given by:
\begin{align}
    f_m = \epsilon_{mi} \sigma^{(b)}_{ij} d_j 
    = \frac{(1-\nu)}{2\pi r^2} \bigg\{& B \left[ r_k \epsilon_{kn} b_n  \epsilon_{mj} d_j +r_k \epsilon_{ki} d_i \epsilon_{mk} b_k +  r_m b_i d_j \left(  \frac{ 2 r_i r_j}{r^2} -\delta_{ij} \right) \right] +\nonumber  \\
    &A\left[ r_m d_i \epsilon_{ij} b_j +\epsilon_{mn} r_n b_i d_j \left( \frac{2 r_i r_j }{r^2} - \delta_{ij}\right) \right]
    \bigg\}
\end{align}
Notice that when $A \to 0$, we recover the usual defect interaction, which is notably symmetric under $b_i \leftrightarrow d_i$ and $r_i \to - r_i$. When $A$ is nonzero, the Peach-Koehler force violates this symmetry and one obtains:
\begin{align}
    f^{b \to d}_m + f^{d \to b}_m = \frac{ (1-\nu) }{\pi r^2} A r_m d_i \epsilon_{ij} b_j
\end{align}
which is notably nonzero. 
When $\Gamma$ and $\Lambda$ are nonzero, the antisymmetric part of the Peach-Koehler force generalizes to
\begin{align} 
f^{b \to d}_m + f^{d \to b}_m = \frac{2}{\pi r^2} d_i \epsilon_{ij} b_j r_m  \frac{ (A- \Lambda) [\mu^2 + (K^o)^2 ] + 2K^o (B \Gamma - A \Lambda ) }{ (B+\mu) (\mu +\Gamma) + (A+ K^o) (K^o - \Lambda) } \label{eq:genforce}
\end{align} 
Notice that Eq.~(\ref{eq:genforce}) indicates that that there is a net force on the dislocation pair pointing along their relative separation vector. Moreover, the strength of the force is proportional the cross product of the two Burgers vectors. If $\Lambda \neq0$, one does not need sources of angular momentum to obtain the nonmutual interaction. 

In Fig.~\ref{fig:edge} of the main text, we illustrate the interaction between two dislocations with parallel glide planes, e.g. $\vb b = b\hat {\vb x} $ and $\vb d= d \hat {\vb x}$. Then the force $f$ on the relative coordinate $r_x$ is given by:
\begin{align}
    f (r_x) = \frac{(1-\nu) bd}{\pi r^4 } ( r_x^2-r_y^2 ) ( Br_x +Ar_y )  
\end{align}
Notice, that $f(r_x)=0$ for $r_x = \pm r_y$ and $r_x = - (A/B) r_y$. For the antiparallel defects in Fig.~\ref{fig:edge}c-h, we take $bd<0$. In this case, the solutions $r_x = \pm r_y$ are stable for $A < B$ and the solutions $r_x = r_y$ and $r_x = - (A/B) r_y$ are stable for $A>B$. In the limit $A/B \to \infty$, only one stable configuration remains. Notice that when $b d>0$, the stability of each solution is inverted. One can integrate the $f(r_x)$ along the glide plane to obtain an effective potential $V_\text{eff}$ felt by the dislocation: 
\begin{align}
V_\text{eff} (r_x) = \frac{(1- \nu)b d }{\pi }\left[ B \log r + \frac{r_y}{r^2} (B r_y - A r_x) \right] 
\end{align}
which is plotted in Fig.~\ref{fig:edge}d,f,h. 

\subsection{Mass conservation condition}\label{sec:mass}

\begin{figure}
    \centering
    \includegraphics[width=0.6\textwidth]{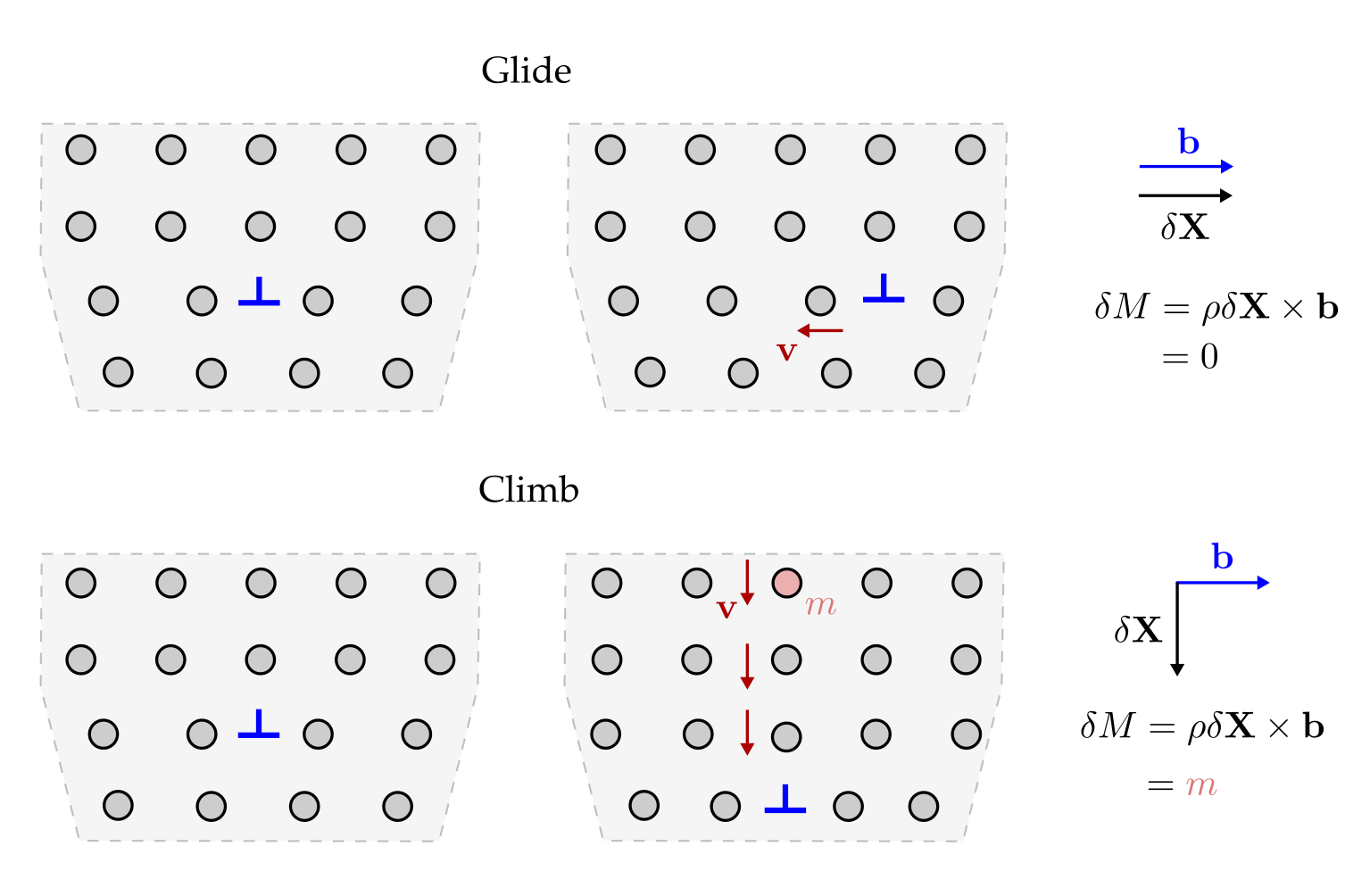}
    \caption{{\bf Glide constraints on dislocations.}~An illustration of the kinematic constraints that restrict dislocation motion to lie along the glide plane. When a dislocation glides, the total mass (i.e. number of particles) within the selected region does not change and the motion is localized to the particles near the core. (In this realization $\vb v$ is the velocity of the particle with the most motion during the rearrangement.) However, when the dislocation climbs, significant motion extends all the way out to the boundary of the sample since a column of particles moves downward. The net change in mass of the highlighted region is $\delta M =\rho  \delta \vb X \times \vb b =m$, where $\vb b= \vb{\hat x}a $ is the Burgers vector, $\delta \vb X = -\vb{\hat y} a$ is the motion of the dislocation, $m$ is the mass of the particle, $\rho=m/a^2$ is the density, and $a$ is the lattice spacing. We empirically find in simulation that climbing events remain almost entirely forbidden even in the presence of broken MBR. }
    \label{fig:mass}
\end{figure}

In practice, dislocations are often constrained to move along their glide planes. Though the origin of this effect arises from microscopic considerations, the underlying assumptions can be formulated within the continuum theory. In analogy to \S\ref{sec:PKforce} consider a dislocation that moves from $\delta \tilde{\vb X}$ to $\tilde{\vb X} + \delta \tilde{ \vb X}$ in the reference material.
The change in mass associated with a fixed region in real space reads:
\begin{align}
    \delta M =&  \int_{\partial \regA } \rho \; \vb{\hat N} \cdot \delta \vb u \; \dd L
\end{align}
where $\rho$ is the density in physical space, and $\regA $ is the patch of material coinciding with the region of real space prior to the motion of the dislocation. We will take $\partial \regA$ to have two contributions, a physical boundary $\mathcal{D}$ and well as the boundary along the branch cut $q^i(\gamma)$ as introduced in \S\ref{sec:PKforce}. Thus we may write:
\begin{align}
    \delta M =& \int_\mathcal{D} \rho \; \vb{\hat N} \cdot \delta \vb u \;   \dd L  + \int_{\gamma> 0}  \rho \; [\vb{\hat N}_+ \cdot \delta \vb u_+ + \vb{\hat N}_- \cdot \delta \vb u_-]  \; \dd L  \\
    =&\int_\mathcal{D} \rho \; \vb{\hat N} \cdot \delta \vb u \; \dd L  + \int_{\gamma> 0}  \rho_0 \;  [\vb{\hat n}_+ \cdot \vb J^{-T} \cdot \delta \vb u_+ + \vb{\hat n}_- \cdot \vb J^{-T} \cdot \delta \vb u_-]
    \; \dd \ell \label{eq:mbone}
    \\
    =& \int_\mathcal{D} \rho \; \vb{\hat N} \cdot \delta \vb u \; \dd L  + \rho_0 \; \delta \tilde{\vb X} \cdot \boldsymbol  \epsilon \cdot \vb J^{-T} \cdot \vb b \label{eq:massbalance}
\end{align}
In Eq.~(\ref{eq:mbone}) $\rho_0= J \rho $ is the density in the reference material, and we have used the fact that  $\vb {\hat N} \; \dd L = J \vb J^{-1} \cdot \vb {\hat n} \; \dd \ell $. To arrive at Eq.~(\ref{eq:massbalance}), we have used the branch cut argument from \S\ref{sec:PKforce} along with the fact that $\vb J$ is continuous across the branch cut for a dislocation.  
The first term Eq.~(\ref{eq:massbalance}) captures the change in material due to the macroscopic flow through the boundary. Now we make a crucial kinematical assumption: this macroscopic mass flow must be the entire mass flow.   
This statement implies that when the dislocation moves, the microscopic particle velocities are only discontinuous at the core, and not along an arbitrary contour located a macroscopic distance away from the core. 
In practice, it is typically safe to assume that the dominant mode of dislocation motion does not require microscopic rearrangements to be coordinated over arbitrarily large macroscopic distances. See Fig.~\ref{fig:mass} for an illustration. 
This assumption implies 
\begin{align} 
\delta\tilde{ \vb  X} \cdot \boldsymbol{\epsilon} \cdot \vb J^{-T} \cdot \vb  b=0 \label{eq:constraint} 
\end{align} 
 Recalling that the dislocation motion in real space is given by $\delta \vb X = \vb  J^T  \cdot \delta \tilde{\vb X} $, the constraint may be written as:
\begin{align}
    \delta \vb X \cdot \boldsymbol{\epsilon} \cdot \vb b =0. \label{eq:constraint2}
\end{align}
Notice that $\vb J^T \cdot \vb b$ points along the glide plane in material space. Hence Eq.~(\ref{eq:constraint}) and Eq.~(\ref{eq:constraint}2) are equivalent statements that the dislocation moves parallel to the glide plane.

\section{Active core force} \label{sec:core}

\begin{figure}
    \centering
    \includegraphics[width=\textwidth]{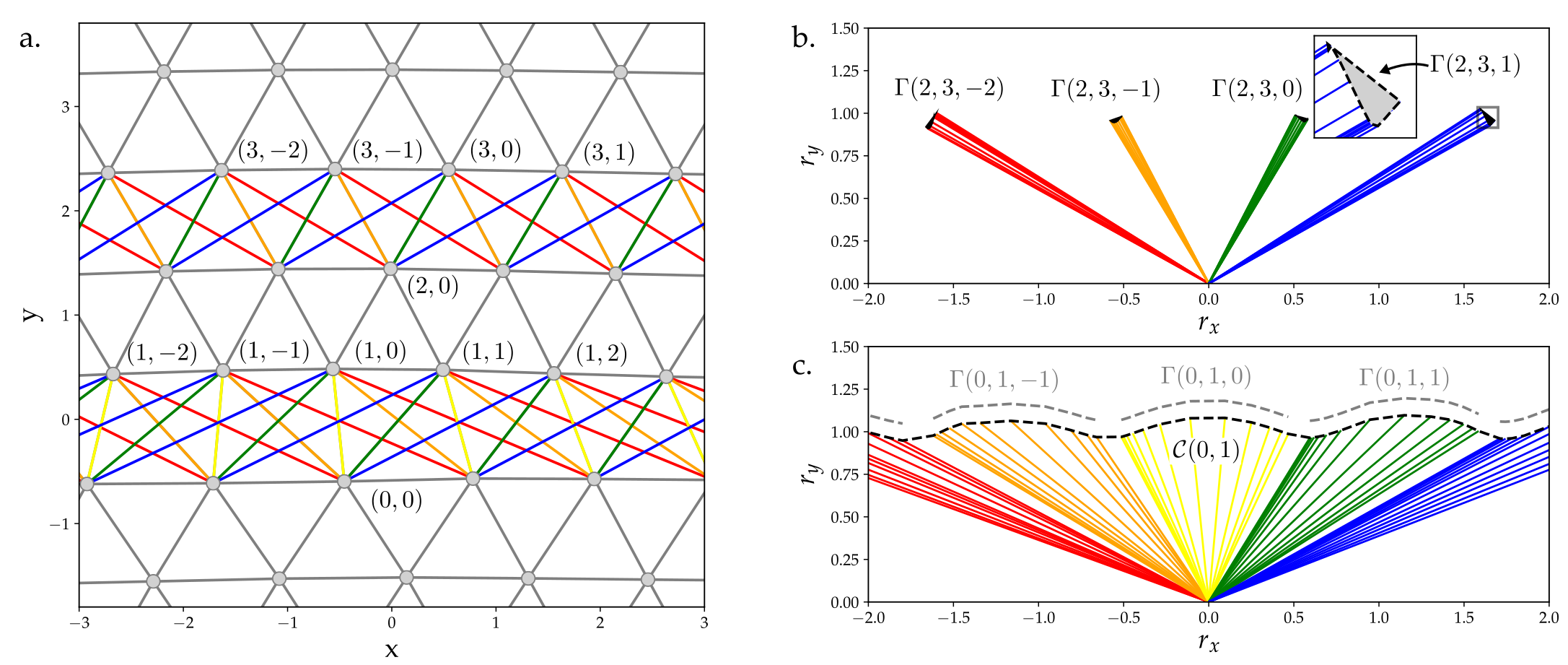}
    \caption{{\bf Schematic of dislocation core.}~{\bf a.~}A rendering of the dislocation core with certain particles highlighted by their index $(R(\alpha), C_R (\alpha) ) $. 
    {\bf b.~} Bonds connecting rows 2 and 3 are plotted in the space of their relative coordinates $(r_x, r_y)$. The inset shows the contour $\Gamma(2,3,1)$ which interpolates between each of the bond positions.  
    {\bf c.~} The bonds connecting rows 0 and 1 are shown. The individual contours $\Gamma(0,1,n)$ do not close since the rows 0 and 1 are on opposite sides of the glide plane. Instead, the bonds can be concatenated into a single continuous contour $\mathcal{C}(0,1)$. The data shown here is take from a simulation of the transverse force $F^\perp_\text{Lub}$ whose parameters are shown in Table~\ref{tab:params} 
    }
    \label{fig:schematic}
\end{figure}

Here we derive the active contribution to the dislocation force that arises from bond reassignment at the core (see Supplemental Movie~S2).  
We begin by considering an infinite lattice with a single isolated dislocation. The quanitity we wish to compute is the total work done by all the interactions if the dislocations glides one lattice spacing. If the interactions are described by a potential energy that depends only on relative coordinates and is symmetric under exchange of particles, then no work is done since the final configuration is a translation of the previous arrangement (up to a relabeling of particles). However, the work is generally nonzero if the interactions are nonconservative, and the total work depends on the microscopic path that interpolates between the two configurations.
Explicitly, the total work done by the interactions is given by:
\begin{align}
    W_\text{glide} = \int_{t_i}^{t_f} \dd t \sum_{\alpha } \sum_{\alpha' > \alpha} \vb F^{\alpha \alpha'} \cdot \dv{\scrb^{\alpha \alpha'}}{t} \label{eq:total_work} 
\end{align}
Here, $\scrb^{\alpha \alpha'}$ is the relative coordinate pointing from particle $\alpha$ to particle $\alpha'$ and  $\vb F^{\alpha \alpha '}$ is the force from particle $\alpha $ on $\alpha '$. 
To arrange the sum, we use a labeling system shown in Fig.~\ref{fig:schematic}a. We first label each horizontal (i.e. parallel to the glide plane) row of atoms by an integer. In Fig.~\ref{fig:schematic}a, we choose $R=0$ to be the row containing the 7-fold and $R=1$  to be the row containing the five fold. 
Let $R(\alpha)$ denote the row containing particle $\alpha$.

Within each row $R$, we label the each particle by an integer $C_R(\alpha)$ that increases from left to right. Then each particle $\alpha$ can be uniquely identified by the ordered pair $(R(\alpha), C_R(\alpha) ) $. We may then write the work in the following form:
\begin{align}
    W_\text{glide} = \int_{t_i}^{t_f} \dd t \sum_{R \in \ZZ }  \sum_{R' \ge  R  } \sum_{ n \in S_{R',R}  } \sum_{m \in \ZZ}  \vb F \cdot \dv{ \scrb^{ (R,m) , ( R', m+n) }   }{t} 
\end{align}
where $S_{R',R} = \ZZ \setminus \{ 0 \}   $ if $R' =R$ and $S_{R',R} = \ZZ$ otherwise. Usage of $S_{R, R'}$ simply precludes self-interaction among the particles. We have suppressed indices on $\vb F$ for simplicity of notation. 

Upon translation of the dislocation by one unit cell, we have $\scrb^{ (R, m), (R', m') } (t_f) =\scrb^{ (R, m-1), (R', m'-1) } (t_i)  $. Hence, we may concatenate the trajectory of individual bonds into the following continuous contours:
\begin{align}
    \Gamma (R, R', n) \equiv & \bigcup_{ m \in \ZZ  } \qty{ \scrb^{ (R, m), (R', m+n)  } (t) : t \in [t_i, t_f]  }. 
\end{align}
Notice that
\begin{align}
\sum_{m \in \ZZ }  \int_{t_i}^{t_f } \vb F \cdot  \dv{ \scrb^{ (R, m), (R', m+n) } }{t} \dd t = \int_{\Gamma(R, R', n) } \vb F \cdot  \dd \scrb . 
\end{align}
Hence, the expression for the work becomes:
\begin{align}
    W_\text{glide} = \sum_{R \in \ZZ } \sum_{R' > R } \sum_{n \in S_{R',R } } \int_{\Gamma(R, R', n)}  \vb F \cdot \dd  \scrb . 
\end{align}
To further simplify the sum, we introduce the following two sets
\begin{align} 
\BB \equiv & \{ (R, R') : R' > R \text{ and either } R \ge  1 \text{ or } 0 \ge R' \} \\
\GG  \equiv & \{ (R, R') : R' \ge 1  \text{ and } 0 \ge R \} 
\end{align} 
Here, $\BB$ is the set of of all pairs of rows that do not straddle the glide plane and $\GG$ is the set of all pairs of rows that do straddle the glide plane. 
For $(R, R') \in \BB$, we have 
\begin{align} 
\scrb^{ (R, m), (R',m+n ) } =\scrb^{ (R, -m), (R',-m+n ) }  \text{ as } m \to \infty. \label{eq:lim}
\end{align} 
Equation~(\ref{eq:lim}) is a consequence of the lattice being undeformed far away from the dislocation. 
From Eq.~(\ref{eq:lim}), we deduce that $\Gamma(R, R',n) $ is a closed contour for $(R,R') \in \BB$. 

However, for $(R, R') \in \GG $, we have 
\begin{align} 
\scrb^{ (R, m) , (R', m+n) } = \scrb^{ (R, -m), (R', -m +n -1 ) }  \text{ as }  m \to \infty. \label{eq:rcond}
\end{align} 
Equation Eq.~(\ref{eq:rcond}) follows from the definition of a dislocation: there is a mismatch in the number of particles in the rows above and below the glide plane. Notice that Eq.~(\ref{eq:rcond}) implies $\Gamma(R, R', n)$ is \emph{not} a closed contour for $ (R, R') \in \GG $. However, Eq.~(\ref{eq:rcond}) implies that we may form a single infinite continuous contour $\mathcal{C}(R',R)$ by concatenating the individual pieces 
\begin{align}
    \mathcal{C} (R, R') \equiv  \bigcup_{ n \in \ZZ } \Gamma(R, R',n) 
\end{align}
Whenever the interactions decay faster than $1/\scr$, we will be able to close the contour $\mathcal{C}(R,R') $ in the upper half plane. 
Hence, the expression for the work becomes:
\begin{align}
    W_\text{glide} =& \sum_{ (R, R') \in \BB   } \sum_{n \in S_{R',R}} \oint_{\Gamma(R,R',n)}  \vb F \cdot  \dd \scrb + \sum_{ (R, R') \in \GG } \oint_{ \mathcal{C} (R,R') }  \vb F \cdot \dd \scrb  \\
    =&\sum_{ (R, R') \in \BB   } \sum_{n \in S_{R',R}} \int_{ V_n (R, R')  } \nabla \times \vb F  \, \dd^2 \scr + \sum_{ (R, R') \in \GG } \int_{ V (R,R') }  \nabla \times \vb F \,  \dd^2 \scr
\end{align}
where $V_n (R,R')$ and $V(R,R')$ are the signed areas enclosed by $\Gamma(R,R',n)$ and $\mathcal{C}(R,R') $

From the point of view of the bonds in set $\BB$, the solid is actually two disjoint sets separated by a cut along the glide plane. One could estimate various contributions from $\BB$ by taking the dot product of the average motion with the boundary stresses on the cut surface. This approach is the essence of the Peach-Koehler force computed in the continuum. 
The contributions from $\GG$, however, involves bond reassignment through the core. This piece is disregarded in the Peach-Koehler calculation when a branch cut discontinuity is introduced into the derivation. 
Assuming short ranged interaction, the dominant contribution from the $\GG$ terms comes from $\mathcal{C}(0,1)$, i.e. the bonds that directly span the glide plane. In this approximation, we have:
\begin{align}
    W_\text{glide} \approx \oint_{\mathcal{C}(0,1)} \vb F \cdot \dd \scrb
\end{align}
The exact shape of $\mathcal{C}(0,1)$ (referred to as $\mathcal{C}$ in the main text) is not \emph{a priori} known. The simplest approximation is to assume that all the atoms move along straight lines parallel to the glide plane, in which case $\mathcal{C}$ is given by the horizontal line $\scr_y = \frac{\sqrt{3}}{2} a$ traversed from right to left.  In this approximation, the work becomes:
\begin{align}
    W_\text{glide} \approx \eval{ \int_{-\infty}^\infty F^\perp_x \dd \scr_x}_{\scr_y = \frac{ \sqrt{3}}2 a} 
\end{align}
Beyond using the straight line $\scr_y = \frac{\sqrt{3}}2 a$, a more refined approximation would be to use the static image of the dislocation to construct an approximation to $\mathcal{C}$ by interpolating the relative bond coordinates. 

\section{Numerics and microscopic models} \label{sec:num}

\subsection{Molecular dynamics simulations} 

To investigate dislocation motion, noncentral pairwise forces were implemented in the molecular dynamics (MD) framework \texttt{HOOMD-Blue} (v 2.9) \cite{Anderson2008, Glaser2015}.
For all MD simulations reported here, the particle position $\vb x^\alpha$ evolves according to a Langevin integrator at zero temperature $\ddot {\vb x}^\alpha + \gamma \dot {\vb x}^\alpha  = \vb F^{\alpha} $, where $\vb F^\alpha$ is the total force on particle $\alpha$ and $\gamma=100$, resulting in highly-damped dynamics. All simulations are conducted in two dimensions and use a time step $dt=10^{-3}$.

\begin{figure}
    \centering
    \includegraphics[width=\textwidth]{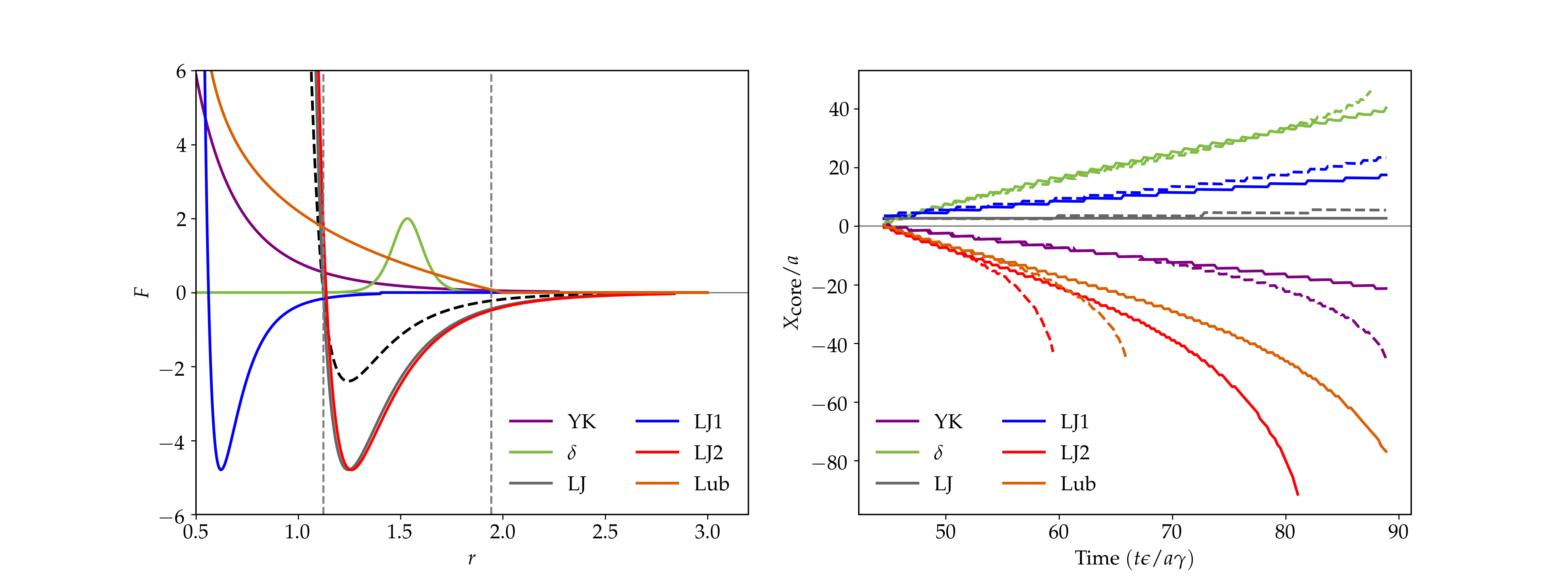}
    \caption{ {\bf Dislocation motion in free floating clusters.}~(Left)~Plots of $F^\perp(r)$ for the functions and parameters listed in Table~\ref{tab:params}. The dashed black curve indicates $F^\parallel (r)$, which is the same for all simulations. The vertical grey dashed lines indicate the first and second neighbor shells.  (Right)~The core of dislocations is tracked as a function of time for cluster sizes of $R=100a$ (solid) and $R=200a$ (dashed).  
    }
    \label{fig:motfull}
\end{figure}

Our interactions between the particles are of the form $\vb F (\vb r ) = F^\parallel (r) \, \vb {\hat r}  - F^\perp(r) \, \bph $. The functions $F^\parallel(r)$ and $F^\perp(r) $ are drawn from the following four families of curves:
\begin{align} 
            F_\text{LJ} (r)  = &  
            \begin{cases}
             -4\epsilon\left(6\frac{\sigma^6}{r^7} - 12\frac{\sigma^{12}}{r^{13}}\right) & r \le r_\text{cut} \\
             0 &r > r_\text{cut} 
            \end{cases} \label{eq:LJf}
            \\
            F_\text{Lub} (r) =& 
            \begin{cases}
             \epsilon \log\left(\frac{r_{cut}}{r-D}\right) & r-D \le r_\text{cut} \\
             0 &r-D > r_\text{cut} 
            \end{cases}
            \\ 
            F_\text{YK}(r) =   &
            \begin{cases}
             \epsilon e^{-\kappa r} \qty( \frac{1}{r^2}+ \frac{\kappa}{r} ) & r \le r_\text{cut} \\
             0 &r > r_\text{cut} 
            \end{cases}
            \\ 
            F_\delta (r) =  &
            \begin{cases}
             4\epsilon \frac{e^{\sigma(r - \delta)}}{ \qty(1+e^{\sigma(r-\delta)})^2} & r \le r_\text{cut} \\
             0 &r > r_\text{cut} \label{eq:deltaf} 
            \end{cases} 
\end{align} 
Here, $F_\text{LJ}$ is the gradient of the Lennard-Jones potential \cite{Jones1924}, $F_\text{Lub}$ is a model for the hydrodynamically-mediated forces between spinning spheres \cite{Kim2013microhydrodynamics}, and $F_\text{YK}$ is gradient of the Yukawa potential \cite{Yukawa1935}.
For all simulations, we use  $F^\parallel (r) = F_\text{LJ} (r) $ with parameters $\sigma = \epsilon =1$.
This radial force has a zero at $a= 2^{1/6}$, which sets the lattice spacing of the clusters. We cut off the radial interaction at $r_\text{cut} = 2.5 a$. This interaction is shown as the dashed black line in Fig.~\ref{fig:motfull}. Below, we describe in detail the methodology used for each of the simulations appearing in the main text. Notice that $F_\text{YK}(r)$ and $F_\text{Lub} (r)$ are generic, monotonically decaying functions of a single sign. The force $F_\text{LJ}$ has a zero crossing with a sign change, while $F_\delta$ has a single isolated peak of single sign.

\subsubsection{Cluster data} \label{sec:freecluster}

The data in Fig.~\ref{fig:motion} and Fig.~\ref{fig:motfull} consist of a single dislocation in a free-floating cluster. 
To prepare the cluster, an initially perfect crystal domain was initialized, and particle displacements consistent with a dislocation of Burgers vector $\vb b = a \vb {\hat x}$ were applied.
The resulting distorted crystallite was cut into a circle and placed within a periodic simulation domain three times the width of the cluster.
In general, the transverse pairwise interactions give rise to a solid-body rotation of the cluster. 
To allow each cluster to reach its terminal rotation speed before measuring dislocation motion, a network of harmonic bonds was created between neighboring particles in a region surrounding the core of the dislocation.
(The rest lengths of these auxiliary bonds were chosen based on an interpolation of the continuum  displacements field near the dislocation core. The spring constants $k$ were chosen to be as small as possible to stabilize the dislocation, see Table~\ref{tab:params}.)
With the Hookean bonds in place, the cluster was allowed to relax for $5000$ simulation time units, after which the Hookean bonds were disabled and the system evolved according to their pairwise interactions for an additional $5000$ time units.
To measure the dislocation position, a Voronoi tessellation was performed and and the dislocation position was defined as the geometric center of the $5$-sided and $7$-sided Voronoi cells.
Dislocation core position is  tracked in the reference frame co-rotating with the cluster.
The parameters for the interactions are described below:

\begin{enumerate} 
\item {\bf  Comparison of functional forms.} 
We perform the cluster simulations described above with the $F^\perp$ taking the function form listed in Table~\ref{tab:params}. For reference, the simulations are given the labels LJ, LJ1, LJ2, $\delta$, YK, Lub. 
The functional forms and the results of all simulation runs are presented in Fig.~\ref{fig:motfull}, and LJ, YK, and Lub are highlighted in main-text Fig.~\ref{fig:motfull}a-b as representative examples. In Table~\ref{tab:params}, we list the ambient torque density ($\tau \approx \sqrt{3} F^\perp(a)/a$) for each interaction. For LJ, LJ1, LJ2, YK, and Lub, the Peach-Koehler force $f_i^\text{PK} = - \tau b_i$ gives the observed direction of travel listed in Table~\ref{tab:params}. [Since $\vb b = a \vb {\hat x}$, the force $\vb f^\text{PK}$  points to the left (right) for $\tau>0$ ($\tau <0$) in the reference frame co-rotating with the cluster.] However, the simulation $\delta$ travel opposite the direction predicted the sign of $\vb f^\text{PK} $. To diagnose the origin of the sign change, we compare the active core force $\vb f^\text{core}$ derived in \S\ref{sec:core} to the Peach-Koehler interaction, as described below: 

\item {\bf Variation of $F_\delta^\perp$ interaction.} For this set of simulations we use the transverse force $F^\perp(r) = F_\delta (r)$ with parameters $\epsilon=2$, $\sigma=20$, $k=100$, and $\delta$ taking 7 equally spaced values in the interval $[1.0795 a , 1.1830 a ]$. This range of $\delta$ corresponds to moving the central peak of the function $F_\delta$ from roughly 10\% to 25\% of the way between the first and second neighbor shells. 
As illustrated in Fig.~\ref{fig:motion}e-f of the main text, we observe a transition between right traveling at larger $\delta$ and left traveling at smaller $\delta$. 

From Eq.~(\ref{eq:coreapprox}) in the main text, we obtain
\begin{align}
   f^\text{core} = \frac1a  W \approx \frac1a \int_{-\infty}^\infty 4 \epsilon \frac{r_y}{r} \frac{ e^{\sigma(r-\delta ) }}{ [1 + e^{\sigma(r-\delta )}]^2 }     \dd r_x   \label{eq:estimate} 
\end{align}
where $r= \sqrt{r_x^2 + r_y^2}$, $r_y = \frac{\sqrt{3}} 2 a$, $a=2^{1/6}$, $\sigma=20$, and $\epsilon=2$.  The Peach-Koehler force due to the torque density is given by $f^\text{PK} =-\sqrt{3} F_\delta (a)$. The quantities $\abs{f^\text{core}}$ and $\abs{f^\text{PK} }$  are plotted in Fig.~\ref{fig:motion}f, and their crossover coincides with the change in direction. 
\end{enumerate}

\subsubsection{Uniaxial compressions}

\begin{figure}
    \centering
    \includegraphics{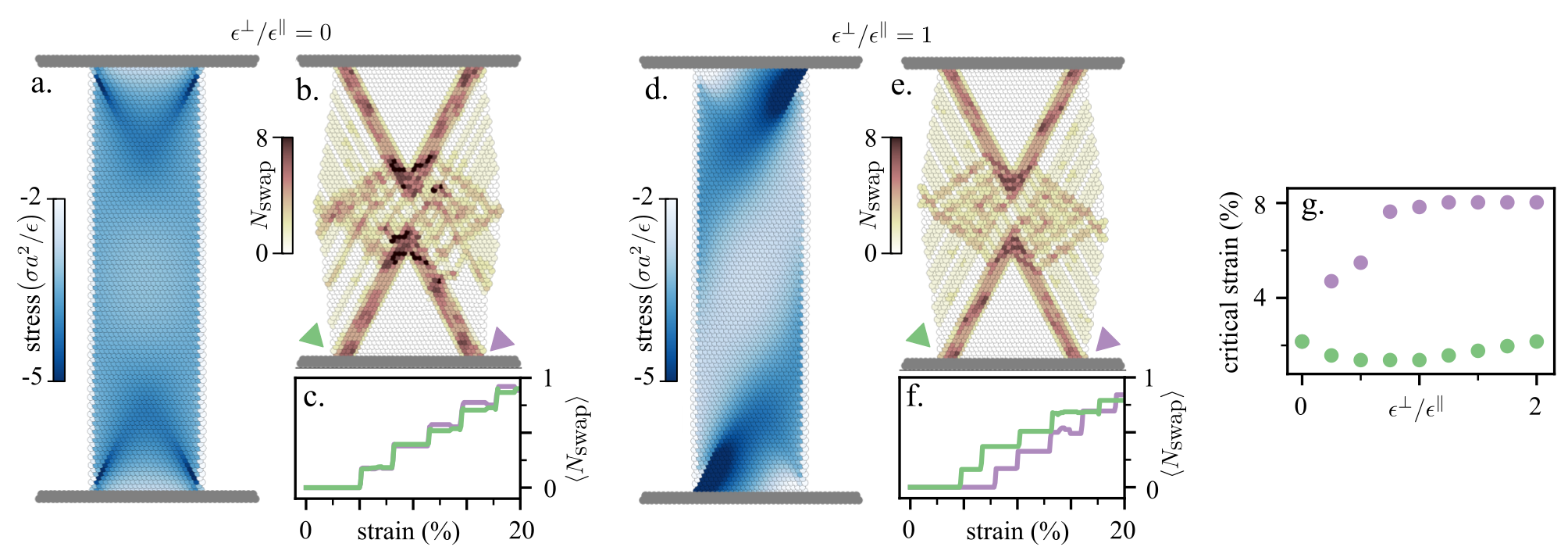}
    \caption{{\bf Uniaxial compression without ambient torque density.} 
    The plots and labels for this figure are identical to that of Fig.~\ref{fig:plastic}a-g. In contrast to Fig.~\ref{fig:plastic}, the transverse force here is taken to be $F^\perp(r) = F_\text{LJ} (r) $ with $\sigma =1$ and $\epsilon = \epsilon^\perp$. This choice of transverse force does not entail an ambient torque density prior to deformation. Nonetheless, $F^\perp(r) = F_\text{LJ} (r) $ gives rise to odd elastic moduli which are responsible for biasing the stress distribution in response to applied strain. Hence, many of the qualitative features here, such as the biased dislocation nucleation, are shared with Fig.~\ref{fig:plastic}a-g in the main text. 
    }
    \label{fig:plasticSI}
\end{figure}

Fig.~\ref{fig:plastic}a-f show snapshots from simulations of uniaxial compressions. The simulations consist of three groups of particles: material particles and two groups of wall particles (in grey). 
A rectangular monocrystalline beam of material particles was initialized between two thin monocrystalline slabs of wall particles.
The orientations of the wall and material crystals were epitaxial, with a lattice vector pointing in the $\vb{\hat x}$  direction (parallel to the interface).
All particles (including wall particles) interact via identical pairwise force laws. The material particles evolve according to their interactions, while trajectories of the wall particles follow a prescribed path regardless of the forces they experience. 
To simulate compression, the two walls were moved closer together in small discrete steps followed by a relaxation time.
For the data shown, a compression of total distance $30a$ was broken into steps of size $0.01a$ and spaced evenly over a time of $20,000$ simulation time units.
For beams of initial length $100a$, this produced a strain rate of $\dot \eta = 1.5\times 10^{-5}$ (inverse simulation time units).

In Fig.~\ref{fig:plastic}a,d, a projection of the stress tensor is shown. The per-particle stress tensors were computed using the virial equation of microscopic stress:
\begin{equation}
\sigma_{ij}^{\alpha} = -[\dot x_i^\alpha -  v_i (\vb x ^\alpha)  ][\dot x_j^\alpha - v_j (\vb x^\alpha ) ] - \frac{1}{2}\sum_{\beta, \beta \neq \alpha } (x_i^\alpha - x_i^\beta )F_j^{\beta \alpha } 
\end{equation}
where $\vb x^\alpha$ is the position of particle $\alpha$, $\vb v(\vb x)$ is the local average velocity field, $\vb F^{\alpha \beta}$ is the force from particle $\beta$ on particle $\alpha $, and the summation over $\beta$ includes all neighbors of $\alpha$ within the cutoff radius of pair interactions.
This corresponds to Eq.~(\ref{eq:coarsestress}) with a uniform smoothing function. 
In Fig.~\ref{fig:plastic}a,d we plot as a color map the following component of the stress tensor:
\begin{align}
    \sigma^\alpha =- (a^2_i \epsilon_{jk} a^2_k +a^2_j \epsilon_{ik} a^2_k + a^3_i \epsilon_{jk} a^3_k +a^3_j \epsilon_{ik} a^3_k )  \sigma_{ij}^\alpha = -\tau^3_{ij} \sigma_{ij}^\alpha
\end{align}
where $\vb a^1 = \qty( -\frac12, \frac{\sqrt3}2 )^T$ and $\vb a^3 = \qty(-\frac12, -\frac{\sqrt3}2)^T$.
Physically, $\sigma^\alpha$ is a proxy for the sum of the shear stresses along the $[11]$ and $[\bar{1}1]$ crystal planes. 

In Fig.~\ref{fig:plastic}b,e, we visualize the plastic deformation by coloring each particle by the number of neighbor swaps $N_\text{swap}$.
The quantity $N_\text{swap}$ is defined as follows.
For every particle a unique label is created as well as a list of the labels of all neighbors within a distance $1.6$.
The value of $N_\text{swap}$ for each particle is incremented each time a new unique label appears, an old label disappears, or a swap of labels in the list occurs.

In Fig.~\ref{fig:plastic} of the main text, we show data with $F^\perp=0$ (panels a-c) and $F^\perp(r) = F_\text{Lub} (r)$. We use the parameters $r_\text{cut} =1.5$, $D = 0.25$, and $\epsilon$ assuming 9 equally spaced values in the range $[0,2]$.  We note that $F^\perp(r) = F_\text{Lub} (r) $ gives rise to an appreciable ambient torque density even in absence of elastic strains. One may ask whether the torque density or the elastic stresses are responsible for the qualitative features of the plastic deformation. 
In Fig.~\ref{fig:plasticSI}, we present an additional data set with $F^\perp(r) = F_\text{LJ}$ with $\sigma=1$ and $\epsilon $ assuming 9 equally spaced intervals in the range $[0,2]$. This choice of $F^\perp$ does not give rise to an ambient torque density (see Table~\ref{tab:params}). Nonetheless, the plastic deformation displays many of the same qualitative features such as the biased nucleation of dislocations. See also Supplemental Movies S3-4.

\subsubsection{Disk compressions} 

In Fig.~\ref{fig:plastic}h shows dislocation motion in response to compression in a free-floating cluster. 
For these simulations, we use $F^\perp (r) = F_\text{LJ}(r)$ with $r_\text{cut} =2.5 a$, $\sigma=1$, and $\epsilon=2$.
The clusters are prepared as described in \S\ref{sec:freecluster}.
After an equilibration time of $5000$ simulation time units, the cluster is exposed to a dilation. The dilation is applied via a continuous box resizing. The position of each particle, $\vb x^\alpha$, is scaled via $ (1+ \lambda  ) \vb x^\alpha $ each time step without rescaling the pair force interaction parameters.
Clusters were therefore compressed isotropically until pair forces were sufficiently strong to counteract the position rescaling. We used six dilation rates $\lambda$ evenly spaced in the range $ [-5 \times 10^{-5},0]$ on clusters of diameter $200a$.
Comparative tests on clusters of passive particles ($F^\perp=0$) showed no dislocation core motion at any of the dilation rates studied.
Positive dilation rates applied to the same system resulted in crystal fracture before dislocation motion.

\begin{table}[]
    \centering
    \begin{tabular}{| c| c|c| c| c| }
    \hline 
        Label & $F^\perp$ & Parameters &  $\tau$ & behavior\\
         \hline 
         \hline 
         LJ & $F_\text{LJ} $ & $r_\text{cut} = 2.5 a $, $\epsilon=2$, $\sigma=1$, $k=1$ & 0.0 & static \\ \hline 
        LJ1 &  $F_\text{LJ}$&  $r_\text{cut} = 1.25 a $, $\epsilon =2$, $\sigma=0.5$, $k=1$  & $-0.25$  & right moving   \\ \hline 
        LJ2 & $F_\text{LJ}$ & $r_\text{cut}=2.53 a$, $\epsilon = 2$, $\sigma=1.025$, $k=1$ &  $2.79$ & left moving \\  \hline 
        $\delta$ & $F_\delta$ & $r_\text{cut}=2.03$, $\epsilon=2$, $\sigma =20$, $\delta = a \frac{1 + \sqrt 3}2 $, $k=100$  &  $3.3 \times 10^{-3}$ & right moving \\ \hline 
        YK & $F_\text{YK}$ & $r_\text{cut} = 2.28$, $\epsilon = 2$, $\kappa =2$, $k=10$ & 0.84  & left moving \\ \hline 
        Lub & $F_\text{Lub}$ & $r_\text{cut} = 1.5$, $\epsilon=2$, $D =0.25$, $k=10$ & 2.71  & left moving\\ \hline 
    \end{tabular}
    \caption{ {\bf Simulation parameters for cluster data}. This table contains the simulation labels, the corresponding functional forms for $F^\perp(r)$, and the simulation parameters used in the free floating cluster data shown in Fig.~\ref{fig:motfull} and Fig.~\ref{fig:motion} of the main text. The explicit functional forms and the definitions of the parameters are provided in Eqs.~(\ref{eq:LJf}-\ref{eq:deltaf}).  
    Here, $a=2^{1/6}$. The parameter $k$ corresponds to the spring constant of the auxiliary Hookean springs used in the initial relaxation of the lattice before data is acquired. 
    }
    \label{tab:params}
\end{table}

\subsection{Fixed topology simulations} \label{sec:FTS}

\begin{figure}[t!]
    \centering
    \includegraphics[width=1\textwidth]{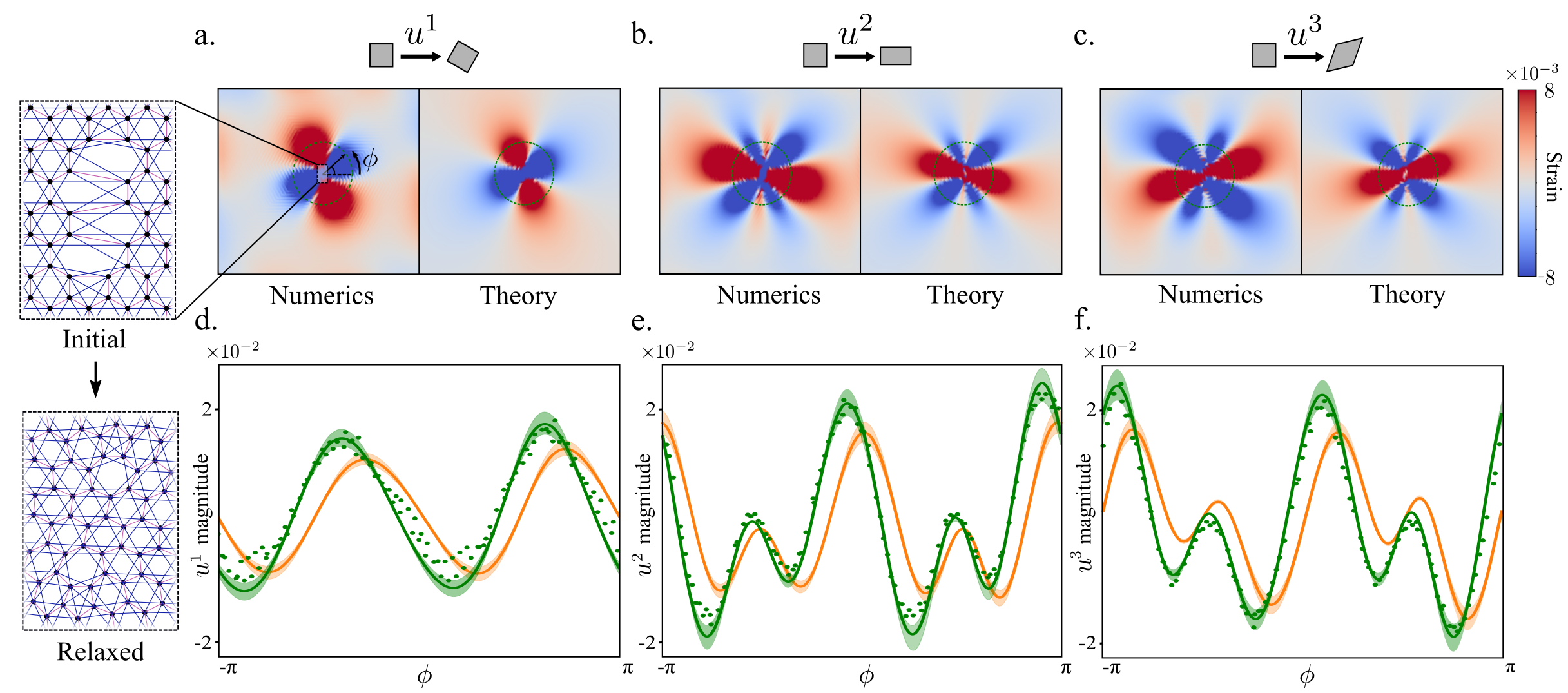}
    \caption{{\bf Strain field of a dislocation pair.}  {\bf a-c.}~The distribution of rotation ($u^1$), shear 1 ($u^2$), and shear 2 ($u^3$) surrounding a pair of dislocations in an odd elastic solid with $\nu =0.8$ and $\nu^o = -0.88$. The right panels visualize the continuum theory while the left panel is the result of numerics. The inset to panel (a) renders the individual masses and bonds comprising the dislocation pair before (top) and after (bottom) relaxation. {\bf d-f.}~We quantitatively compare numerics and experiments by sampling the strain (green dots) at points between 8.0 and 9.2 lattice spacings from the center. The green line is the theoretical curve with $\nu =0.8$ and $\nu^o = -0.88$. The shaded background accounts in the variation in distance from the center. 
    The orange lines, provided for reference, are theoretical curves for a passive solid with $\nu =0.8$ and $\nu^o = 0$. We note that the dilation ($u^0$) is too small for numerical validation.}
    \label{fig:numerics}
\end{figure}

\begin{figure}[t!]
    \centering
    \includegraphics[width=0.5\textwidth]{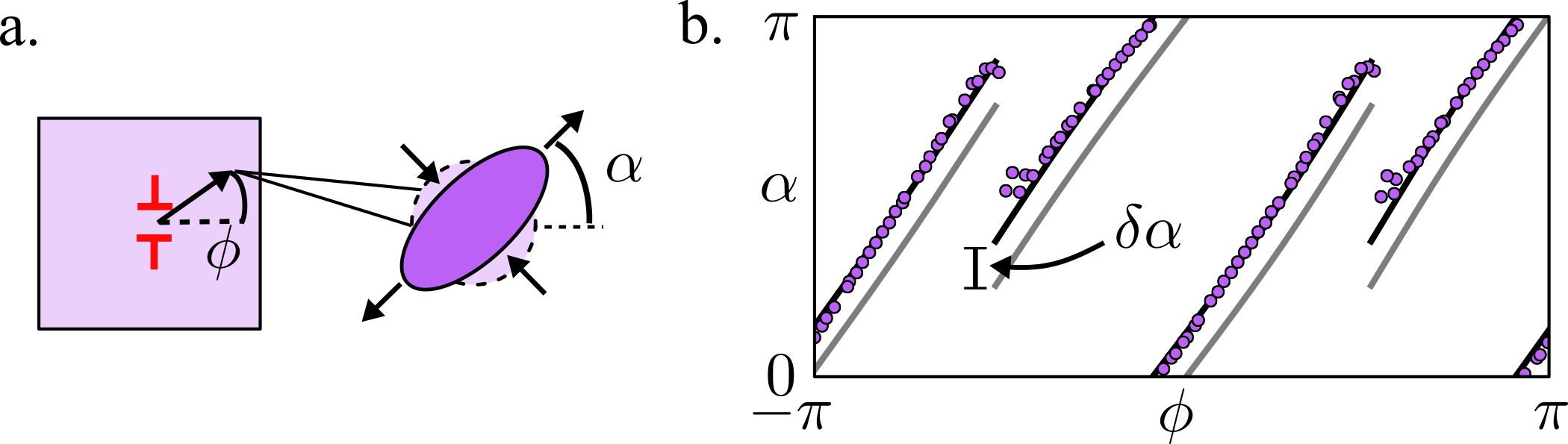}
    \caption{  {\bf Chiral strain distribution around defects in odd elastic solids.}  {\bf a.}~Two antialigned dislocations are placed in the center of a honeycomb lattice of generalized Hookean springs, see Fig.~\ref{fig:numerics}. {\bf b.}~The orientation of the local shear strain is captured by an angle $\alpha$, plotted as a function of the polar angle $\phi$. The grey lines are theoretical curves for $\nu^o=0$, the black lines (colored points) are theoretical curves (numerical data) for $\nu^o=-0.88$. The vertical separation between the grey and black lines is denoted $\delta \alpha$ and is given in Eq.~(\ref{eq:shearrot}) of the main text.  }
    \label{fig:strain}
\end{figure}

To validate the static strain field, we use a microscopic model with fixed bond topology. 
This model consists of a mass spring network with generalized Hookean springs 
for which $F^\parallel(r) = - k (r-\ell)$, $F^\perp(r) = -k^a (r-\ell) $, 
and $\ell$ is the rest length of the bond (see  Fig.~\ref{fig:coretorque}a and Ref.~\cite{scheibner2020}).
We use a honeycomb lattice with nearest-neighbor and next-nearest-neighbor bonds. Such a lattice permits an analytical coarsegraining thereby allowing the expression of macroscopic elastic coefficients in terms of microscopic parameters~\cite{scheibner2020}. Specifically, we have
\begin{align} \label{displacement solution}
    &B = \frac{k_\NN + 6k_\NNN}{2\sqrt{3}} &A = \frac{k_\NN^a + 6k^a_\NNN}{2\sqrt{3}} &&\mu = \frac{\sqrt{3}k_\NNN}{2} &&K^o = \frac{\sqrt{3}k_\NNN^a}{2}
\end{align}
where the subscripts $\NN$ and $\NNN$ denote the nearest-neighbor and next-nearest-neighbor parameters, respectively.

For the data shown in Fig.~\ref{fig:strain}, we simulate a $60\times60$ unit cell rectangle with free boundary conditions. As shown in the inset of Fig.~\ref{fig:numerics}a, we introduce a dislocation pair vertically separated by four unit cells by removing removing a column of 8 particles from the center of the lattice and locally reassigning bonds. We then allow the lattice to relax under first order dynamics, integrated using second-order Runge-Kutta. Once the lattice equilibrates, we compute the strain at each node by performing a linear regression from an undeformed template hexagon to the hexagon defined by next-nearest-neighbor connections.  
For the data shown in Fig.~\ref{fig:strain} and Fig.~\ref{fig:numerics}, we have $k_\NN=0.92$, $k_\NN^a = 0.39$, $k_\NNN = 0.027$, and $k_\NNN^a=0.012$ which corresponds to $\nu =0.8$ and $\nu^o = -0.88$.

As noted in Eq.~(\ref{eq:shearrot}) of the main text, the qualitative effect of the parameter $\nu^o$ is to rotate the local axis of shear strain. This rotation angle is measured around the dislocation dipole in Fig.~\ref{fig:strain}b.  
The theoretical curves in Fig.~\ref{fig:strain}b are produced by superimposing two copies of the exact continuum solutions in Eq.~(\ref{eq:udisl}) then computing the angle $\alpha$ corresponding to the local axis of shear elongation. The theoretical curve takes a simple analytical form in the dipole approximation. Consider a pair of dislocations with Burgers vectors $\vb b_{\pm} = \pm b \vb {\hat x}  $ at locations $(0, \pm \delta r/2)$. When $\nu^o=0$, the shear strain $\tilde u_{ij} = \frac12( \partial_i u_j + \partial_j u_i - \partial_k u_k \delta_{ij}) $  to leading order in $\delta r$ is given by:
\begin{align}
    \tilde u_{ij} \propto  \frac{b \delta r}{2r^2} \mqty( \cos 2 \phi + \cos 4\phi & \sin 2 \phi + \sin 4 \phi \\ \sin 2 \phi + \sin 4 \phi  & -\cos 2\phi- \cos 4\phi ) 
\end{align}
where we have set aside constant prefactors. 
Hence, we have 
\begin{align}
u^2=& \tau^2_{ij} \tilde u_{ij} \propto  \frac{b \delta r}{r^2} \qty( \cos 2 \phi  + \cos 4 \phi) \\
u^3 =& \tau^3_{ij} \tilde u_{ij} \propto \frac{b \delta r }{r^2} \qty (\sin 2 \phi  + \sin 4 \phi)
\end{align}
Finally, the angle $\alpha$ is defined as
\begin{align}
    R e^{2 i \alpha } = u^2 + i u^3 
\end{align}
with a positive amplitude $R$ that represents the total magnitude of the shear strain. We find 
\begin{align}
    \alpha = \frac32 \phi - \frac{\pi}{4}(1 - \operatorname{sgn} (\cos \phi )  )
\end{align}
Physically, the angle $\alpha$ specifies the axis of elongation for the local shear and is therefore periodic modulo $\pi$. Notice that $\alpha$ contains discontinuities at $\phi = \pm \pi/2 $. However, the amplitude $R  \propto \cos^2 (\phi) / r^2$ vanishes at these angles, so the physical strain field is continuous.  Finally, when $\nu^o$ is nonzero, the expression for $\alpha$ simply receives an additive correction $\delta \alpha = -\arctan[2 \nu^o / (1+\nu)] /2$ yields:
\begin{align}
    \alpha = \frac32 \phi - \frac{\pi}{4}[1 - \operatorname{sgn} (\cos \phi )  ] - \frac12 \arctan( \frac{2 \nu^o } { 1+\nu } )  \label{eq:dipole}
\end{align}

\section{Coarsegraining procedures} \label{sec:coarse}

In this section, we discuss the relationship between microscopic systems and the continuum fields.
Suppose that each particle $\alpha$ is a complex object and has a mass distribution in space $\rho^\alpha (\vb R, t) $ with an underlying microscopic velocity field $\vb v^\alpha (\vb R, t)$. We express the conservation of mass via 
$\dot \rho^\alpha = - \nabla \cdot (\rho^\alpha \vb v^\alpha) $  
and we represent linear momentum exchange via 
\begin{align}
 \rho^\alpha (\vb R)  [\dot {\vb v}^\alpha (\vb R)  + \vb v^\alpha (\vb R) \cdot \nabla {\vb  v}^\alpha (\vb R)  ] = \vb f^{\ext,\alpha} (\vb R) + \sum_{\beta } \int  \vb f^{\beta \alpha } (\vb R', \vb R)  \; \dd^2 R' \label{eq:microfull} 
\end{align}
where $\vb f^{\ext, \alpha } (\vb R)$ is the external force density and $\vb f^{\beta\alpha } (\vb R', \vb R) = - \vb f^{\alpha \beta} (\vb R, \vb R') $ is the force from the constituents of particle $\beta$ to the constituents of particle $\alpha$. This force, for example, could be be the hydrodynamic force mediated by a fluid in between two spinning colloidal particles. 
In principle, $\vb f^{\alpha \beta}$ and $\vb f^\text{ext}$ are intractably complicated functions of the state of the system. Often, however, one is only interested in the center-of-mass translation and spinning of the individual particles. 
The mass $m$, center of mass $\vb x^\alpha (t) $, linear momentum $\vb p^\alpha (t)$, moment of inertia $I$, angular velocity $\omega^\alpha(t) $, and angular momentum $\ell^\alpha(t)$ are given by
\begin{align}
m =& \int  \rho^\alpha (\vb R,t ) \; \dd^2 R &
\vb x^\alpha (t) =&\frac1{m} \int \vb R \rho^\alpha (\vb R,t) \; \dd^2 R \\
I =& \int \abs{ \vb R - \vb x^\alpha(t) }^2 \rho^\alpha (\vb R,t) \; \dd^2 R & 
\omega^\alpha(t) =& \frac1I \int  [\vb R -\vb x^\alpha(t) ] \times \vb v^\alpha (\vb R,t) \rho^\alpha(\vb R, t) \; \dd^2 R \\
\vb p^\alpha =& \int \vb v^\alpha(\vb R,t) \rho^\alpha (\vb R,t) \; \dd^2 R = m \dot{ \vb x}^\alpha(t) &
\ell^\alpha(t) =&  \int \vb R \times \vb v^\alpha(\vb R,t) \rho(\vb R,t) \; \dd^2 R = \ell^\alpha_\text{cm}(t) + \ell^\alpha_\text{int}(t)
\end{align} 
where $\ell^\alpha_\text{cm} = \vb x^\alpha \times \vb p^\alpha  $ is the center-of-mass (i.e. orbital) angular momentum and $\ell^\alpha_\text{int} = I \omega^\alpha  $ is the internal (i.e. spin) angular momentum. We have assumed for simplicity that $m$ and $I$ are the same for all particles and are time independent.
The forces an torques governing these degrees of freedom are given by:
\begin{align}
\vb F^{\ext, \alpha } =& \int \vb f^{\ext,\alpha} (\vb R) \; \dd^2 R &    \vb F^{\alpha \beta } =& \int  \vb f^{\alpha \beta } (\vb R, \vb R') \;  \dd^2 R \; \dd^2 R' \\
T^{\ext, \alpha} =& \int \vb R \times \vb f^{\ext, \alpha} (\vb R) \; \dd^2 R &  T^{\alpha \beta } =& \int    \vb R' \times \vb f^{\alpha \beta } (\vb R, \vb R') \; \dd^2 R \; \dd^2 R'
\end{align}
It is useful to divide the torques into center-of-mass and internal contributions $\tau$
\begin{align}
T^{\ext, \alpha} =&   \vb x^\alpha \times \vb F^{\ext, \alpha}  + \tau^{\ext, \alpha} &  &\text{ where }  &  \tau^{\ext, \alpha } =& \int (\vb R- \vb x^\alpha) \times \vb f^{\ext, \alpha }(\vb R)  \; \dd^2 R \\ 
T^{\alpha \beta} =& \vb x^\beta \times \vb F^{\alpha \beta} + \tau^{\alpha \beta} & &\text{ where } &  \tau^{\alpha \beta} =& \int  (\vb R' - \vb x^\beta) \times \vb f^{\alpha \beta} (\vb R, \vb R') \; \dd^2 R \; \dd^2 R'
\end{align}
Finally, the transition to continuum fields entails a notion of homogenization for which there are many possible approaches~\cite{yang2012generalized}. For example, let $g(\vb R)$ be a smoothing function with $\int g(\vb R) \dd^2 R =1$. We define the density $\rho(\vb R)$, the momentum $\vb p (\vb R)$, velocity $\vb v(\vb R)$, and the external forces $\vb F^\ext (\vb R) $ as follows
\begin{align}
    \rho(\vb R) =& \sum_\alpha m g(\vb R - \vb x^\alpha) & \vb p (\vb R) =&  \sum_\alpha \vb p^\alpha g( \vb R - \vb x^\alpha)  \\
    \vb v(\vb R) =& \frac{\vb p(\vb R)}{ \rho(\vb R) } &
    \vb F^\ext(\vb R) =& \sum_\alpha \vb F^{\ext ,\alpha} g(\vb R- \vb x^\alpha) 
\end{align}
as well as the Cauchy stress tensor
\begin{align}
    \sigma_{ij} (\vb R)  =& - \sum_{\alpha } [\dot x_i^\alpha - v_i(\vb R) ] p^\alpha_j g(\vb R-\vb x^\alpha) -\frac12 \sum_{\alpha, \beta } r^{\alpha \beta}_i F_j^{\alpha \beta } B(\vb R, \vb x^\alpha, \vb x^\beta) \\
    \text{where} \quad B (\vb R, \vb x^\alpha, \vb x^\beta) =& \int_0^1 g[\vb R - \vb x^\alpha + \lambda(  \vb x^\alpha - \vb x^\beta)  ] \;  \dd \lambda  \label{eq:coarsestress}
\end{align}
Notice that this construction yields 
\begin{align} 
\dot p_j +  \partial_i ( v_i p_j )= \partial_i \sigma_{ij} + F_j^\ext \label{eq:linmot}
\end{align}
consistent with microscopic and macroscopic conservation laws. Next, given only information about the velocity field, it is useful to refer to $L(\vb R) = \vb R \times \vb p(\vb R)$ as the angular momentum density. In this case, the linear momentum conservation equation (\ref{eq:linmot}) directly yields:
\begin{align}
    \dot L + \partial_i (v_i L ) = \partial_i ( r_k \epsilon_{kj} \sigma_{ij} ) - \epsilon_{kj} \sigma_{kj} + r_i \epsilon_{ij} F_j^\ext \label{eq:L}
\end{align}
In Eq.~(\ref{eq:L}), the antisymmetric stress enters explicitly as a source term. The antisymmetric stress is proportional to the noncentral components of the interaction forces
\begin{align}
    \epsilon_{ij} \sigma_{ij} = - \frac12 \sum_{\alpha \beta} r_i^{\alpha \beta} \epsilon_{ij} F_j^{\alpha \beta} B(\vb R, \vb x^\alpha, \vb x^\beta) 
\end{align}
However, $L$ does not represent to full angular momentum density of the system. The total angular momentum density $\ell (\vb R) $, angular momentum current $\vb M(\vb R)$, and total external torques $T^\ext(\vb R)$ are given by 
\begin{align}
    \ell( \vb R) =& \sum_\alpha \ell^\alpha g (\vb R - \vb x^\alpha) & \vb M (\vb R) =& - \frac12 \sum_{\alpha, \beta }\vb r^{\alpha \beta} T^{\alpha \beta} B(\vb R, \vb x^\alpha,\vb x^\beta) & T^\ext(\vb R) = \sum_\alpha T^{\ext,\alpha} g(\vb R - \vb x^\alpha) 
\end{align}
which together obey the conservation equation 
\begin{align} 
\dot \ell +  \partial_i ( v_i \ell )= \partial_i M_i + T^\ext \label{eq:consL}
\end{align}
A generic situation is one in which the individual particles are driven to spin at a given speed via external torques $\tau^{\ext,\alpha}$, while applying negligible or zero external forces on each particle. The torques $\tau^{\ext, \alpha}$ enter Eq.~(\ref{eq:consL}) through $T^\ext$, but then are transduced into the linear momentum equation (\ref{eq:linmot}) via the transverse forces that create the antisymmetric stress.


\begin{thebibliography}{10}
\expandafter\ifx\csname url\endcsname\relax
  \def\url#1{\texttt{#1}}\fi
\expandafter\ifx\csname urlprefix\endcsname\relax\def\urlprefix{URL }\fi
\providecommand{\bibinfo}[2]{#2}
\providecommand{\eprint}[2][]{\url{#2}}

\bibitem{Nelson2002}
\bibinfo{author}{Nelson, D.~R.}
\newblock \emph{\bibinfo{title}{Defects and geometry in condensed matter
  physics}} (\bibinfo{publisher}{Cambridge University Press},
  \bibinfo{address}{Cambridge ; New York}, \bibinfo{year}{2002}).

\bibitem{Weertman1964}
\bibinfo{author}{Weertman, J.}
\newblock \emph{\bibinfo{title}{Elementary dislocation theory}}.
\newblock Macmillan series in materials science
  (\bibinfo{publisher}{Macmillan}, \bibinfo{address}{New York},
  \bibinfo{year}{1964}).

\bibitem{Chaikin1995}
\bibinfo{author}{Chaikin, P.~M.}
\newblock \emph{\bibinfo{title}{Principles of condensed matter physics}}
  (\bibinfo{publisher}{Cambridge University Press}, \bibinfo{address}{Cambridge
  ; New York, NY, USA}, \bibinfo{year}{1995}).

\bibitem{Paulose2015}
\bibinfo{author}{Paulose, J.}, \bibinfo{author}{Chen, B. G.-g.} \&
  \bibinfo{author}{Vitelli, V.}
\newblock \bibinfo{title}{{Topological modes bound to dislocations in
  mechanical metamaterials}}.
\newblock \emph{\bibinfo{journal}{Nat Phys}} \textbf{\bibinfo{volume}{11}},
  \bibinfo{pages}{153–--156} (\bibinfo{year}{2015}).

\bibitem{mietke2020anyonic}
\bibinfo{author}{Mietke, A.} \& \bibinfo{author}{Dunkel, J.}
\newblock \bibinfo{title}{Anyonic defect braiding and spontaneous chiral
  symmetry breaking in dihedral liquid crystals}.
\newblock \emph{\bibinfo{journal}{arXiv:2011.04648}}  (\bibinfo{year}{2020}).

\bibitem{Pretko2018fracton}
\bibinfo{author}{Pretko, M.} \& \bibinfo{author}{Radzihovsky, L.}
\newblock \bibinfo{title}{Fracton-elasticity duality}.
\newblock \emph{\bibinfo{journal}{Phys. Rev. Lett.}}
  \textbf{\bibinfo{volume}{120}}, \bibinfo{pages}{195301}
  (\bibinfo{year}{2018}).

\bibitem{Bowick2009}
\bibinfo{author}{Bowick, M.~J.} \& \bibinfo{author}{Giomi, L.}
\newblock \bibinfo{title}{Two-dimensional matter: order, curvature and
  defects}.
\newblock \emph{\bibinfo{journal}{Advances in Physics}}
  \textbf{\bibinfo{volume}{58}}, \bibinfo{pages}{449--563}
  (\bibinfo{year}{2009}).

\bibitem{Nelson1979}
\bibinfo{author}{Nelson, D.} \& \bibinfo{author}{Halperin, B.}
\newblock \bibinfo{title}{{Dislocation-mediated melting in two dimensions}}.
\newblock \emph{\bibinfo{journal}{Phys. Rev. B}} \textbf{\bibinfo{volume}{19}},
  \bibinfo{pages}{2457--2484} (\bibinfo{year}{1979}).

\bibitem{Wachtel2016Electrodynamic}
\bibinfo{author}{Wachtel, G.}, \bibinfo{author}{Sieberer, L.~M.},
  \bibinfo{author}{Diehl, S.} \& \bibinfo{author}{Altman, E.}
\newblock \bibinfo{title}{Electrodynamic duality and vortex unbinding in
  driven-dissipative condensates}.
\newblock \emph{\bibinfo{journal}{Phys. Rev. B}} \textbf{\bibinfo{volume}{94}},
  \bibinfo{pages}{104520} (\bibinfo{year}{2016}).

\bibitem{shankar2020topological}
\bibinfo{author}{Shankar, S.}, \bibinfo{author}{Souslov, A.},
  \bibinfo{author}{Bowick, M.~J.}, \bibinfo{author}{Marchetti, M.~C.} \&
  \bibinfo{author}{Vitelli, V.}
\newblock \bibinfo{title}{Topological active matter}.
\newblock \emph{\bibinfo{journal}{arXiv:2010.00364}}  (\bibinfo{year}{2020}).

\bibitem{Giomi2013}
\bibinfo{author}{Giomi, L.}, \bibinfo{author}{Hawley-Weld, N.} \&
  \bibinfo{author}{Mahadevan, L.}
\newblock \bibinfo{title}{Swarming, swirling and stasis in sequestered
  bristle-bots}.
\newblock In \emph{\bibinfo{booktitle}{Proceedings of the Royal Society of
  London A: Mathematical, Physical and Engineering Sciences}}, vol.
  \bibinfo{volume}{469}, \bibinfo{pages}{20120637} (\bibinfo{organization}{The
  Royal Society}, \bibinfo{year}{2013}).

\bibitem{Duclos2020topological}
\bibinfo{author}{Duclos, G.} \emph{et~al.}
\newblock \bibinfo{title}{Topological structure and dynamics of
  three-dimensional active nematics}.
\newblock \emph{\bibinfo{journal}{Science}} \textbf{\bibinfo{volume}{367}},
  \bibinfo{pages}{1120--1124} (\bibinfo{year}{2020}).

\bibitem{Colen2021machine}
\bibinfo{author}{Colen, J.} \emph{et~al.}
\newblock \bibinfo{title}{Machine learning active-nematic hydrodynamics}.
\newblock \emph{\bibinfo{journal}{Proceedings of the National Academy of
  Sciences}} \textbf{\bibinfo{volume}{118}} (\bibinfo{year}{2021}).

\bibitem{vafa2020multidefect}
\bibinfo{author}{Vafa, F.}, \bibinfo{author}{Bowick, M.~J.},
  \bibinfo{author}{Marchetti, M.~C.} \& \bibinfo{author}{Shraiman, B.~I.}
\newblock \bibinfo{title}{Multi-defect dynamics in active nematics}.
\newblock \emph{\bibinfo{journal}{arXiv:2007.02947}}  (\bibinfo{year}{2020}).

\bibitem{pearce2020programming}
\bibinfo{author}{Pearce, D. J.~G.}, \bibinfo{author}{Gat, S.},
  \bibinfo{author}{Livne, G.}, \bibinfo{author}{Bernheim-Groswasser, A.} \&
  \bibinfo{author}{Kruse, K.}
\newblock \bibinfo{title}{Programming active metamaterials using topological
  defects}.
\newblock \emph{\bibinfo{journal}{arXiv:2010.13141}}  (\bibinfo{year}{2020}).

\bibitem{Thijssen2020binding}
\bibinfo{author}{Thijssen, K.} \& \bibinfo{author}{Doostmohammadi, A.}
\newblock \bibinfo{title}{Binding self-propelled topological defects in active
  turbulence}.
\newblock \emph{\bibinfo{journal}{Phys. Rev. Research}}
  \textbf{\bibinfo{volume}{2}}, \bibinfo{pages}{042008} (\bibinfo{year}{2020}).

\bibitem{Rouzaire2021Defect}
\bibinfo{author}{Rouzaire, Y.} \& \bibinfo{author}{Levis, D.}
\newblock \bibinfo{title}{Defect superdiffusion and unbinding in a 2d xy model
  of self-driven rotors}.
\newblock \emph{\bibinfo{journal}{Phys. Rev. Lett.}}
  \textbf{\bibinfo{volume}{127}}, \bibinfo{pages}{088004}
  (\bibinfo{year}{2021}).

\bibitem{Chardac2021Topology}
\bibinfo{author}{Chardac, A.}, \bibinfo{author}{Hoffmann, L.~A.},
  \bibinfo{author}{Poupart, Y.}, \bibinfo{author}{Giomi, L.} \&
  \bibinfo{author}{Bartolo, D.}
\newblock \bibinfo{title}{Topology-driven ordering of flocking matter}.
\newblock \emph{\bibinfo{journal}{Phys. Rev. X}} \textbf{\bibinfo{volume}{11}},
  \bibinfo{pages}{031069} (\bibinfo{year}{2021}).

\bibitem{Fruchart2021phase}
\bibinfo{author}{Fruchart, M.}, \bibinfo{author}{Hanai, R.},
  \bibinfo{author}{Littlewood, P.~B.} \& \bibinfo{author}{Vitelli, V.}
\newblock \bibinfo{title}{Non-reciprocal phase transitions}.
\newblock \emph{\bibinfo{journal}{Nature}} \textbf{\bibinfo{volume}{592}},
  \bibinfo{pages}{363--369} (\bibinfo{year}{2021}).

\bibitem{Whitfield2017}
\bibinfo{author}{Whitfield, C.~A.} \emph{et~al.}
\newblock \bibinfo{title}{Hydrodynamic instabilities in active cholesteric
  liquid crystals}.
\newblock \emph{\bibinfo{journal}{The European Physical Journal E}}
  \textbf{\bibinfo{volume}{40}}, \bibinfo{pages}{50} (\bibinfo{year}{2017}).

\bibitem{Kole2021Layered}
\bibinfo{author}{Kole, S.~J.}, \bibinfo{author}{Alexander, G.~P.},
  \bibinfo{author}{Ramaswamy, S.} \& \bibinfo{author}{Maitra, A.}
\newblock \bibinfo{title}{Layered chiral active matter: Beyond odd elasticity}.
\newblock \emph{\bibinfo{journal}{Phys. Rev. Lett.}}
  \textbf{\bibinfo{volume}{126}}, \bibinfo{pages}{248001}
  (\bibinfo{year}{2021}).

\bibitem{Maitra2020}
\bibinfo{author}{Maitra, A.}, \bibinfo{author}{Lenz, M.} \&
  \bibinfo{author}{Voituriez, R.}
\newblock \bibinfo{title}{Chiral active hexatics: Giant number fluctuations,
  waves, and destruction of order}.
\newblock \emph{\bibinfo{journal}{Phys. Rev. Lett.}}
  \textbf{\bibinfo{volume}{125}}, \bibinfo{pages}{238005}
  (\bibinfo{year}{2020}).

\bibitem{gupta2020active}
\bibinfo{author}{Gupta, R.~K.}, \bibinfo{author}{Kant, R.},
  \bibinfo{author}{Soni, H.}, \bibinfo{author}{Sood, A.~K.} \&
  \bibinfo{author}{Ramaswamy, S.}
\newblock \bibinfo{title}{Active nonreciprocal attraction between motile
  particles in an elastic medium}.
\newblock \emph{\bibinfo{journal}{arXiv:2007.04860}}  (\bibinfo{year}{2020}).

\bibitem{kumar2014flocking}
\bibinfo{author}{Kumar, N.}, \bibinfo{author}{Soni, H.},
  \bibinfo{author}{Ramaswamy, S.} \& \bibinfo{author}{Sood, A.~K.}
\newblock \bibinfo{title}{Flocking at a distance in active granular matter}.
\newblock \emph{\bibinfo{journal}{Nature Communications}}
  \textbf{\bibinfo{volume}{5}}, \bibinfo{pages}{4688} (\bibinfo{year}{2014}).

\bibitem{VanSaders2019}
\bibinfo{author}{VanSaders, B.} \& \bibinfo{author}{Glotzer, S.~C.}
\newblock \bibinfo{title}{Designing active particles for colloidal
  microstructure manipulation via strain field alchemy}.
\newblock \emph{\bibinfo{journal}{Soft Matter}} \textbf{\bibinfo{volume}{15}},
  \bibinfo{pages}{6086--6096} (\bibinfo{year}{2019}).

\bibitem{VanSaders2021Sculpting}
\bibinfo{author}{VanSaders, B.} \& \bibinfo{author}{Glotzer, S.~C.}
\newblock \bibinfo{title}{Sculpting crystals one burgers vector at a time:
  Toward colloidal lattice robot swarms}.
\newblock \emph{\bibinfo{journal}{Proceedings of the National Academy of
  Sciences}} \textbf{\bibinfo{volume}{118}} (\bibinfo{year}{2021}).

\bibitem{digregorio2021unified}
\bibinfo{author}{Digregorio, P.}, \bibinfo{author}{Levis, D.},
  \bibinfo{author}{Cugliandolo, L.~F.}, \bibinfo{author}{Gonnella, G.} \&
  \bibinfo{author}{Pagonabarraga, I.}
\newblock \bibinfo{title}{Unified analysis of topological defects in 2d systems
  of active and passive disks}.
\newblock \emph{\bibinfo{journal}{arXiv:2106.03454}}  (\bibinfo{year}{2021}).

\bibitem{Poncet2021When}
\bibinfo{author}{Poncet, A.} \& \bibinfo{author}{Bartolo, D.}
\newblock \bibinfo{title}{When soft crystals defy newton's third law:
  Non-reciprocal mechanics and dislocation motility} (\bibinfo{year}{2021}).
\newblock \eprint{2110.02897}.

\bibitem{Yan2015}
\bibinfo{author}{Yan, J.}, \bibinfo{author}{Bae, S.~C.} \&
  \bibinfo{author}{Granick, S.}
\newblock \bibinfo{title}{{Rotating crystals of magnetic Janus colloids}}.
\newblock \emph{\bibinfo{journal}{Soft Matter}} \textbf{\bibinfo{volume}{11}},
  \bibinfo{pages}{147--153} (\bibinfo{year}{2015}).

\bibitem{Bililign2021Motile}
\bibinfo{author}{Bililign, E.~S.} \emph{et~al.}
\newblock \bibinfo{title}{Motile dislocations knead odd crystals into whorls}.
\newblock \emph{\bibinfo{journal}{Nature Physics}}  (\bibinfo{year}{2021}).

\bibitem{vanZuiden2016}
\bibinfo{author}{van Zuiden, B.~C.}, \bibinfo{author}{Paulose, J.},
  \bibinfo{author}{Irvine, W. T.~M.}, \bibinfo{author}{Bartolo, D.} \&
  \bibinfo{author}{Vitelli, V.}
\newblock \bibinfo{title}{Spatiotemporal order and emergent edge currents in
  active spinner materials}.
\newblock \emph{\bibinfo{journal}{Proc. Natl. Acad. Sci. USA}}
  \textbf{\bibinfo{volume}{113}}, \bibinfo{pages}{12919--12924}
  (\bibinfo{year}{2016}).

\bibitem{scheibner2020}
\bibinfo{author}{Scheibner, C.} \emph{et~al.}
\newblock \bibinfo{title}{Odd elasticity}.
\newblock \emph{\bibinfo{journal}{Nature Physics}}
  \textbf{\bibinfo{volume}{16}}, \bibinfo{pages}{475--480}
  (\bibinfo{year}{2020}).

\bibitem{scheibner2020non}
\bibinfo{author}{Scheibner, C.}, \bibinfo{author}{Irvine, W. T.~M.} \&
  \bibinfo{author}{Vitelli, V.}
\newblock \bibinfo{title}{Non-hermitian band topology and skin modes in active
  elastic media}.
\newblock \emph{\bibinfo{journal}{Phys. Rev. Lett.}}
  \textbf{\bibinfo{volume}{125}}, \bibinfo{pages}{118001}
  (\bibinfo{year}{2020}).

\bibitem{tan2021development}
\bibinfo{author}{Tan, T.~H.} \emph{et~al.}
\newblock \bibinfo{title}{Development drives dynamics of living chiral
  crystals}.
\newblock \emph{\bibinfo{journal}{arXiv:2105.07507}}  (\bibinfo{year}{2021}).

\bibitem{Chen2021Realization}
\bibinfo{author}{Chen, Y.}, \bibinfo{author}{Li, X.},
  \bibinfo{author}{Scheibner, C.}, \bibinfo{author}{Vitelli, V.} \&
  \bibinfo{author}{Huang, G.}
\newblock \bibinfo{title}{Realization of active metamaterials with odd
  micropolar elasticity}.
\newblock \emph{\bibinfo{journal}{Nature Communications}}
  \textbf{\bibinfo{volume}{12}}, \bibinfo{pages}{5935} (\bibinfo{year}{2021}).

\bibitem{brandenbourger2021active}
\bibinfo{author}{Brandenbourger, M.}, \bibinfo{author}{Scheibner, C.},
  \bibinfo{author}{Veenstra, J.}, \bibinfo{author}{Vitelli, V.} \&
  \bibinfo{author}{Coulais, C.}
\newblock \bibinfo{title}{Active impact and locomotion in robotic matter with
  nonlinear work cycles}.
\newblock \emph{\bibinfo{journal}{arXiv:2108.08837}}  (\bibinfo{year}{2021}).

\bibitem{Wang2015}
\bibinfo{author}{Wang, P.}, \bibinfo{author}{Lu, L.} \&
  \bibinfo{author}{Bertoldi, K.}
\newblock \bibinfo{title}{{Topological Phononic Crystals with One-Way Elastic
  Edge Waves.}}
\newblock \emph{\bibinfo{journal}{Physical review letters}}
  \textbf{\bibinfo{volume}{115}}, \bibinfo{pages}{104302}
  (\bibinfo{year}{2015}).

\bibitem{zhao2020gyroscopic}
\bibinfo{author}{Zhao, Y.}, \bibinfo{author}{Zhou, X.} \&
  \bibinfo{author}{Huang, G.}
\newblock \bibinfo{title}{Non-reciprocal rayleigh waves in elastic gyroscopic
  medium}.
\newblock \emph{\bibinfo{journal}{Journal of the Mechanics and Physics of
  Solids}} \textbf{\bibinfo{volume}{143}}, \bibinfo{pages}{104065}
  (\bibinfo{year}{2020}).

\bibitem{Carta2014dispersion}
\bibinfo{author}{Carta, G.}, \bibinfo{author}{Brun, M.},
  \bibinfo{author}{Movchan, A.}, \bibinfo{author}{Movchan, N.} \&
  \bibinfo{author}{Jones, I.}
\newblock \bibinfo{title}{Dispersion properties of vortex-type monatomic
  lattices}.
\newblock \emph{\bibinfo{journal}{International Journal of Solids and
  Structures}} \textbf{\bibinfo{volume}{51}}, \bibinfo{pages}{2213 -- 2225}
  (\bibinfo{year}{2014}).

\bibitem{Hassanpour2014dynamics}
\bibinfo{author}{Hassanpour, S.}
\newblock \bibinfo{title}{Dynamics of gyroelastic continua}
  (\bibinfo{year}{2014}).

\bibitem{Carta2017deflecting}
\bibinfo{author}{Carta, G.}, \bibinfo{author}{Jones, I.~S.},
  \bibinfo{author}{Movchan, N.~V.}, \bibinfo{author}{Movchan, A.~B.} \&
  \bibinfo{author}{Nieves, M.~J.}
\newblock \bibinfo{title}{``deflecting elastic prism''and unidirectional
  localisation for waves in chiral elastic systems}.
\newblock \emph{\bibinfo{journal}{Scientific Reports}}
  \textbf{\bibinfo{volume}{7}}, \bibinfo{pages}{26} (\bibinfo{year}{2017}).

\bibitem{Nash2015}
\bibinfo{author}{Nash, L.~M.} \emph{et~al.}
\newblock \bibinfo{title}{{Topological mechanics of gyroscopic metamaterials.}}
\newblock \emph{\bibinfo{journal}{Proc. Natl. Acad. Sci. USA}}
  \textbf{\bibinfo{volume}{112}}, \bibinfo{pages}{14495--500}
  (\bibinfo{year}{2015}).

\bibitem{Mitchell2018}
\bibinfo{author}{Mitchell, N.~P.}, \bibinfo{author}{Nash, L.~M.} \&
  \bibinfo{author}{Irvine, W. T.~M.}
\newblock \bibinfo{title}{Tunable band topology in gyroscopic lattices}.
\newblock \emph{\bibinfo{journal}{Phys. Rev. B}} \textbf{\bibinfo{volume}{98}},
  \bibinfo{pages}{174301} (\bibinfo{year}{2018}).

\bibitem{Mitchell2018Realization}
\bibinfo{author}{Mitchell, N.~P.}, \bibinfo{author}{Nash, L.~M.} \&
  \bibinfo{author}{Irvine, W. T.~M.}
\newblock \bibinfo{title}{Realization of a topological phase transition in a
  gyroscopic lattice}.
\newblock \emph{\bibinfo{journal}{Phys. Rev. B}} \textbf{\bibinfo{volume}{97}},
  \bibinfo{pages}{100302} (\bibinfo{year}{2018}).

\bibitem{Mitchell2018Nature}
\bibinfo{author}{Mitchell, N.~P.}, \bibinfo{author}{Nash, L.~M.},
  \bibinfo{author}{Hexner, D.}, \bibinfo{author}{Turner, A.~M.} \&
  \bibinfo{author}{Irvine, W. T.~M.}
\newblock \bibinfo{title}{Amorphous topological insulators constructed from
  random point sets}.
\newblock \emph{\bibinfo{journal}{Nature Physics}}
  \textbf{\bibinfo{volume}{14}}, \bibinfo{pages}{380--385}
  (\bibinfo{year}{2018}).

\bibitem{Brun2012vortex}
\bibinfo{author}{Brun, M.}, \bibinfo{author}{Jones, I.~S.} \&
  \bibinfo{author}{Movchan, A.~B.}
\newblock \bibinfo{title}{Vortex-type elastic structured media and dynamic
  shielding}.
\newblock \emph{\bibinfo{journal}{Proceedings of the Royal Society A:
  Mathematical, Physical and Engineering Sciences}}
  \textbf{\bibinfo{volume}{468}}, \bibinfo{pages}{3027--3046}
  (\bibinfo{year}{2012}).

\bibitem{Sonin1987vortex}
\bibinfo{author}{Sonin, E.~B.}
\newblock \bibinfo{title}{Vortex oscillations and hydrodynamics of rotating
  superfluids}.
\newblock \emph{\bibinfo{journal}{Rev. Mod. Phys.}}
  \textbf{\bibinfo{volume}{59}}, \bibinfo{pages}{87--155}
  (\bibinfo{year}{1987}).

\bibitem{Gifford2008dislocation}
\bibinfo{author}{Gifford, S.~A.} \& \bibinfo{author}{Baym, G.}
\newblock \bibinfo{title}{Dislocation-mediated melting in superfluid vortex
  lattices}.
\newblock \emph{\bibinfo{journal}{Phys. Rev. A}} \textbf{\bibinfo{volume}{78}},
  \bibinfo{pages}{043607} (\bibinfo{year}{2008}).

\bibitem{Nguyen2020Fracton}
\bibinfo{author}{Nguyen, D.~X.}, \bibinfo{author}{Gromov, A.} \&
  \bibinfo{author}{Moroz, S.}
\newblock \bibinfo{title}{{Fracton-elasticity duality of two-dimensional
  superfluid vortex crystals: defect interactions and quantum melting}}.
\newblock \emph{\bibinfo{journal}{SciPost Phys.}} \textbf{\bibinfo{volume}{9}},
  \bibinfo{pages}{76} (\bibinfo{year}{2020}).

\bibitem{Moroz2018effective}
\bibinfo{author}{Moroz, S.}, \bibinfo{author}{Hoyos, C.},
  \bibinfo{author}{Benzoni, C.} \& \bibinfo{author}{Son, D.~T.}
\newblock \bibinfo{title}{{Effective field theory of a vortex lattice in a
  bosonic superfluid}}.
\newblock \emph{\bibinfo{journal}{SciPost Phys.}} \textbf{\bibinfo{volume}{5}},
  \bibinfo{pages}{39} (\bibinfo{year}{2018}).

\bibitem{Fetter2009rotating}
\bibinfo{author}{Fetter, A.~L.}
\newblock \bibinfo{title}{Rotating trapped bose-einstein condensates}.
\newblock \emph{\bibinfo{journal}{Rev. Mod. Phys.}}
  \textbf{\bibinfo{volume}{81}}, \bibinfo{pages}{647--691}
  (\bibinfo{year}{2009}).

\bibitem{Blatter1994vortices}
\bibinfo{author}{Blatter, G.}, \bibinfo{author}{Feigel'man, M.~V.},
  \bibinfo{author}{Geshkenbein, V.~B.}, \bibinfo{author}{Larkin, A.~I.} \&
  \bibinfo{author}{Vinokur, V.~M.}
\newblock \bibinfo{title}{Vortices in high-temperature superconductors}.
\newblock \emph{\bibinfo{journal}{Rev. Mod. Phys.}}
  \textbf{\bibinfo{volume}{66}}, \bibinfo{pages}{1125--1388}
  (\bibinfo{year}{1994}).

\bibitem{Tkachenko1968elasticity}
\bibinfo{author}{Tkachenko, V.~K.}
\newblock \bibinfo{title}{Elasticity of vortex lattices}.
\newblock \emph{\bibinfo{journal}{JETP}} \textbf{\bibinfo{volume}{29}},
  \bibinfo{pages}{945} (\bibinfo{year}{1969}).

\bibitem{tkachenko1966vortex}
\bibinfo{author}{Tkachenko, V.}
\newblock \bibinfo{title}{On vortex lattices}.
\newblock \emph{\bibinfo{journal}{Sov. Phys. JETP}}
  \textbf{\bibinfo{volume}{22}}, \bibinfo{pages}{1282--1286}
  (\bibinfo{year}{1966}).

\bibitem{tkachenko1966stability}
\bibinfo{author}{Tkachenko, V.}
\newblock \bibinfo{title}{Stability of vortex lattices}.
\newblock \emph{\bibinfo{journal}{Sov. Phys. JETP}}
  \textbf{\bibinfo{volume}{23}}, \bibinfo{pages}{1049--1056}
  (\bibinfo{year}{1966}).

\bibitem{Benzoni2021Rayleigh}
\bibinfo{author}{Benzoni, C.}, \bibinfo{author}{Jeevanesan, B.} \&
  \bibinfo{author}{Moroz, S.}
\newblock \bibinfo{title}{Rayleigh edge waves in two-dimensional crystals with
  lorentz forces: From skyrmion crystals to gyroscopic media}.
\newblock \emph{\bibinfo{journal}{Phys. Rev. B}}
  \textbf{\bibinfo{volume}{104}}, \bibinfo{pages}{024435}
  (\bibinfo{year}{2021}).

\bibitem{Huang2020melting}
\bibinfo{author}{Huang, P.} \emph{et~al.}
\newblock \bibinfo{title}{Melting of a skyrmion lattice to a skyrmion liquid
  via a hexatic phase}.
\newblock \emph{\bibinfo{journal}{Nature Nanotechnology}}
  \textbf{\bibinfo{volume}{15}}, \bibinfo{pages}{761--767}
  (\bibinfo{year}{2020}).

\bibitem{Ochoa2017Gyrotropic}
\bibinfo{author}{Ochoa, H.}, \bibinfo{author}{Kim, S.~K.},
  \bibinfo{author}{Tchernyshyov, O.} \& \bibinfo{author}{Tserkovnyak, Y.}
\newblock \bibinfo{title}{Gyrotropic elastic response of skyrmion crystals to
  current-induced tensions}.
\newblock \emph{\bibinfo{journal}{Phys. Rev. B}} \textbf{\bibinfo{volume}{96}},
  \bibinfo{pages}{020410} (\bibinfo{year}{2017}).

\bibitem{Muhlbauer2009skyrmion}
\bibinfo{author}{M{\"u}hlbauer, S.} \emph{et~al.}
\newblock \bibinfo{title}{Skyrmion lattice in a chiral magnet}.
\newblock \emph{\bibinfo{journal}{Science}} \textbf{\bibinfo{volume}{323}},
  \bibinfo{pages}{915--919} (\bibinfo{year}{2009}).

\bibitem{Yu2010Real}
\bibinfo{author}{Yu, X.~Z.} \emph{et~al.}
\newblock \bibinfo{title}{Real-space observation of a two-dimensional skyrmion
  crystal}.
\newblock \emph{\bibinfo{journal}{Nature}} \textbf{\bibinfo{volume}{465}},
  \bibinfo{pages}{901--904} (\bibinfo{year}{2010}).

\bibitem{Brearton2021Magnetic}
\bibinfo{author}{Brearton, R.} \emph{et~al.}
\newblock \bibinfo{title}{Deriving the skyrmion hall angle from skyrmion
  lattice dynamics}.
\newblock \emph{\bibinfo{journal}{Nature Communications}}
  \textbf{\bibinfo{volume}{12}}, \bibinfo{pages}{2723} (\bibinfo{year}{2021}).

\bibitem{Grzybowski2000Dynamic}
\bibinfo{author}{Grzybowski, B.~A.}, \bibinfo{author}{Stone, H.~A.} \&
  \bibinfo{author}{Whitesides, G.~M.}
\newblock \bibinfo{title}{Dynamic self-assembly of magnetized, millimetre-sized
  objects rotating at a liquid--air interface}.
\newblock \emph{\bibinfo{journal}{Nature}} \textbf{\bibinfo{volume}{405}},
  \bibinfo{pages}{1033--1036} (\bibinfo{year}{2000}).

\bibitem{Han2021Fluctuating}
\bibinfo{author}{Han, M.} \emph{et~al.}
\newblock \bibinfo{title}{Fluctuating hydrodynamics of chiral active fluids}.
\newblock \emph{\bibinfo{journal}{Nature Physics}}
  \textbf{\bibinfo{volume}{17}}, \bibinfo{pages}{1260--1269}
  (\bibinfo{year}{2021}).

\bibitem{Yeo2015Collective}
\bibinfo{author}{Yeo, K.}, \bibinfo{author}{Lushi, E.} \&
  \bibinfo{author}{Vlahovska, P.~M.}
\newblock \bibinfo{title}{Collective dynamics in a binary mixture of
  hydrodynamically coupled microrotors}.
\newblock \emph{\bibinfo{journal}{Phys. Rev. Lett.}}
  \textbf{\bibinfo{volume}{114}}, \bibinfo{pages}{188301}
  (\bibinfo{year}{2015}).

\bibitem{Scholz2018Rotating}
\bibinfo{author}{Scholz, C.}, \bibinfo{author}{Engel, M.} \&
  \bibinfo{author}{P{\"o}schel, T.}
\newblock \bibinfo{title}{Rotating robots move collectively and self-organize}.
\newblock \emph{\bibinfo{journal}{Nature Communications}}
  \textbf{\bibinfo{volume}{9}}, \bibinfo{pages}{931} (\bibinfo{year}{2018}).

\bibitem{Goldman1967Brenner}
\bibinfo{author}{Goldman, A.}, \bibinfo{author}{Cox, R.} \&
  \bibinfo{author}{Brenner, H.}
\newblock \bibinfo{title}{Slow viscous motion of a sphere parallel to a plane
  wall---i motion through a quiescent fluid}.
\newblock \emph{\bibinfo{journal}{Chemical Engineering Science}}
  \textbf{\bibinfo{volume}{22}}, \bibinfo{pages}{637--651}
  (\bibinfo{year}{1967}).

\bibitem{Aragones2016}
\bibinfo{author}{Aragones, J.~L.}, \bibinfo{author}{Steimel, J.~P.} \&
  \bibinfo{author}{Alexander-Katz, A.}
\newblock \bibinfo{title}{{Elasticity-induced force reversal between active
  spinning particles in dense passive media.}}
\newblock \emph{\bibinfo{journal}{Nat Commun}} \textbf{\bibinfo{volume}{7}},
  \bibinfo{pages}{11325} (\bibinfo{year}{2016}).
\newblock \eprint{1512.02562}.

\bibitem{Note1}
\bibinfo{note}{While the quantity $P$ corresponds to the power exerted by the
  interaction forces in purely mechanical systems, this physical interpretation
  does not carry over in gyroscopic or vortex crystals~\cite
  {scheibner2020non}}.

\bibitem{Truesdell1963Meaning}
\bibinfo{author}{Truesdell, C.~A.}
\newblock \bibinfo{title}{The meaning of betti's reciprocal theorem}.
\newblock \emph{\bibinfo{journal}{Journal of Research of the National Bureau of
  Standards Section B Mathematics and Mathematical Physics}}
  \bibinfo{pages}{85} (\bibinfo{year}{1963}).

\bibitem{zubov2008nonlinear}
\bibinfo{author}{Zubov, L.}
\newblock \emph{\bibinfo{title}{Nonlinear Theory of Dislocations and
  Disclinations in Elastic Bodies}}.
\newblock Lecture Notes in Physics Monographs (\bibinfo{publisher}{Springer
  Berlin Heidelberg}, \bibinfo{year}{2008}).

\bibitem{Storm2005Nonlinear}
\bibinfo{author}{Storm, C.}, \bibinfo{author}{Pastore, J.~J.},
  \bibinfo{author}{MacKintosh, F.~C.}, \bibinfo{author}{Lubensky, T.~C.} \&
  \bibinfo{author}{Janmey, P.~A.}
\newblock \bibinfo{title}{Nonlinear elasticity in biological gels}.
\newblock \emph{\bibinfo{journal}{Nature}} \textbf{\bibinfo{volume}{435}},
  \bibinfo{pages}{191--194} (\bibinfo{year}{2005}).

\bibitem{Lubensky2002Symmetries}
\bibinfo{author}{Lubensky, T.~C.}, \bibinfo{author}{Mukhopadhyay, R.},
  \bibinfo{author}{Radzihovsky, L.} \& \bibinfo{author}{Xing, X.}
\newblock \bibinfo{title}{Symmetries and elasticity of nematic gels}.
\newblock \emph{\bibinfo{journal}{Phys. Rev. E}} \textbf{\bibinfo{volume}{66}},
  \bibinfo{pages}{011702} (\bibinfo{year}{2002}).

\bibitem{antman2013non}
\bibinfo{author}{Antman, S.}, \bibinfo{author}{Truesdell, C.} \&
  \bibinfo{author}{Noll, W.}
\newblock \emph{\bibinfo{title}{The Non-Linear Field Theories of Mechanics}}
  (\bibinfo{publisher}{Springer Berlin Heidelberg}, \bibinfo{year}{2013}).

\bibitem{Ogden2001nonlinear}
\bibinfo{editor}{Ogden, R.~W.} (ed.) \emph{\bibinfo{title}{Nonlinear elasticity
  : theory and applications}}.
\newblock London Mathematical Society lecture note series.
  (\bibinfo{publisher}{Cambridge University Press},
  \bibinfo{address}{Cambridge, U.K. ; New York}, \bibinfo{year}{2001}).

\bibitem{Note0}
\bibinfo{note}{See the Supplemental Material for additional discussion,
  including Refs.~\cite
  {Born1954dynamical,Fruchart2020Symmetries,Nelson1987,Anderson2008,Glaser2015,Jones1924,Kim2013microhydrodynamics,Yukawa1935,yang2012generalized}}.

\bibitem{Landau7}
\bibinfo{author}{Landau, L.} \emph{et~al.}
\newblock \emph{\bibinfo{title}{Theory of Elasticity}}.
\newblock Course of theoretical physics (\bibinfo{publisher}{Elsevier Science},
  \bibinfo{year}{1986}).

\bibitem{Born1954dynamical}
\bibinfo{author}{Born, M.}
\newblock \emph{\bibinfo{title}{Dynamical theory of crystal lattices}}.
\newblock The International series of monographs on physics
  (\bibinfo{publisher}{Clarendon Press}, \bibinfo{address}{Oxford},
  \bibinfo{year}{1954}).

\bibitem{Fruchart2020Symmetries}
\bibinfo{author}{Fruchart, M.} \& \bibinfo{author}{Vitelli, V.}
\newblock \bibinfo{title}{Symmetries and dualities in the theory of
  elasticity}.
\newblock \emph{\bibinfo{journal}{Phys. Rev. Lett.}}
  \textbf{\bibinfo{volume}{124}}, \bibinfo{pages}{248001}
  (\bibinfo{year}{2020}).

\bibitem{Nelson1987}
\bibinfo{author}{{Nelson, D.R.}} \& \bibinfo{author}{{Peliti, L.}}
\newblock \bibinfo{title}{Fluctuations in membranes with crystalline and
  hexatic order}.
\newblock \emph{\bibinfo{journal}{J. Phys. France}}
  \textbf{\bibinfo{volume}{48}}, \bibinfo{pages}{1085--1092}
  (\bibinfo{year}{1987}).

\bibitem{Anderson2008}
\bibinfo{author}{Anderson, J.~A.}, \bibinfo{author}{Lorenz, C.~D.} \&
  \bibinfo{author}{Travesset, A.}
\newblock \bibinfo{title}{General purpose molecular dynamics simulations fully
  implemented on graphics processing units}.
\newblock \emph{\bibinfo{journal}{Journal of Computational Physics}}
  \textbf{\bibinfo{volume}{227}}, \bibinfo{pages}{5342--5359}
  (\bibinfo{year}{2008}).

\bibitem{Glaser2015}
\bibinfo{author}{Glaser, J.} \emph{et~al.}
\newblock \bibinfo{title}{Strong scaling of general-purpose molecular dynamics
  simulations on {GPUs}}.
\newblock \emph{\bibinfo{journal}{Computer Physics Communications}}
  \textbf{\bibinfo{volume}{192}}, \bibinfo{pages}{97--107}
  (\bibinfo{year}{2015}).

\bibitem{Jones1924}
\bibinfo{author}{Jones, J.~E.}
\newblock \bibinfo{title}{On the determination of molecular fields.—i. from
  the variation of the viscosity of a gas with temperature}.
\newblock \emph{\bibinfo{journal}{Proceedings of the Royal Society of London.
  Series A, Containing Papers of a Mathematical and Physical Character}}
  \textbf{\bibinfo{volume}{106}}, \bibinfo{pages}{441--462}
  (\bibinfo{year}{1924}).

\bibitem{Kim2013microhydrodynamics}
\bibinfo{author}{Kim, S.} \& \bibinfo{author}{Karrila, S.~J.}
\newblock \emph{\bibinfo{title}{Microhydrodynamics: principles and selected
  applications}} (\bibinfo{publisher}{Courier Corporation},
  \bibinfo{year}{2013}).

\bibitem{Yukawa1935}
\bibinfo{author}{Yukawa, H.}
\newblock \bibinfo{title}{On the interaction of elementary particles. i}.
\newblock \emph{\bibinfo{journal}{Proceedings of the Physico-Mathematical
  Society of Japan. 3rd Series}} \textbf{\bibinfo{volume}{17}},
  \bibinfo{pages}{48--57} (\bibinfo{year}{1935}).

\bibitem{yang2012generalized}
\bibinfo{author}{Yang, J.~Z.}, \bibinfo{author}{Wu, X.} \& \bibinfo{author}{Li,
  X.}
\newblock \bibinfo{title}{A generalized irving--kirkwood formula for the
  calculation of stress in molecular dynamics models}.
\newblock \emph{\bibinfo{journal}{The Journal of Chemical Physics}}
  \textbf{\bibinfo{volume}{137}}, \bibinfo{pages}{134104}
  (\bibinfo{year}{2012}).

\end{thebibliography}

\end{document}